\documentclass[runningheads]{llncs}
\usepackage[T1]{fontenc}
\usepackage{graphicx}
\usepackage{caption}
\usepackage{subcaption}
\usepackage[linesnumbered,ruled]{algorithm2e}
\usepackage{sidecap}
\usepackage[switch]{lineno}

\usepackage{tikz}
\usetikzlibrary{calc,3d}
\usepackage{multirow}
\usepackage{amsmath}
\usepackage{amssymb}

\usepackage{color}
\definecolor{green}{rgb}{0,1,0}
\definecolor{blue}{rgb}{0,0,1}
\definecolor{color1}{rgb}{0,0.9655,1}
\definecolor{color2}{rgb}{1,0.4828,0.8621}
\definecolor{color3}{rgb}{1,0.8621,0.4828}
\definecolor{color4}{rgb}{0,0.5517,1}
\definecolor{color5}{rgb}{0,0.4138,0.0345}
\newif\iffull

\newcommand{\CB}{\color{black}}

\newcommand{\RNU}[1]{\uppercase\expandafter{\romannumeral #1 \relax}}
\newcommand{\RNL}[1]{\lowercase\expandafter{\romannumeral #1 \relax}}
\newcommand{\CiteFullArXiV}{~\cite{}} 
\usepackage{cases}

\fulltrue
\begin{document}
\title{Generalized Relative Neighborhood Graph (GRNG) for Similarity Search\thanks{The support of NSF award 1910530 is gratefully acknowledged.}}
%
%
%
\author{Cole Foster \and
Berk Sevilmis \and
Benjamin Kimia}
%
%
\authorrunning{C. Foster et al.}
%
%
\institute{Brown University, Providence RI 02912, USA\\
\email{\{cole\_foster,benjamin\_kimia\}@brown.edu}}
\maketitle              
%
\begin{abstract}
Similarity search is a fundamental building block for information retrieval on a variety of datasets. The notion of a neighbor is often based on binary considerations, such as the $k$ nearest neighbors. However, considering that data is often organized as a manifold with low intrinsic dimension, the notion of a neighbor must recognize higher-order relationship, to capture neighbors in all directions. Proximity graphs, such as the Relative Neighbor Graphs (RNG), use trinary relationships which capture the notion of direction and have been successfully used in a number of applications. However, the current algorithms for computing the RNG, despite widespread use, are approximate and not scalable. This paper proposes a novel type of graph, the Generalized Relative Neighborhood Graph (GRNG) for use in a pivot layer that then guides the efficient and exact construction of the RNG of a set of exemplars. It also shows how to extend this to a multi-layer hierarchy which significantly improves over the state-of-the-art methods which can only construct an approximate RNG.
\end{abstract}


\section{Introduction} \label{sec:introduction}
The vast majority of generated data in our society is now in digital form. The data representation has evolved beyond numbers and strings to complex objects. Organization and retrieval have likewise evolved from cosine similarity in vector spaces through inverted files (Google, Yahoo, Microsoft, etc.), to either embedding complex objects in Euclidean spaces or to the use of similarity metrics. The task of similarity search, namely, finding the “neighbors” of a given query based on similarity, is a fundamental building block in application  domains such as information retrieval (web search engines, e-commerce, museum collections, medical image processing), pattern recognition, data mining, machine learning, and recommendation systems.

Formally, consider the set of all objects of interest $\mathcal{X}$, hereby referred to as points, data points, or exemplars, and let $\mathcal{S} \subset \mathcal{X}$ be a dataset containing $N$ such objects. Let $d\left(x,y\right)$ denote a metric that captures the distance, or the extent of dissimilarity, between $x,y \in \mathcal{X}$. 
It is important to note that the focus of this work is search in a \emph{metric space},\textit{i.e.}, where the metric satisfies $d(x,y)=0 \Leftrightarrow x=y$, $d(x,y)=d(y,x)$, and $d(x,z) \leq d(x,y) + d(y,z)$. Some approaches first embed the metric space in a Euclidean space, such as hashing, quantization, CNN, \emph{etc.}, but this can distort the relative distances: this paper aims to define a hierarchical index structure for a metric space and use it for similarity search. 

%
\begin{figure}[b!]
    \centering
    \begin{subfigure}[b]{0.12\textwidth} 
        \includegraphics[width=\textwidth]{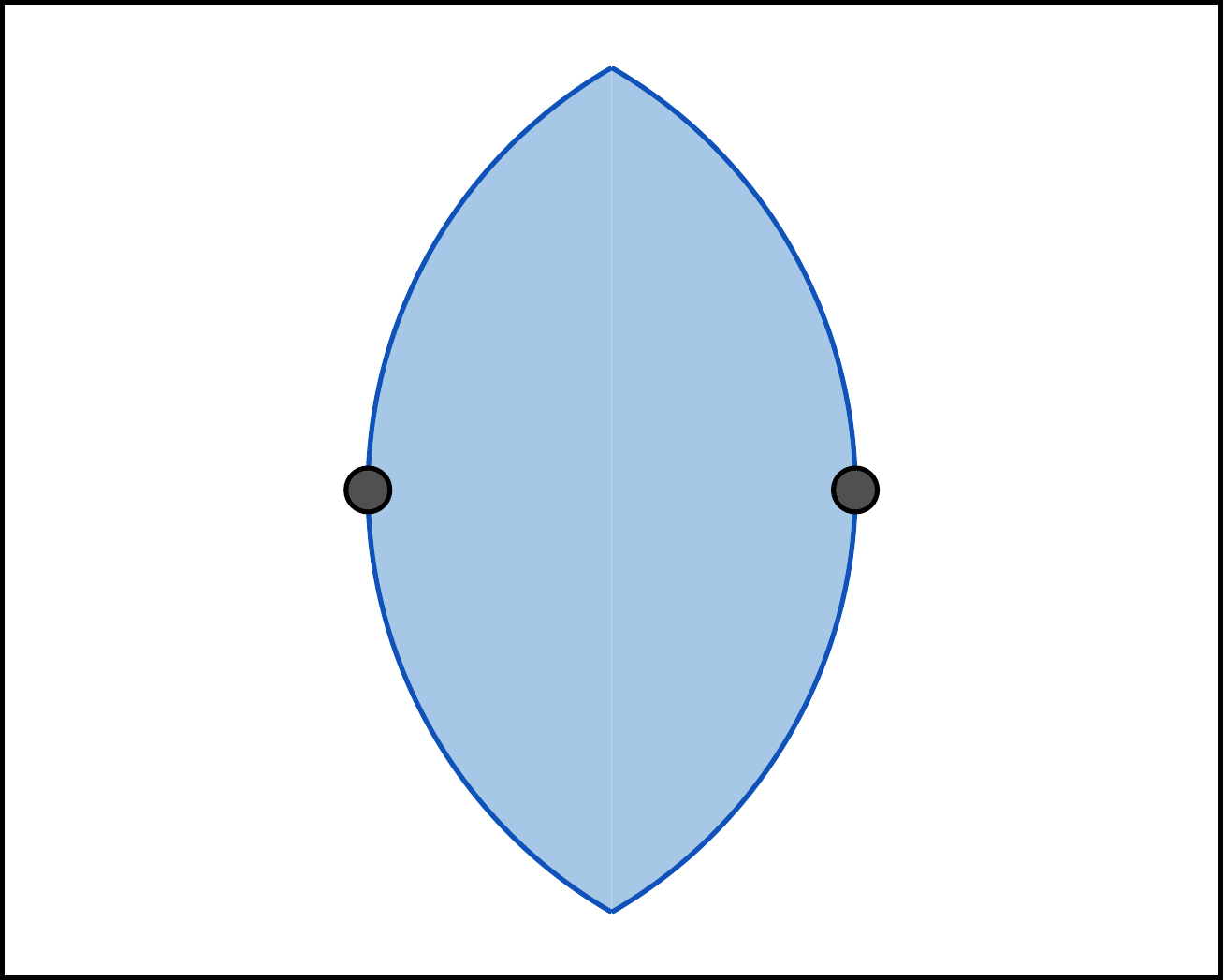}
    \end{subfigure}
    \begin{subfigure}[b]{0.13\textwidth} 
        \includegraphics[width=\textwidth]{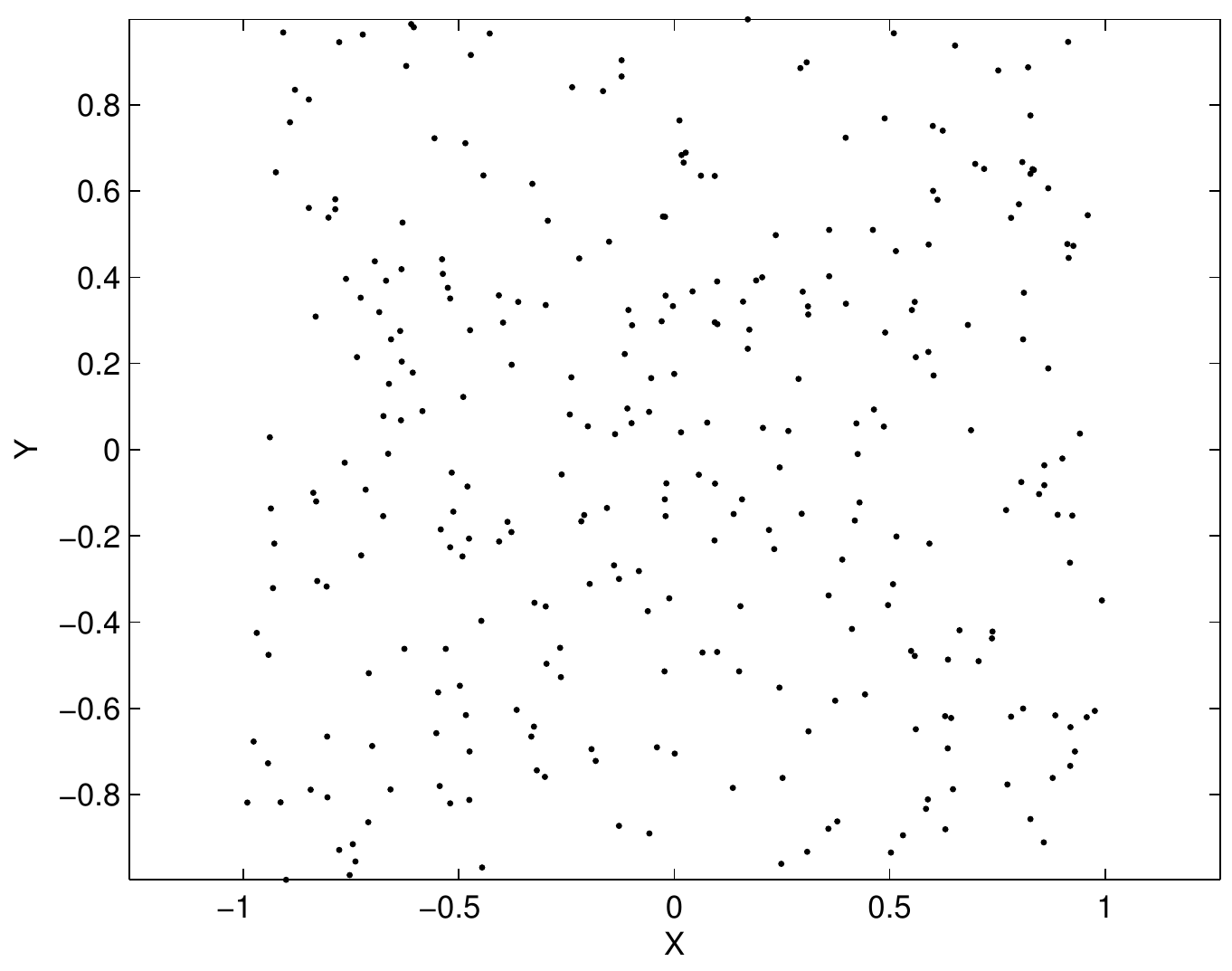}
    \end{subfigure}
    \begin{subfigure}[b]{0.13\textwidth} 
        \includegraphics[width=\textwidth]{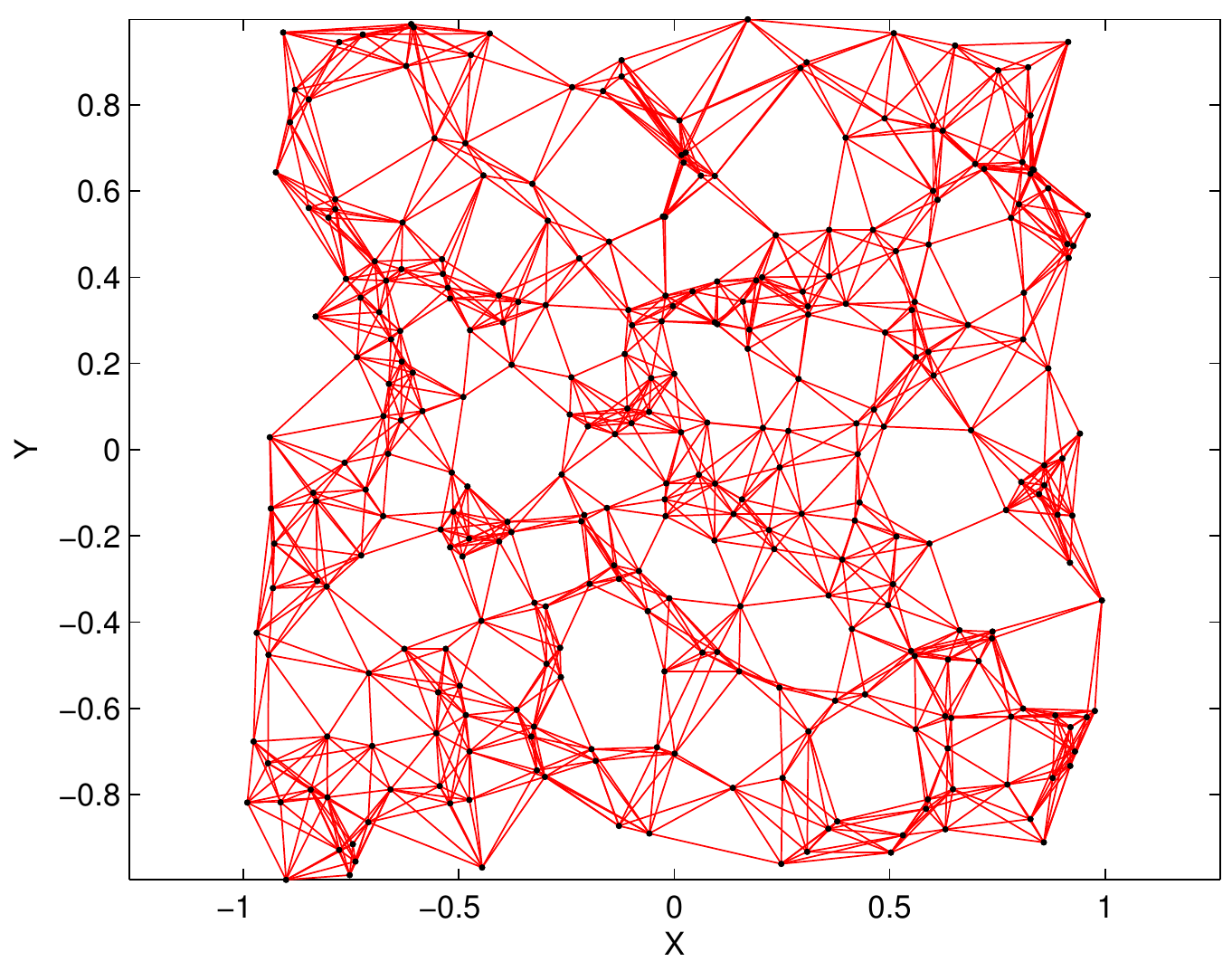}
    \end{subfigure}
    \begin{subfigure}[b]{0.13\textwidth} 
        \includegraphics[width=\textwidth]{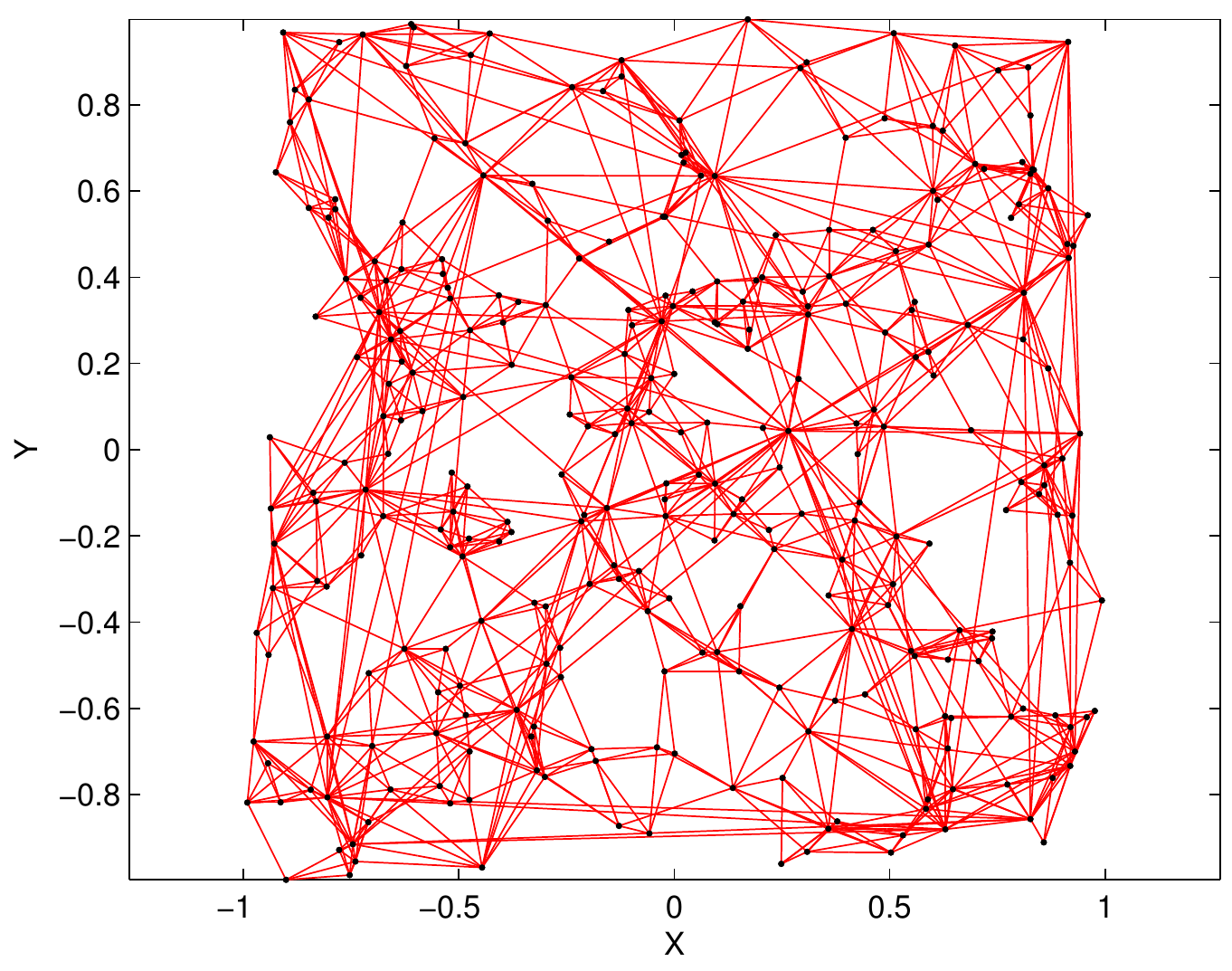}
    \end{subfigure}
    \begin{subfigure}[b]{0.13\textwidth} 
        \includegraphics[width=\textwidth]{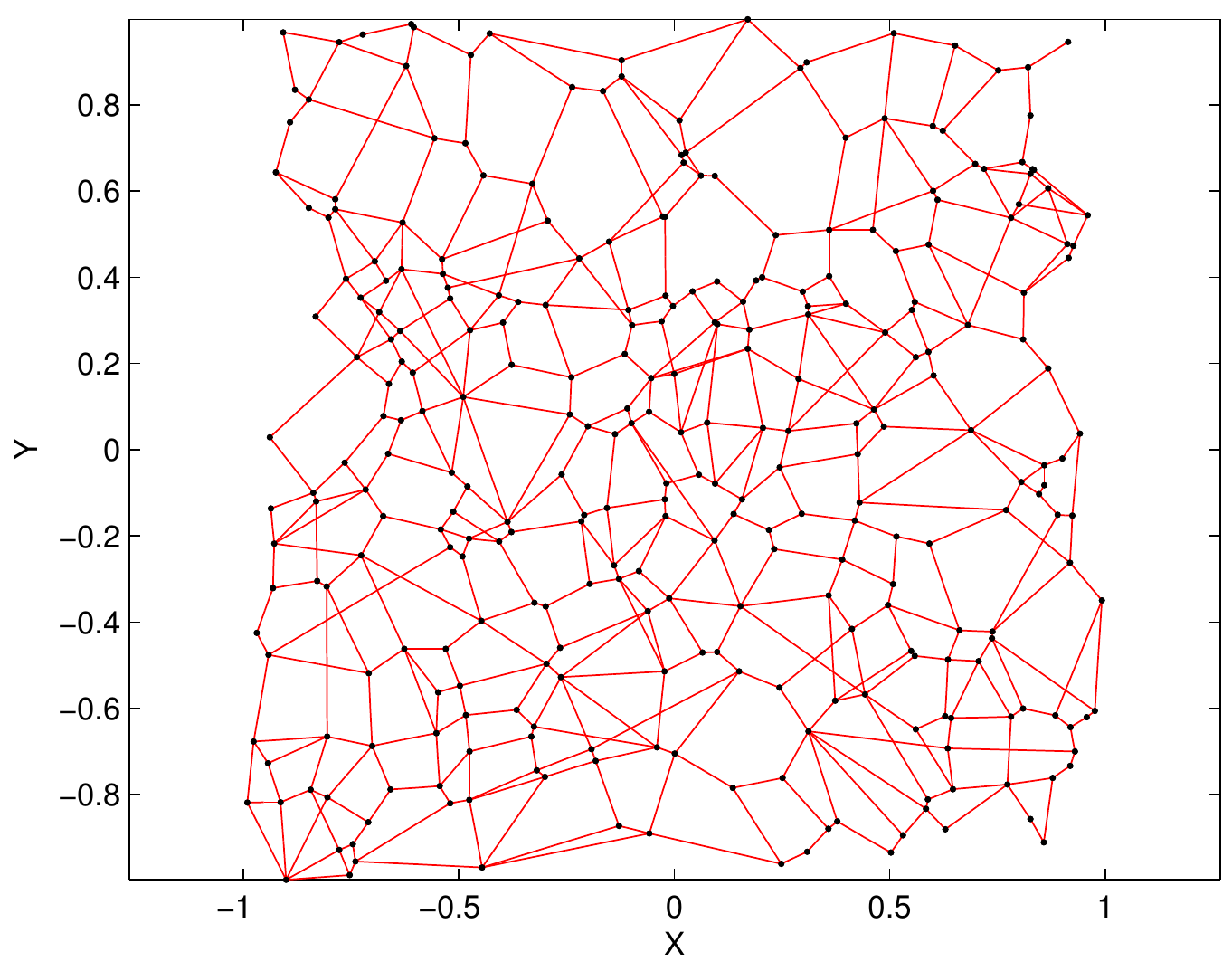}
    \end{subfigure}
    \begin{subfigure}[b]{0.13\textwidth} 
        \includegraphics[width=\textwidth]{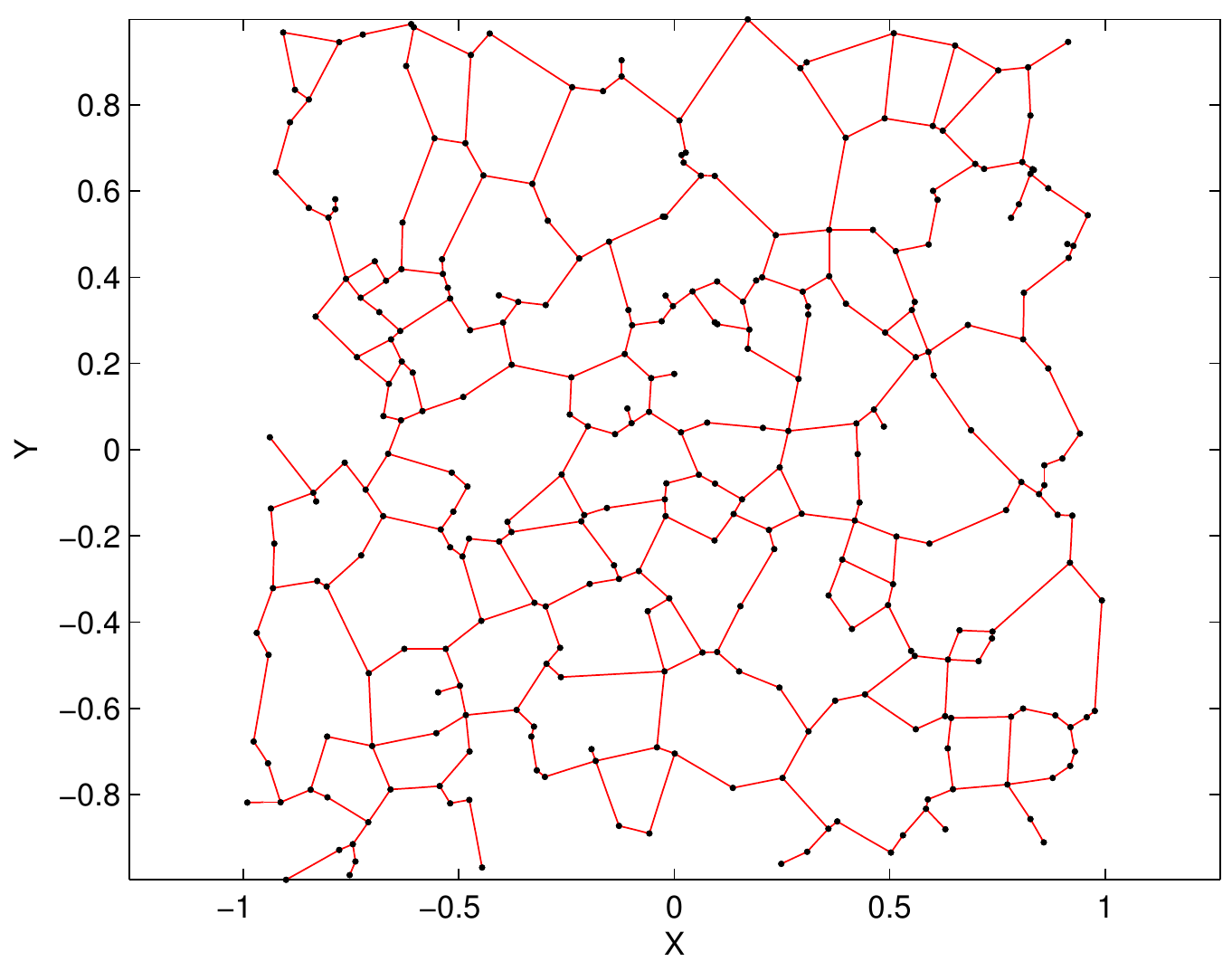}
    \end{subfigure}
    \begin{subfigure}[b]{0.13\textwidth} 
        \includegraphics[width=\textwidth]{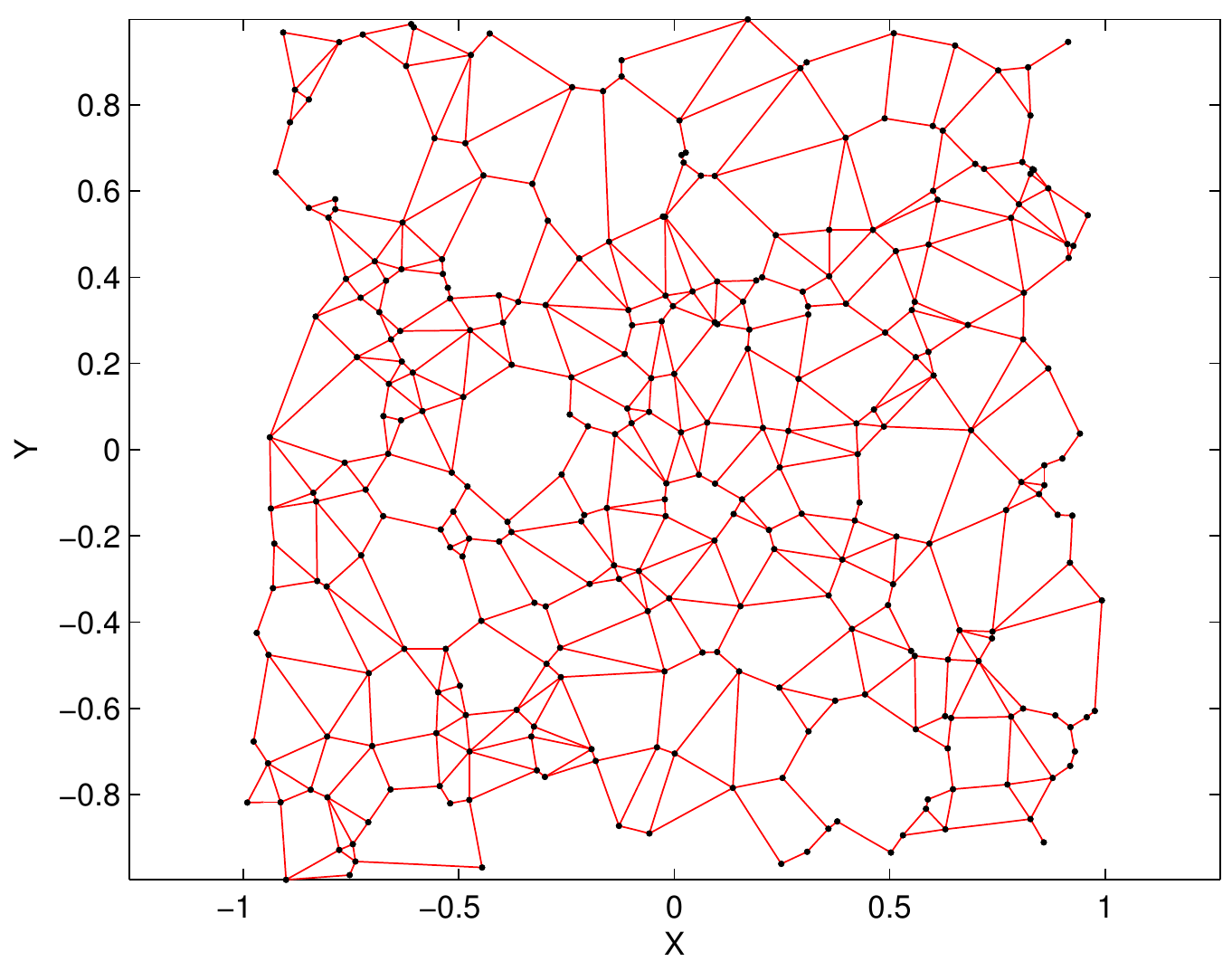}
    \end{subfigure}

    \begin{subfigure}[b]{0.12\textwidth} 
        \includegraphics[width=\textwidth]{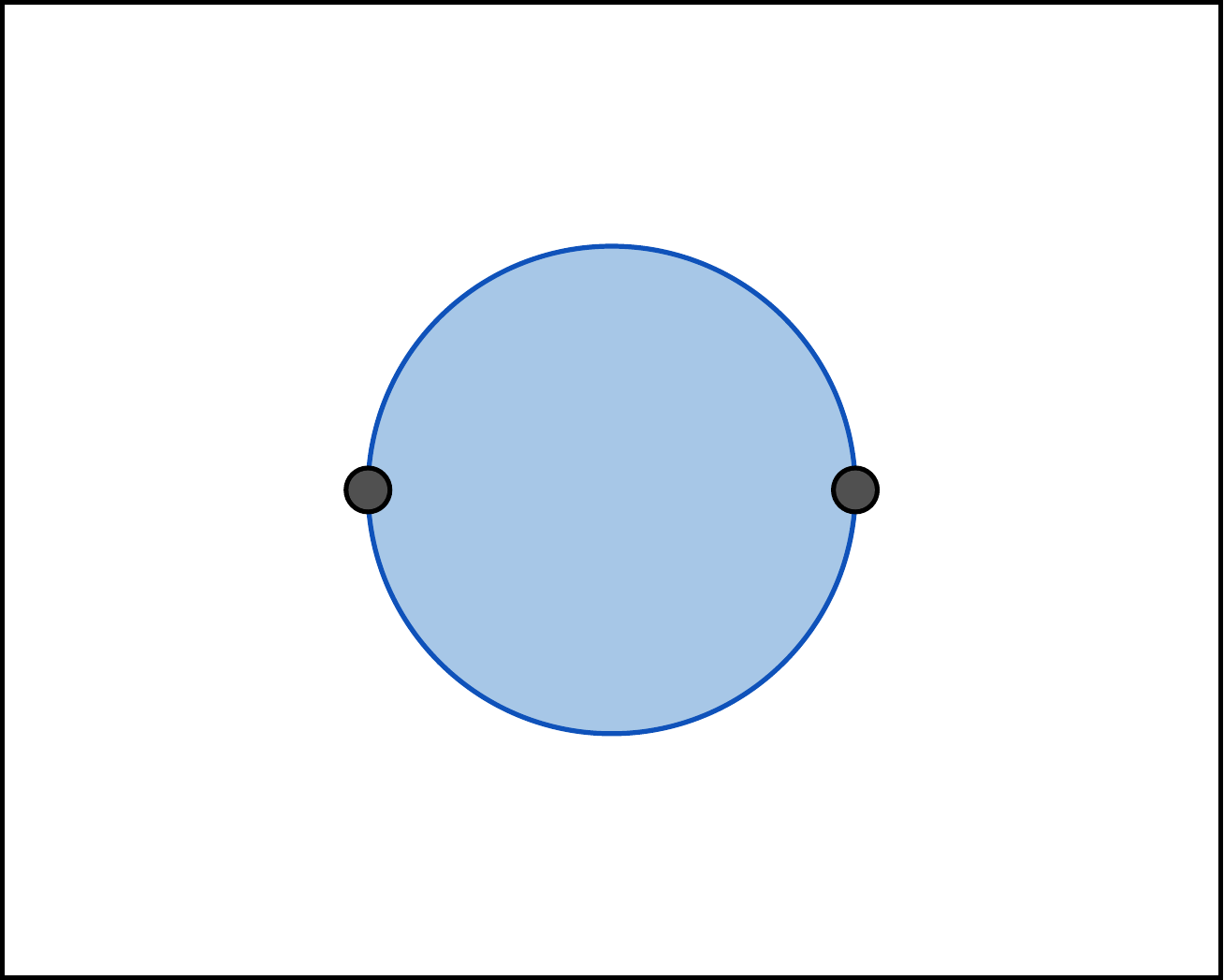}
        \caption{}
    \end{subfigure}
    \begin{subfigure}[b]{0.13\textwidth} 
        \includegraphics[width=\textwidth]{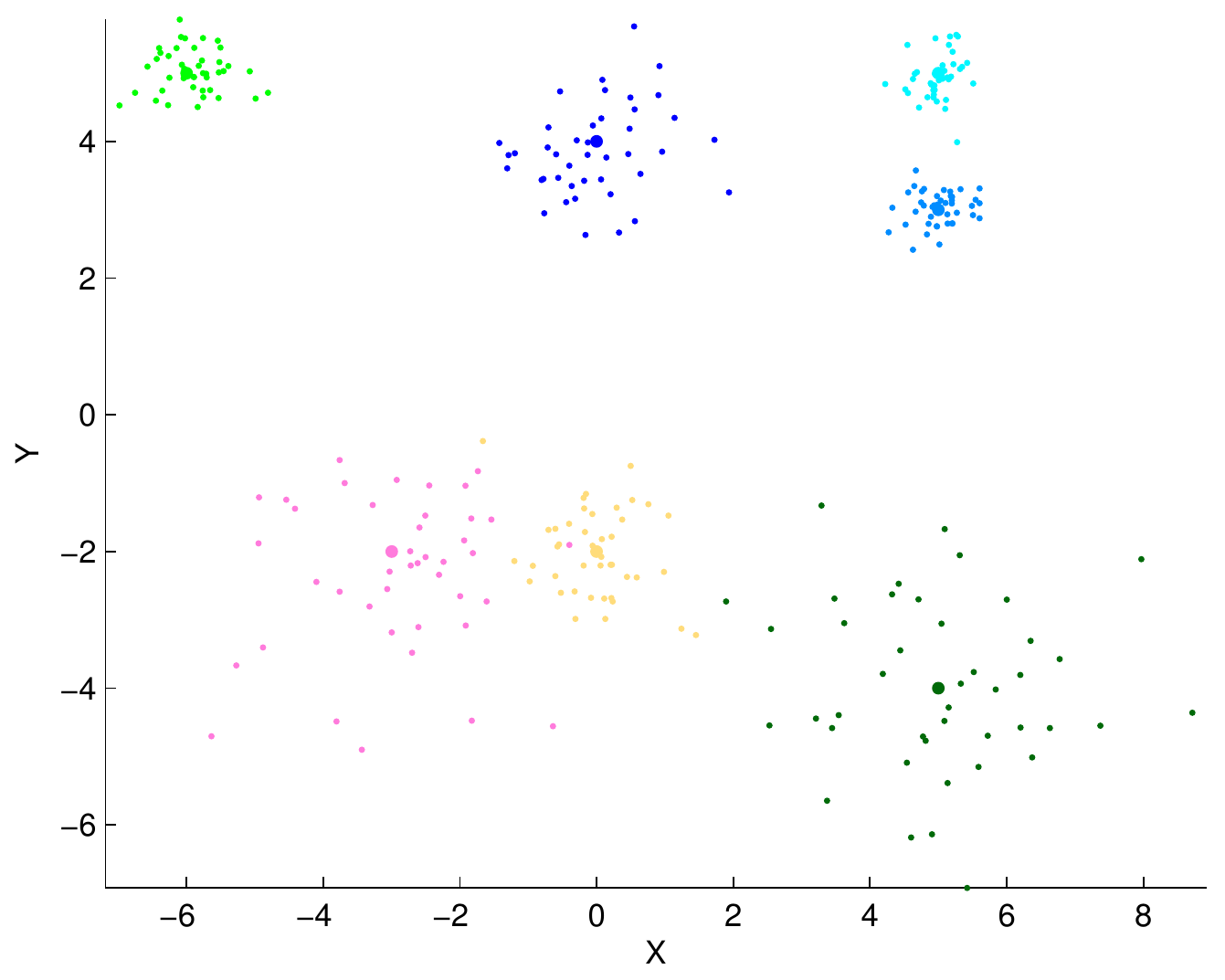}
        \caption{}
    \end{subfigure}
    \begin{subfigure}[b]{0.13\textwidth} 
        \includegraphics[width=\textwidth]{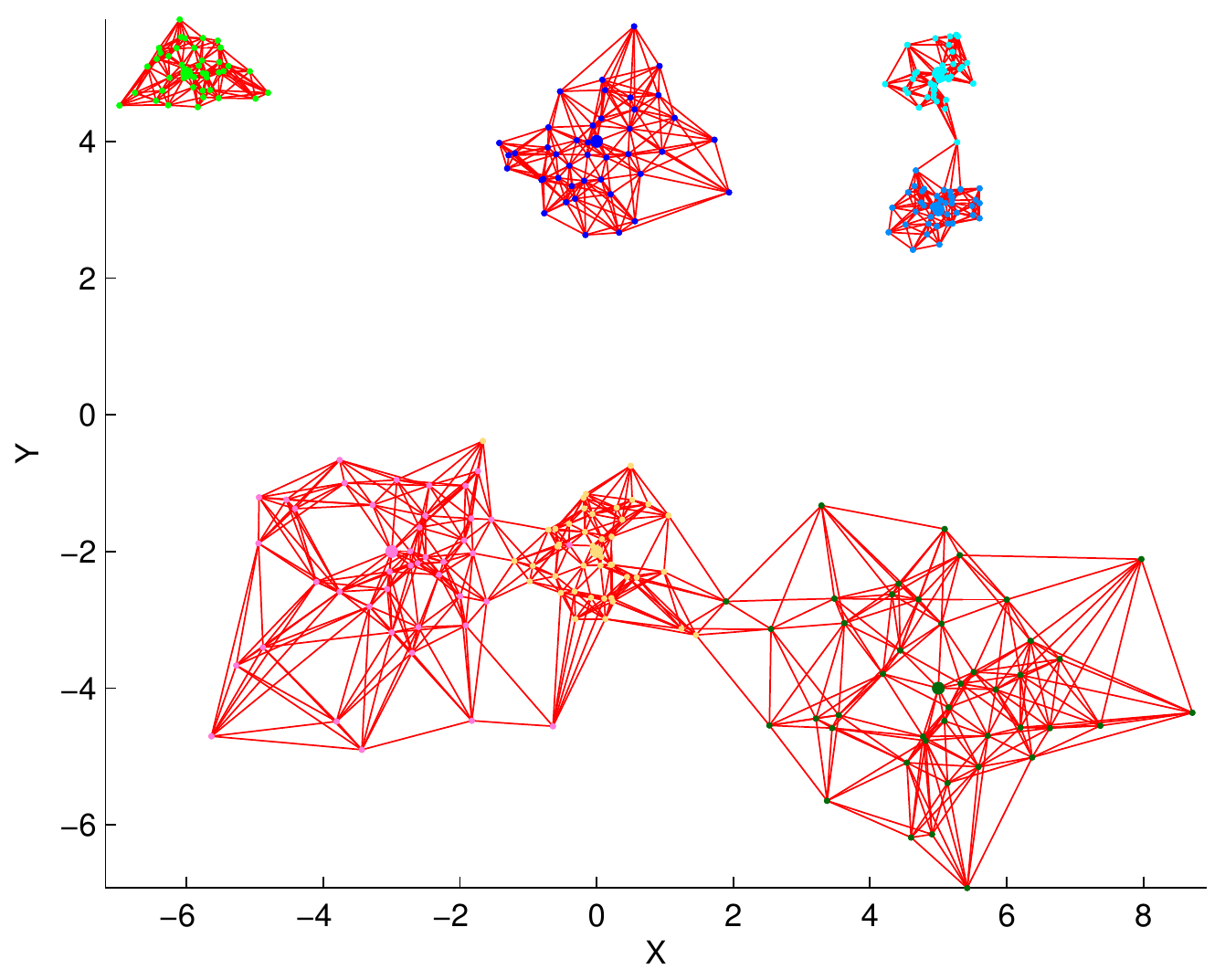}
        \caption{}
    \end{subfigure}
    \begin{subfigure}[b]{0.13\textwidth} 
        \includegraphics[width=\textwidth]{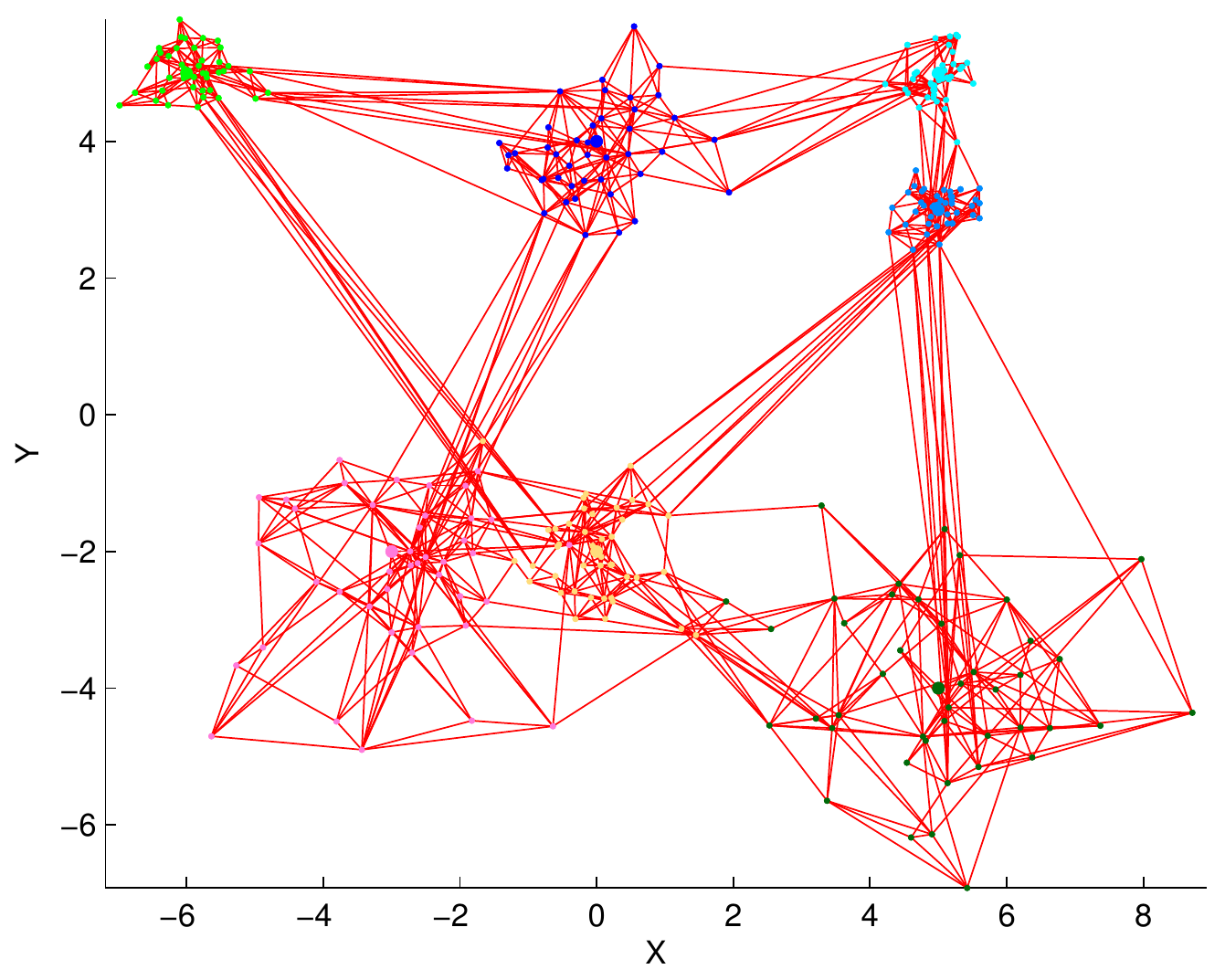}
        \caption{}
    \end{subfigure}
    \begin{subfigure}[b]{0.13\textwidth} 
        \includegraphics[width=\textwidth]{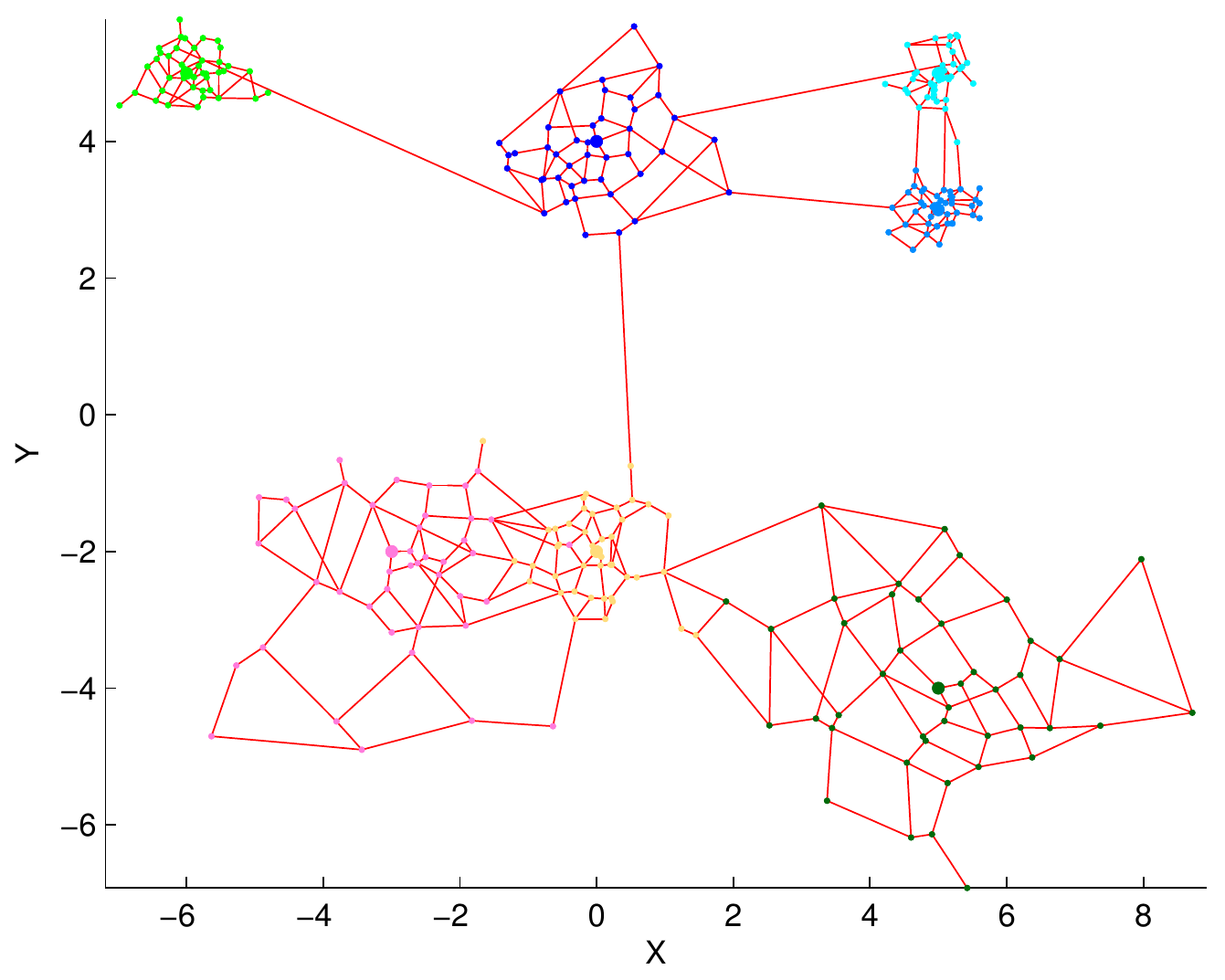}
        \caption{}
    \end{subfigure}
    \begin{subfigure}[b]{0.13\textwidth} 
        \includegraphics[width=\textwidth]{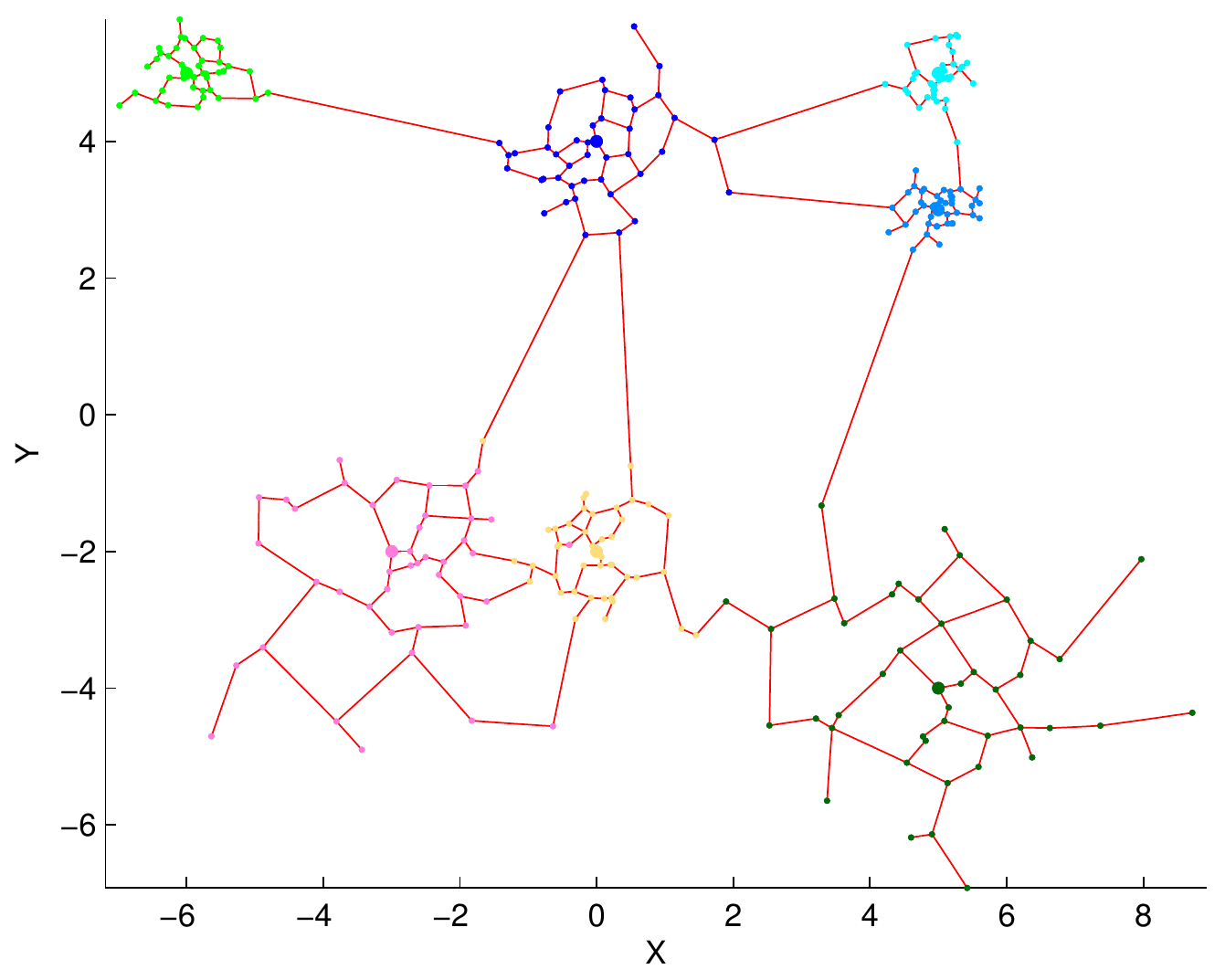}
        \caption{}
    \end{subfigure}
    \begin{subfigure}[b]{0.13\textwidth} 
        \includegraphics[width=\textwidth]{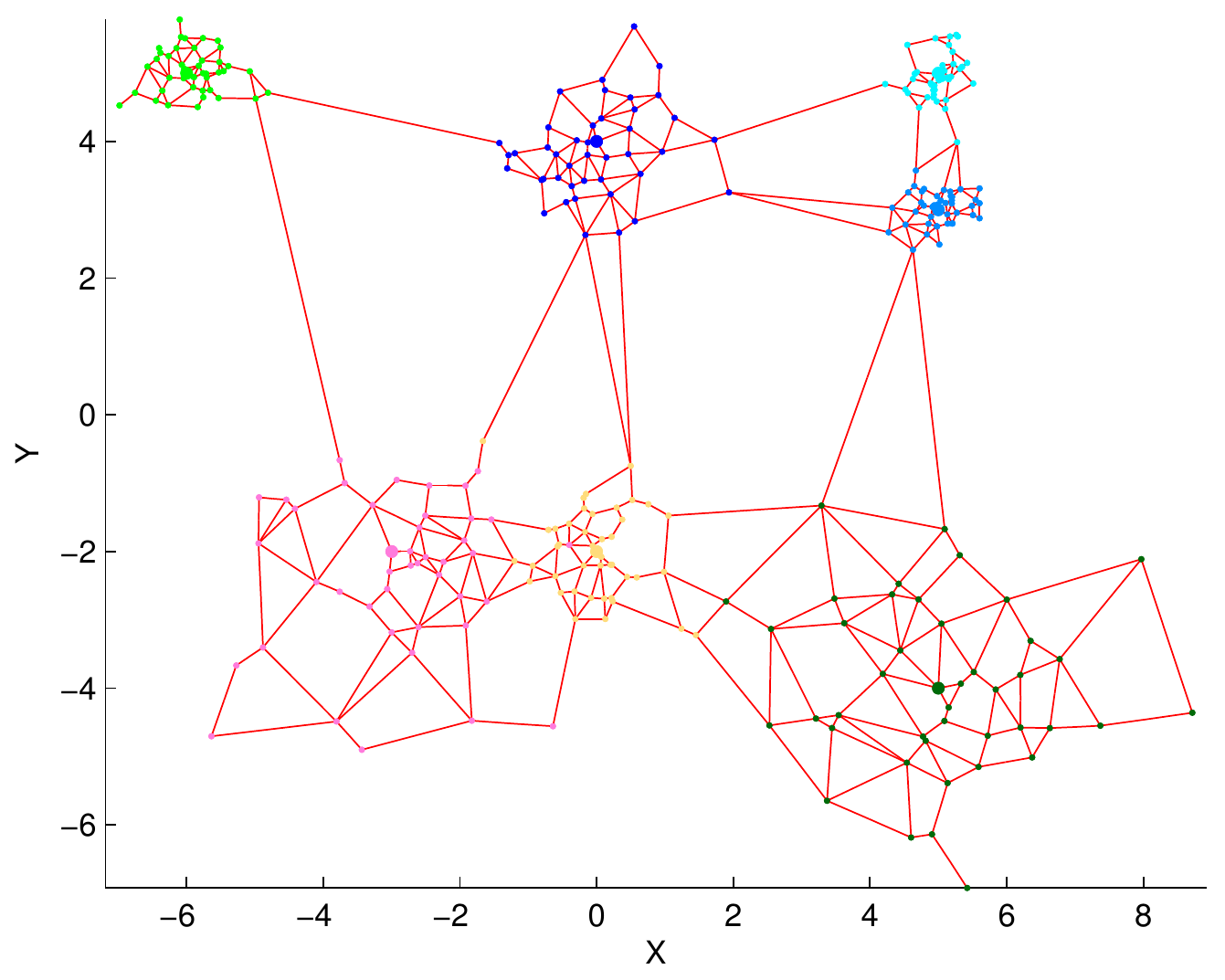}
        \caption{}
    \end{subfigure}
    
    \vspace{-0.25cm}
    \caption{\footnotesize (a) (top) Two points have an RNG connection if the "lune" between them does not contain other points. (bottom) Two points have a Gabriel Graph (GG) connection if the circle with the line segment between the points as diameter is empty. A comparison of graphs for representing both uniformly distrbuted points in $\mathcal{R}^2$ (top) and clustered data (bottom). (b) Points in 2D, (c) kNN, k=8, (d) Tellez \cite{tellez2017local} b=4, t=4, (e) NSG \cite{fu2019fast}, R=8, (f) RNG, and (f) GG. }
    \label{fig:proximityGraphExamples}
\end{figure}

\begin{figure}[t!]
    \centering
    (a)
 \begin{minipage}{0.2\textwidth} 
        \includegraphics[width=\textwidth]{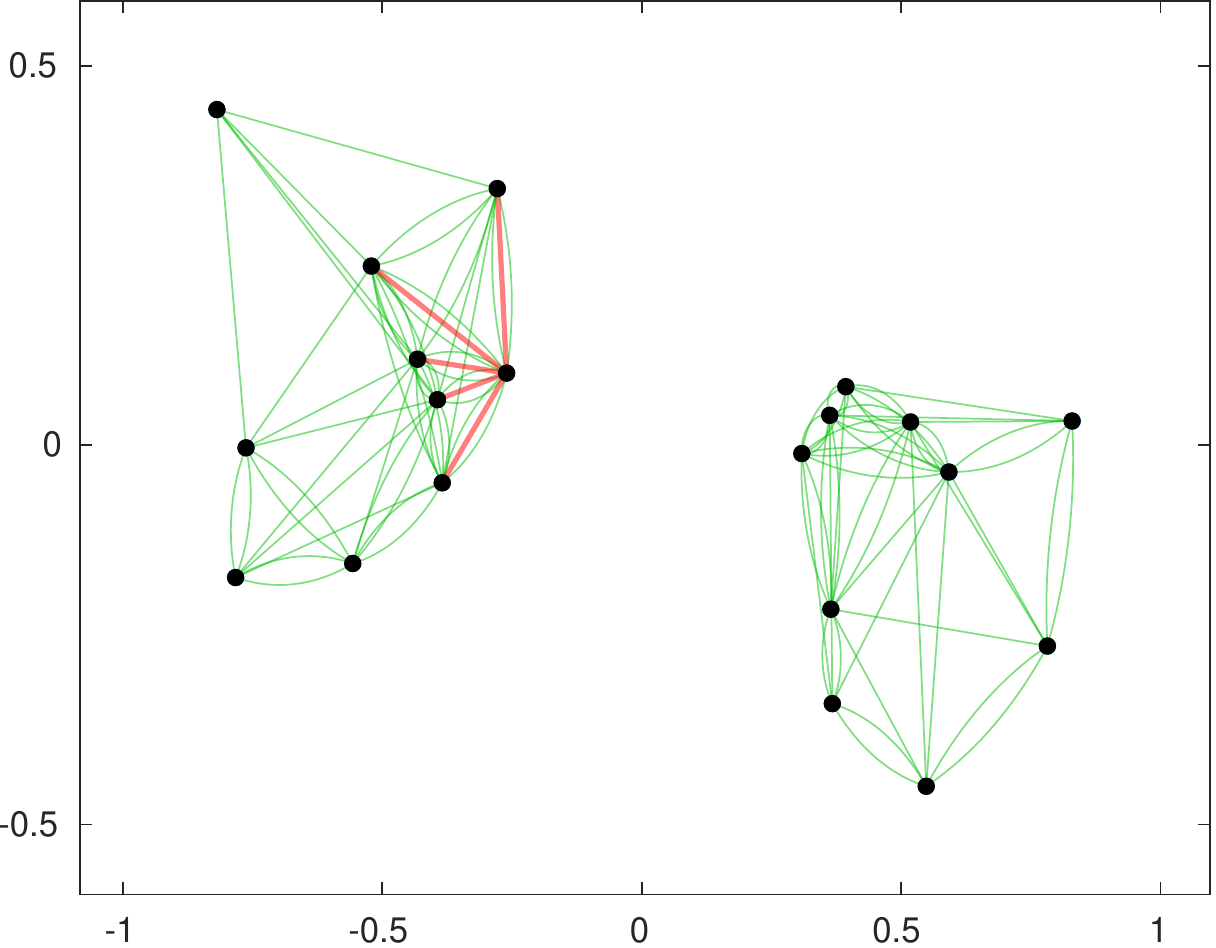}
        \label{fig:Proximity2Cluster:knn5}
    \end{minipage}
    (b)
    \begin{minipage}{0.2\textwidth} 
        \includegraphics[width=\textwidth]{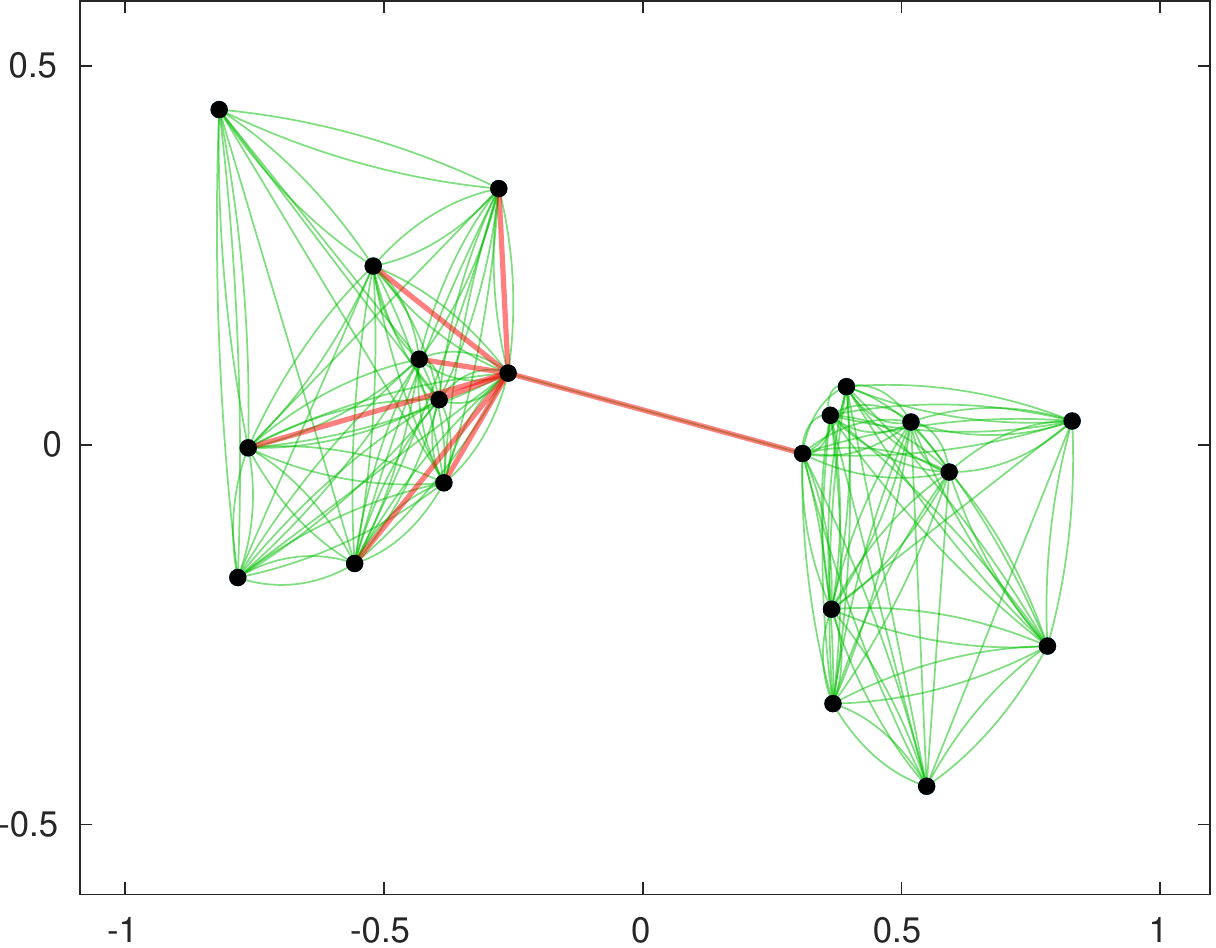}
        \label{fig:Proximity2Cluster:knn8}
    \end{minipage}
    (c)
    \begin{minipage}{0.2\textwidth} 
        \includegraphics[width=\textwidth]{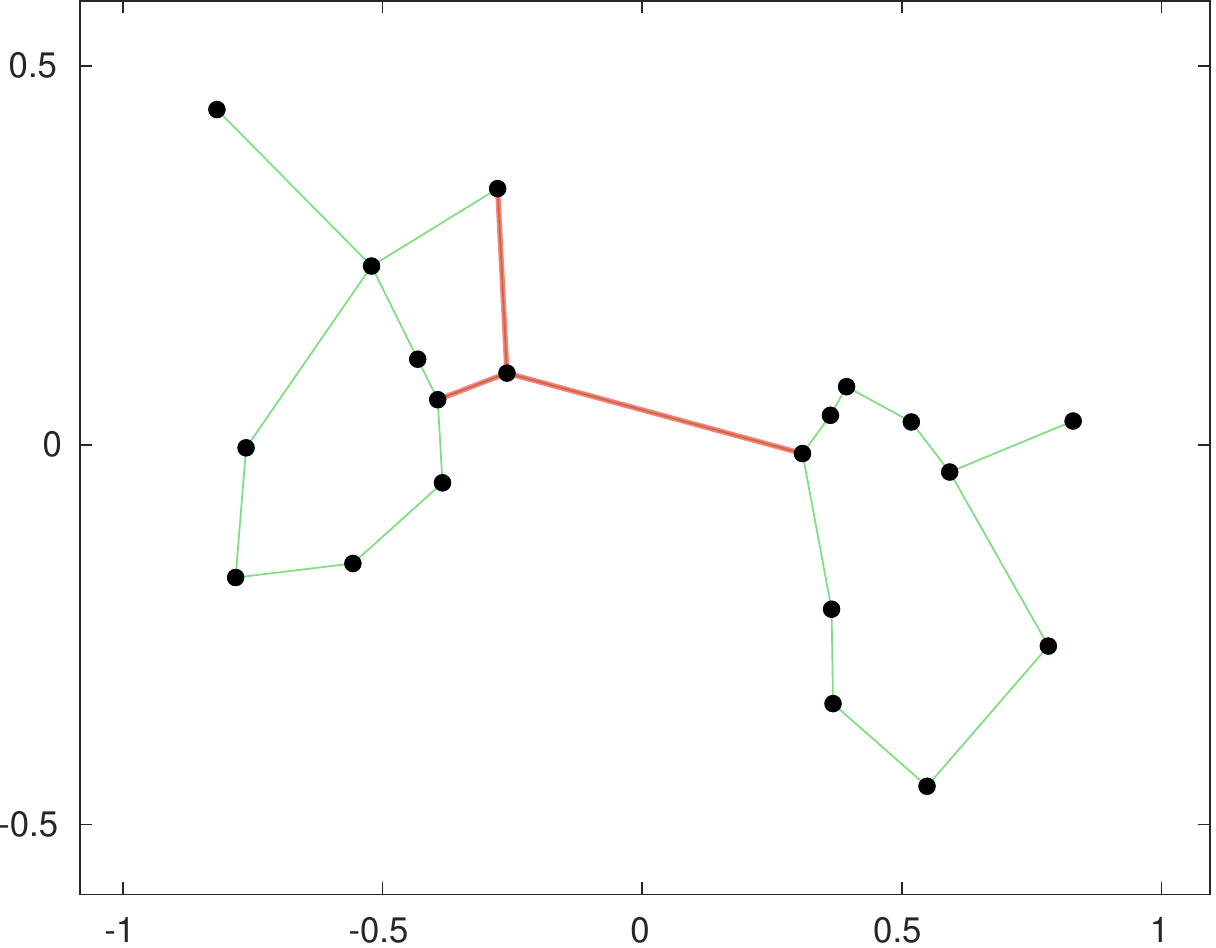}
        \label{fig:Proximity2Cluster:RNG}
    \end{minipage}
    (d)
    \begin{minipage}{0.2\textwidth} 
        \includegraphics[width=\textwidth]{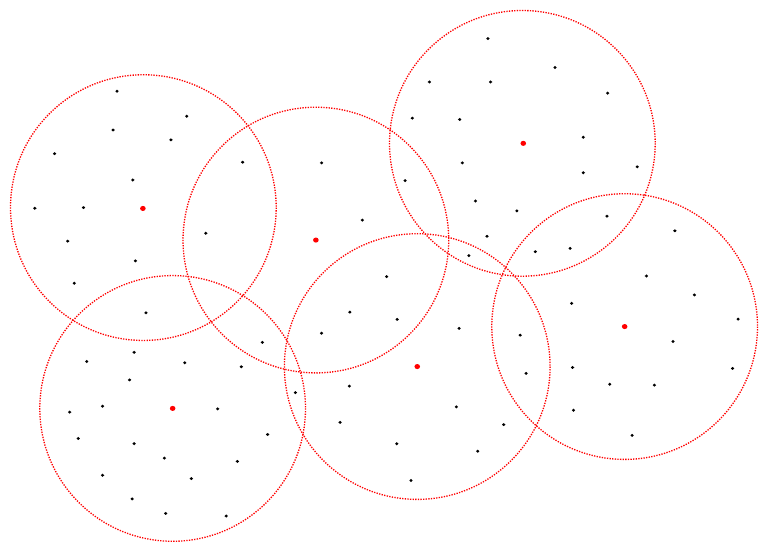}
        \label{fig:pivotDomainsCoverSpace}
    \end{minipage}
    
    \vspace{-0.5cm}

    \caption{\footnotesize The $k$NN connectivity is only based on distance between two elements and not on geometric distribution, (a) k=5 and (b) k=8. In contrast, the RNG (c) captures local geometry without regard to distance and requires no parameters. (d) Pivots (red dots) and associated radii define a pivot domain (red discs). }
    \label{fig:notation_fig}
\end{figure}

In absence of an embedding space, notions of proximity, neighborhood, and topology are constructed through a graph. The two most popular graphs are the $k$NN graph~\cite{dong2011efficient}, where each element
is connected to its $k$ nearest neighbors, and the Minimum Spanning Tree (MST) which is the spanning tree (connected tree involving all nodes) that has the least cumulative sum of distances over all links. However, the $k$NN graph is not necessarily connected: in clustered data, the $k$ closest neighbors may be to one side of an element so that the $k$NN may not faithfully represent the spatial neighborhood, Figure~\ref{fig:proximityGraphExamples}(c), in that only connections to one side are represented. Connectivity can be achieved with a sufficiently high choice of $k$, but that is at the expense of over-representing neighboring connections elsewhere, Figure~\ref{fig:notation_fig}(a,b). A much better choice that captures the spatial layout in all ``directions'' is using a class of \emph{proximity graphs}, which define a spatial neighborhood for every pair of points $x_1$ and $x_2$, and a connection is made if this spatial neighborhood does not contain any other points (also referred to as \emph{empty-neighborhood graphs}). For example, a 
\emph{Gabriel Graph} (GG)~\cite{Gabriel&Sokal} connects two points $x_1,x_2 \in \mathcal{S}$ if the sphere with diameter $x_1x_2$ is empty, or $d^{2}\left(x_{3},x_{1}\right)+d^{2}\left(x_{3},x_{2}\right)\geq d^{2}\left(x_{1},x_{2}\right),\,\forall x_{3}\in\mathcal{S}$. Another important example is the \emph{Relative Neighborhood Graph} (RNG) \cite{jaromczyk1992relative,TOUSSAINT1980261} which connects $x_1$ and $x_2 \in S$ if the \emph{lune}($x_1,x_2$), namely, the intersection of the two spheres of radius $x_1x_2$ through centers $x_1$ and $x_2$, is empty, \emph{i.e.}, if

\begin{equation} \label{eq:RNGdefinitionEquation1}
    \max\left(d\left(x_{3},x_{1}\right),d\left(x_{3},x_{2}\right)\right)\geq d\left(x_{1},x_{2}\right),\,\forall x_{3}\in\mathcal{S}.
\end{equation}
Other proximity graphs of interest include the Half-Space Graph (HSG), which is a superset of RNG and a $t$-spanner~\cite{chavez2005half}, the \emph{Delaunay Triangulation} (DT) graph~\cite{DelaunayGraph}, and the $\beta$-skeleton graph~\cite{KIRKPATRICK1985217}. Proximity graphs generally require consideration of all members $x_3$ of $S$ for each pair $(x_1,x_2)$ of $S$, and as such require $O(N^3)$ for naive construction. Note that $\text{1NN} \subset \text{MST} \subset \text{RNG} \subset \text{GG} \subset \text{DT}$. See Figure \ref{fig:proximityGraphExamples}.

We adopt the use of RNG not only because \emph{(\RNL{1})} it is connected, but also because \emph{(\RNL{2})} it is parameter free, in contrast to $k$NN, where $k$ has to be specified, Tellez~\cite{tellez2017local} where ``b'' and ``t'' have to be defined, and NSG~\cite{fu2019fast} where ``R'' has to be defined, also \emph{(\RNL{3})} the RNG is a relatively sparse graph, unlike other choices presented in Figure \ref{fig:proximityGraphExamples}.  Figure \ref{fig:results_synthetic}(e) shows the out degree of RNG is small and grows very slowly with intrinsic dimension. 

There are a large number of applications that use the RNG. The RNG is used in graph-based visualization of large image datasets for browsing and interactive exploration and is viewed as the smallest proximity graph that captures the local structure of the manifold~\cite{rayar2015approximate,rayar2016incremental,rayar2018viewable}. In urban planning theory, RNGs have been used to model topographical arrangements of cities and the road networks. In internet networks, Escalante \textit{et al.}~\cite{escalante2005rng} found that broadcasting over the RNG network is superior to blind flooding. De Vries \textit{et al.}~\cite{de2016relative} propose to use the RNG to reveal related dynamics of page-level social media metrics. Han \textit{et al.}~\cite{han2008pre} aims to improve the efficiency of a Support Vector Machine (SVM) classifier by using the RNG to extract probable support vectors from all the training samples. Goto \textit{et al.}~\cite{goto2015preselection} use the RNG to reduce a training dataset consisting of handwritten digits to $10\%$ of its original size. A related and more recent area is the selection of training data for Convolutional Neural Networks (CNNs) where the RNG is used to reduce the underlying redundancy of the dataset~\cite{rayar2018cnn}. 

%


Despite such widespread use of RNG, there is not a large literature on efficient construction of the RNG in metric spaces. In Euclidean spaces, the notions of angle and direction allow for an efficient implementation, \textit{e.g.}, an $O(N\log N)$ for $N$ points in $\mathcal{R}^2$~\cite{supowit1983relative}, an $O(N)$ for uniformly distributed points in a rectangle~\cite{katajainen1987linear}, and an $O(N^2)$ for higher dimensions~\cite{supowit1983relative}. The construction of the RNG for general metric spaces, however, has been more challenging, limited to two groups of papers. First, Hacid \textit{et al.}~\cite{hacid2007incremental} propose an approximate incremental RNG construction algorithm for data mining and visualization purposes. This approximate construction defines the set of potential RNG neighbors and the set of potentially invalidated RNG links by only considering dataset items that fall within a hypersphere around the query's nearest neighbor, where its radius is proportional to the distance from the query to its nearest neighbor plus the distance from the nearest neighbor to its furthest RNG neighbor. Second, Rayar \textit{et al.}~\cite{rayar2015approximate} proposed an improvement over Hacid's algorithm by defining the set of potentially invalidated RNG links by the $L^{\text{th}}$ edge neighbors of the query. While both these methods work in any metric space and provide significant speed-up over naive construction, they are \emph{approximate} and thus lose all guarantees provided by the RNG, and make a significant number of errors, as will be shown by Table \ref{tab:results_real}.

The main computational challenge in searching metric spaces is to reduce the number of distance computations which are expensive, in contrast to vector spaces where the aim is to reduce I/O. The general approach is to build an \emph{index} which effectively builds a set of equivalence classes so that some classes can be discarded leaving others to be exhaustively searched, either through compact partitioning or through pivoting~\cite{chavez2001searching}. The notion of a \emph{pivot} arises as a way to capture a group of exemplars. Define the \emph{pivot domain}, Figure \ref{fig:notation_fig}(d), $\mathcal{D}$ of pivot $p_{i}$ and domain radius $r_{i}$ as, 

\begin{equation} \label{eq:pivotMembership}
    \mathcal{D}(p_i,r_i)=\left\{ x\in\mathcal{S}\,|\,d\left(x,p_{i}\right)\leq r_{i}\right\}.
\end{equation}
While pivots do not necessarily need to be members of $S$, in a metric space which cannot generate new members a pivot is also an exemplar/data point. A sufficient number of pivots $\mathcal{P}=\left\{ p_{1},p_{2},...,p_{M}\right\}\subset\mathcal{S}$ are required to cover $\mathcal{S}$, \emph{i.e.}, $\mathcal{S}=\bigcup_{i=1}^{M}\mathcal{D}(p_i,r_i)$. 

Observe that the knowledge of $d(x,p_i)$ bounds $d(x,y)$ for $y \in S_i$ as $d(y,p_i) - r_i \leq d(x,y) \leq d(y,p_i) + r_i$ using the triangle inequality.  In the absence of an embedding Euclidean structure the triangle inequality is the only constraint available for relative ranking of distances between triplets of points. For simplicity we take $r_i = r$ in this paper.

The key aim of this paper is to design a hierarchical index that allows for the construction of the exact RNG and allows for efficient search of RNG neighbors of a given query $Q$. The contribution of the paper is to show that in a two-layer configuration of pivots and exemplars (data points) a novel graph structure, the Generalized Relative Neighborhood Graph (GRNG), allows for efficient and exact construction of RNG of the data points. Note that the RNG is a special case of GRNG when its parameter $r=0$. In addition, we also show that the GRNG of any coarse-layer of pivots can guide the exact construction of the GRNG of any fine-layer pivots. This allows for a highly efficient, scalable, hierarchical construction involving multiple layers (for a dataset of 26 million points in $\mathcal{R}^2$ ten layers is optimal\CiteFullArXiV). Observe that construction is incremental so that the index can be dynamically updated. Given a query, a search process locates it in the hierarchy by examining the coarsest layers, discarding all the exemplar domains for a majority of the pivots and then moving on to the next layers where finer-scale pivot children of a few select coarse-scale pivots need to be considered. This process is then repeated to the lowest layer, the exemplar domain. The query is then located in the RNG and its RNG neighbors are identified. The search process is highly efficient and logarithmic in the number of exemplars in all dimensions, Figure~\ref{fig:results_synthetic}(b,d). 

The incremental construction of the index relies on the search component described above to locate the query in each layer, but in addition, in each layer new connections must be made and existing connections must be validated. The construction is done off-line in contrast to search which is typically done on-line. While the construction is exponential in both the number of exemplars and dimensions for uniformly distributed data, for practical applications where the data is clustered, the construction cost behaves much better. The experimental results summarized in Table~\ref{tab:results_real} show that while our method gives the exact RNG neighbors, it is substantially faster in both constructing the RNG and in searching it.

\section{Incremental Construction of the RNG} \label{sec:build-rng}

The incremental approach to constructing RNG assumes that RNG($\mathcal{S}$) is available and computes RNG($\mathcal{S} \cup {Q}$) from it. The query $Q$ is the newest element: \emph{(\RNL{1}) Localize $Q$ within $\mathcal{S}$:} finding the RNG Neighbors of $Q$. The naive approach would consider for all $x_i \in \mathcal{S}$ whether $\exists x_j \in lune(Q,x_i)$; all $x_i$ with empty lunes are RNG neighbors of $Q$. Note that this involves $O(N^2)$ operations where $N=|\mathcal{S}|$, and this is clearly not scalable, and \emph{(\RNL{2}) Adding $Q$ to the dataset:} When the task is search, the first step finds the RNG neighbors. If $Q$ needs to be added, additionally all pairs of existing links between $x_i$ and $x_j$ need to be validated, whether $Q \in \text{lune}(x_i,x_j)$ in which case $x_i$ and $x_j$ are no longer RNG neighbors. This operation is on the order of $O(\alpha N)$ where $\alpha$ is the average out degree of the RNG, typically a small number. Thus, the localization step is significantly more computationally intensive than the validation step. 

The remedy to indexing complexity is organization. Specifically, when exemplar groups are represented by pivots, many inferences can take place at the level of pivot domains without computing distances between $Q$ and exemplars. The basic idea in this paper is to construct conditions on pivots that have implications for efficient incremental construction of RNG of exemplars. This is organized in seven stages: \emph{\RNL{1}}) In Stages \RNU{1},\RNU{2}, and \RNU{3} entire pivot domains $\mathcal{D}(p_i,r_i)$  or a significant number of exemplars $x_i$ are discarded from considering RNG neighbor relations with $Q$ by just measuring $d(Q,p_i)$; \emph{\RNL{2}}) Stages \RNU{4},\RNU{5}, and \RNU{6}: pivots are used in invalidating potential RNG links with the remaining exemplars; \emph{\RNL{3}}) Stage \RNU{7}: pivots are used to exclude entire domains during the RNG validation process of existing links. What relationship between $p_i$ and $p_j$ can prevent the formation of a RNG link between $x_i$ and $x_j$?

\begin{theorem} \label{thm:RNG_S1}
    Consider exemplars $x_i \in \mathcal{D}(p_i,r_i)$ and $x_j \in \mathcal{D}(p_j,r_j)$. Then 
    \begin{equation} \label{eq:RNG_S1_equations}
        \left\{\begin{aligned}[c]
        d\left(p_{k},p_{i}\right)<d\left(p_{i},p_{j}\right)-\left(2r_{i}+r_{j}\right)\\
        d\left(p_{k},p_{j}\right)<d\left(p_{i},p_{j}\right)-\left(r_{i}+2r_{j}\right)
        \end{aligned}
        \Rightarrow 
        \begin{aligned}[c]
            \max(d(p_k,x_i),d(p_k,x_j)) < d(x_i,x_j)
        \end{aligned}\right.
    \end{equation}
\end{theorem}
\iffull
\begin{proof}
    Equation~\ref{eq:RNG_S1_equations} requires that $d\left(p_{k},x_{i}\right)<d\left(x_{i},x_{j}\right)$ and $d\left(p_{k},x_{j}\right)<d\left(x_{i},x_{j}\right)$, which can be established by a two-fold application of the triangle inequality, first relating $d\left(p_{k},x_{i}\right)$ and $d\left(p_{k},x_{j}\right)$ to $d\left(p_{k},p_{i}\right)$ and then to $d\left(p_{k},p_{j}\right)$, respectively,
    
    \begin{subnumcases}{}
        \scalebox{0.85}{%
        $d\left(p_{k},p_{i}\right)-r_{i}\leq d\left(p_{k},p_{i}\right)-d\left(p_{i},x_{i}\right)\leq d\left(p_{k},x_{i}\right)\leq d\left(p_{k},p_{i}\right)+d\left(p_{i},x_{i}\right)\leq d\left(p_{k},p_{i}\right)+r_{i}$
        } \label{eq:conditionInequality3} \\
        \scalebox{0.85}{%
        $d\left(p_{k},p_{j}\right)-r_{j}\leq d\left(p_{k},p_{j}\right)-d\left(p_{j},x_{j}\right)\leq d\left(p_{k},x_{j}\right)\leq d\left(p_{k},p_{j}\right)+d\left(p_{j},x_{j}\right)\leq d\left(p_{k},p_{j}\right)+r_{j}.$
        } \label{eq:conditionInequality4}
    \end{subnumcases}
    \noindent Similarly, $d\left(x_{i},x_{j}\right)$ can be related to $d\left(p_{i},p_{j}\right)$ by applying the triangle inequality twice:
    
    \begin{equation}
        \resizebox{\textwidth}{!}{$
        d\left(p_{i},p_{j}\right)-r_{j}-r_{i}\leq d\left(p_{i},x_{j}\right)-r_{i}\leq d\left(p_{i},x_{j}\right)-d\left(p_{i},x_{i}\right)\leq d\left(x_{i},x_{j}\right)\leq d\left(p_{i},x_{j}\right)+d\left(p_{i},x_{i}\right)\leq d\left(p_{i},x_{j}\right)+r_{i}\leq  d\left(p_{i},p_{j}\right)+r_{j}+r_{i}.$}
        \label{eq:condition7}
    \end{equation}
    Thus, if the right side of Equation~\ref{eq:conditionInequality3} is smaller than the left side of Equation~\ref{eq:condition7}, \textit{i.e.,}
    
    \begin{equation}
    d\left(p_{k},p_{i}\right)+r_{i}<d\left(p_{i},p_{j}\right)-r_{j}-r_{i},
    \label{eq:condition8}
    \end{equation}
    then combining Equations~\ref{eq:conditionInequality3}, and~\ref{eq:condition7} gives
    
    \begin{equation}
    d\left(p_{k},x_{i}\right)<d\left(x_{i},x_{j}\right).
    \end{equation}
    Similarly, if we have
    
    \begin{equation}
    d\left(p_{k},p_{j}\right)+r_{j}<d\left(p_{i},p_{j}\right)-r_{j}-r_{i},
    \label{eq:condition10}
    \end{equation}
    then combining Equations~\ref{eq:conditionInequality4}, and~\ref{eq:condition7} gives
    
    \begin{equation}
    d\left(p_{k},x_{j}\right)<d\left(x_{i},x_{j}\right).
    \end{equation}
    Thus, when Equations~\ref{eq:condition8} and~\ref{eq:condition10} hold, or stated differently when
    
    \begin{subnumcases}{\label{eq:RNGFormationTheoremProofConditions}}
    d\left(p_{k},p_{i}\right)<d\left(p_{i},p_{j}\right)-2r_{i}-r_{j}\\
    d\left(p_{k},p_{j}\right)<d\left(p_{i},p_{j}\right)-r_{i}-2r_{j},
    \end{subnumcases}
    then $\max\left(d\left(p_{k},x_{i}\right),d\left(p_{k},x_{j}\right)\right)<d\left(x_{i},x_{j}\right)$, \textit{i.e.,} an RNG connection cannot exist between $x_{i}$ and $x_{j}$.
    \qed
\end{proof} 
Theorem \ref{thm:RNG_S1} 
\fi
states that a pivot $p_k$ that falls in a lune defined by the intersection of the sphere at $p_i$ with radius $d\left(p_{i},p_{j}\right)-\left(2r_{i}+r_{j}\right)$ and the sphere at $p_j$ with radius $d\left(p_{i},p_{j}\right)-\left(r_{i}+2r_{j}\right)$ also falls in the RNG lune of $x_i$ and $x_j$, thereby invalidating the potential RNG link between $x_i$ and $x_j$, \emph{without computing $d(p_k,x_i)$ and $d(p_k,x_j)$}! This is a proximity relationship between $p_i,p_j$, and $p_k$, which effectively defines a novel type of graph.

%
\begin{figure}[t!]
    \centering
    \begin{subfigure}[b]{0.19\textwidth}
        \includegraphics[width=\textwidth]{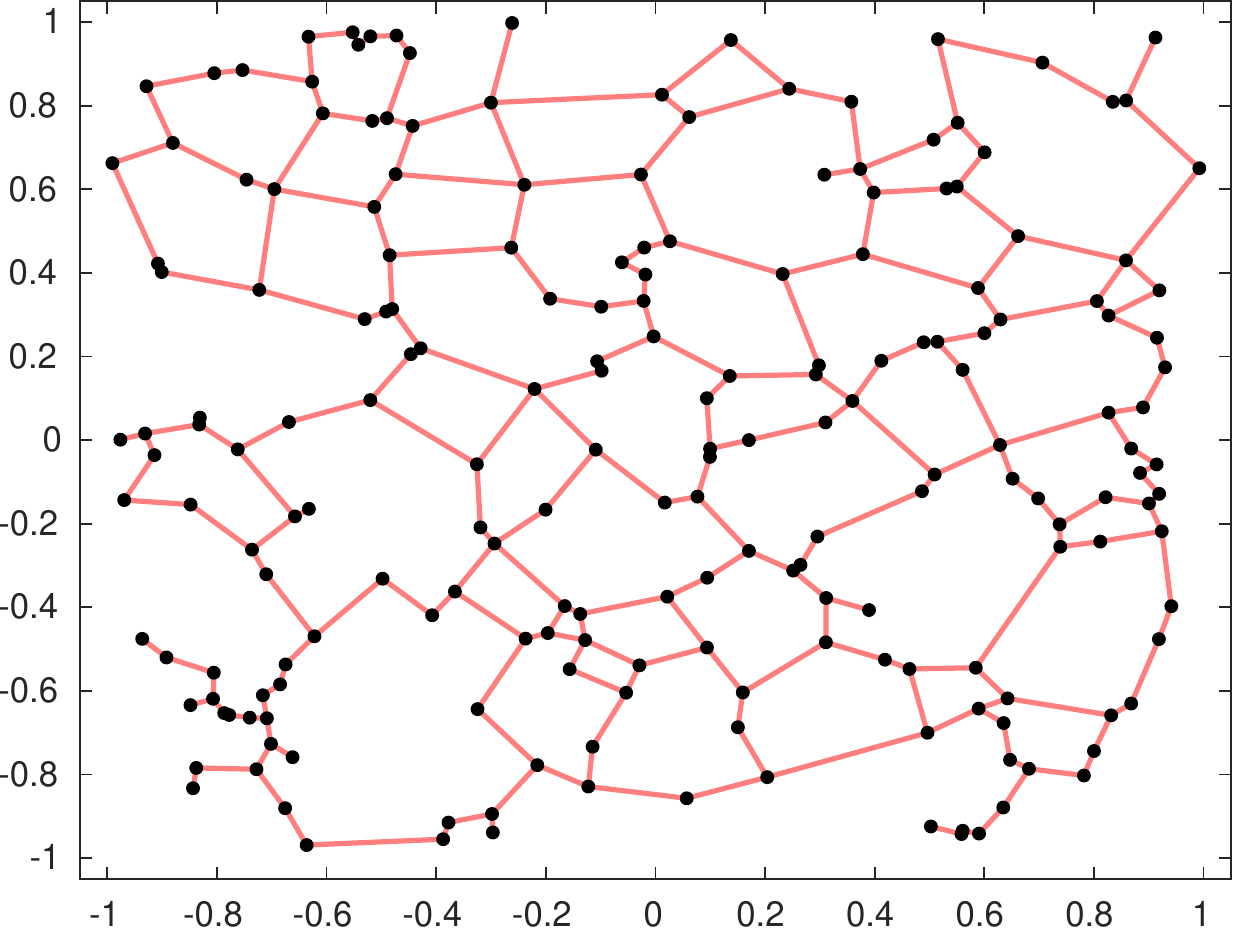}
        \caption{}
    \end{subfigure}
    \begin{subfigure}[b]{0.19\textwidth}
        \includegraphics[width=\textwidth]{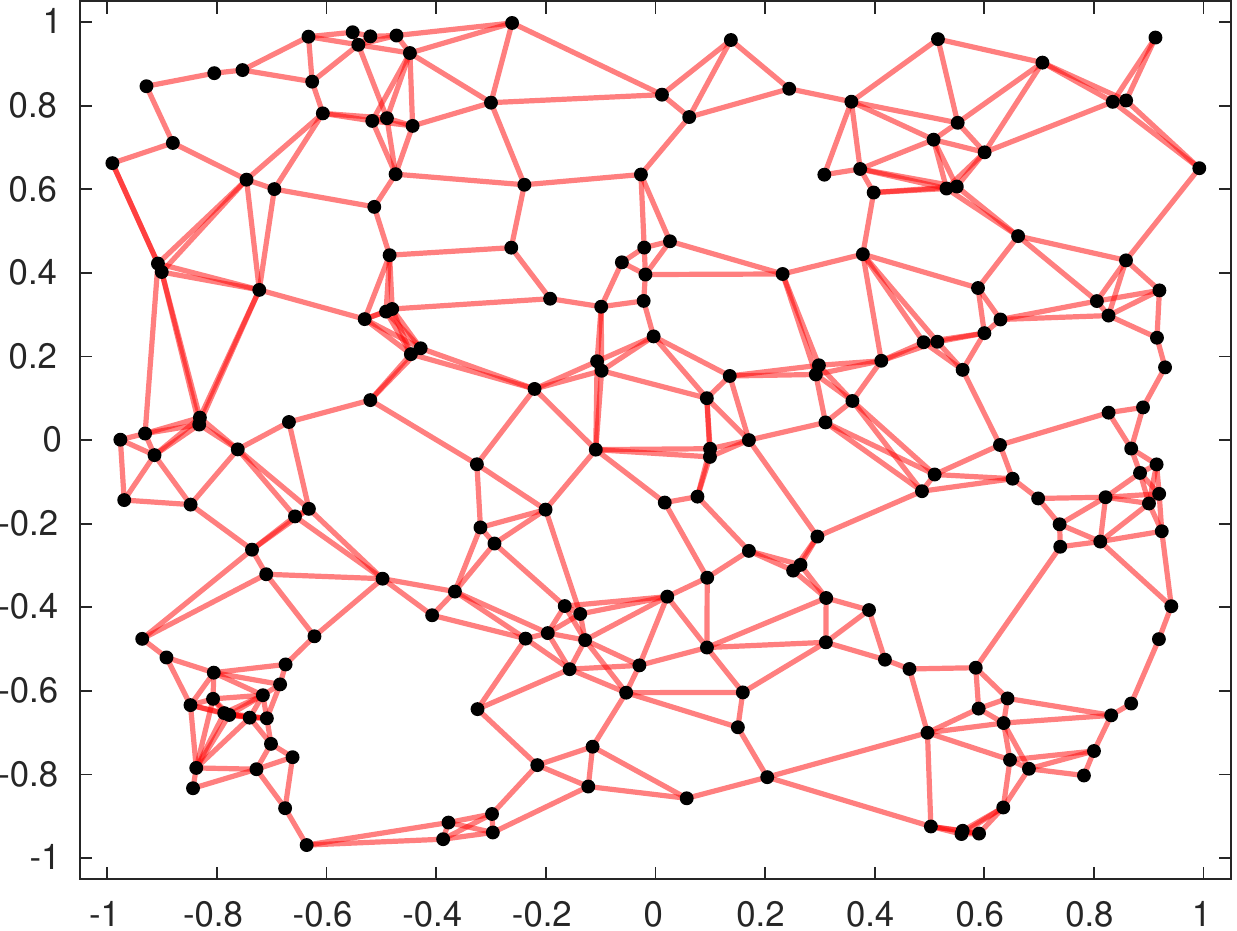}
        \caption{}
    \end{subfigure}
    \begin{subfigure}[b]{0.19\textwidth}
        \includegraphics[width=\textwidth]{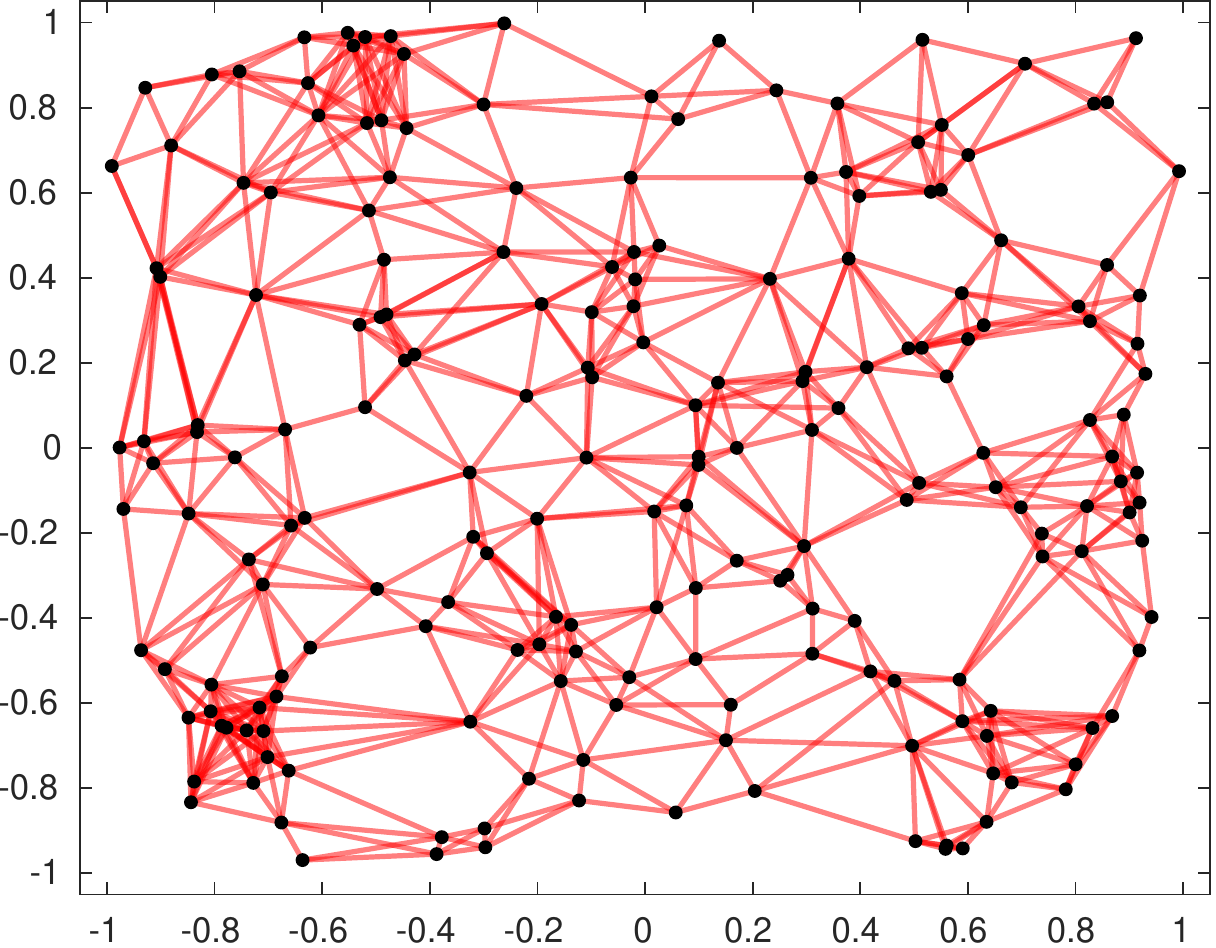}
        \caption{}
    \end{subfigure}
    \begin{subfigure}[b]{0.19\textwidth}
        \includegraphics[width=\textwidth]{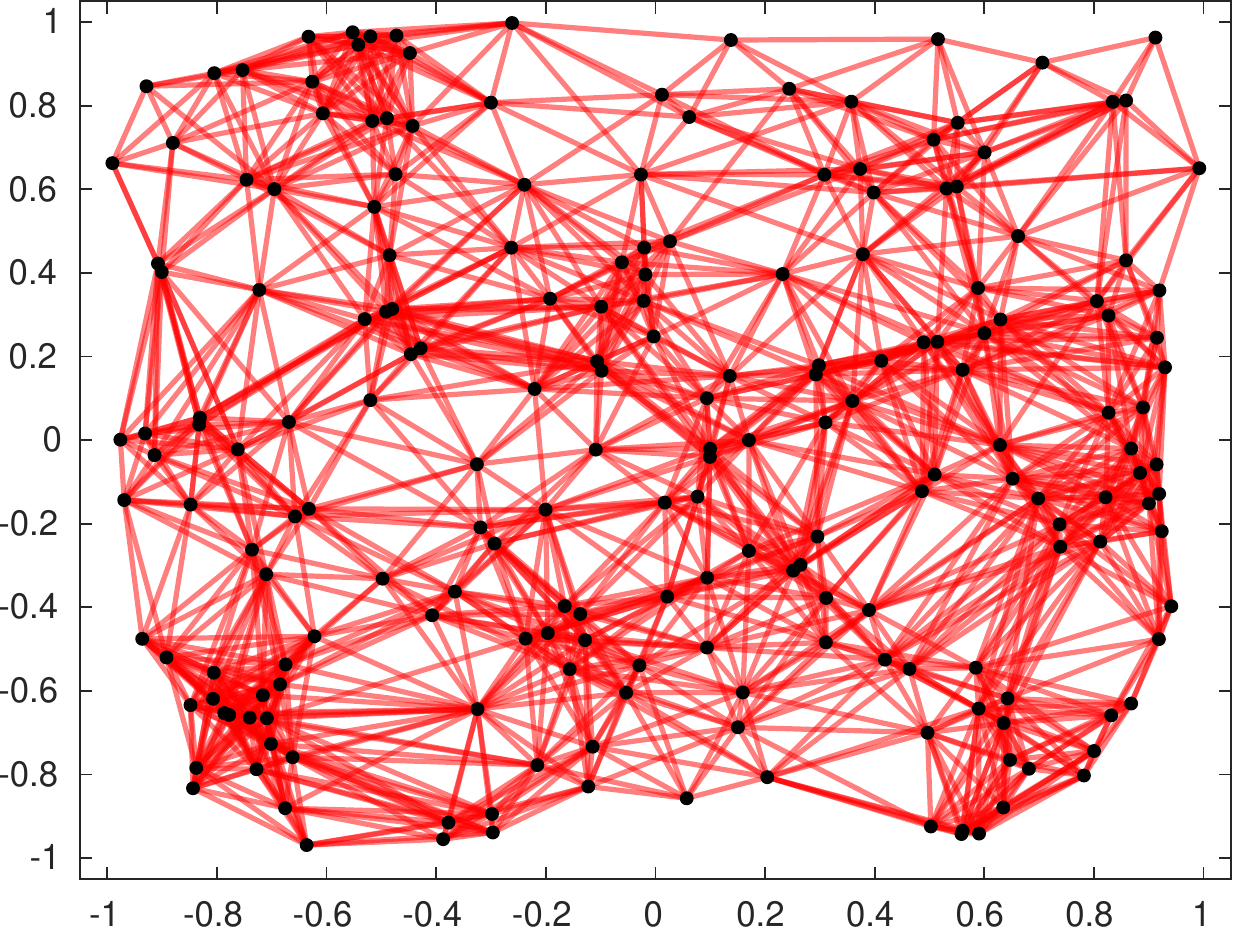}
        \caption{}
    \end{subfigure}
    \begin{subfigure}[b]{0.19\textwidth}
        \includegraphics[width=\textwidth]{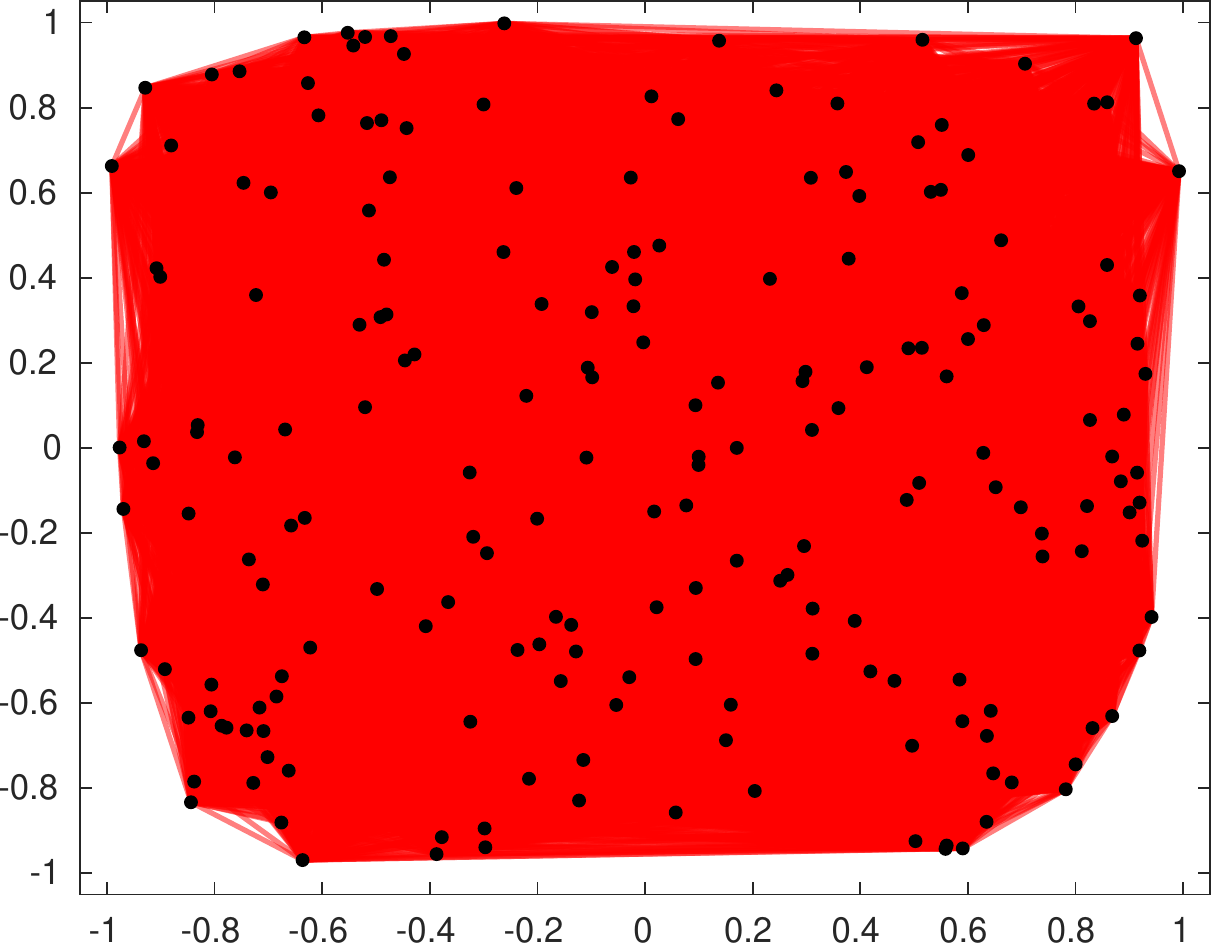}
        \caption{}
    \end{subfigure}
    \vspace{-0.25cm}

    \caption{\footnotesize GRNG of a set of 200 points in $[-1,1]^2$ where all $r_i = r$ and for different selection of $r$: (a) $r=0$, (b) $r=0.01$, (c) $r=0.02$, (d) $r=0.04$, and (e) $r=0.419$. When $r$ exceeds $\frac{1}{6}$ the maximum distance between points it is the complete graph (e).}
    \label{fig:GRNG_Examples}
\end{figure}

%
\iffull
    \begin{figure}[b!]
    \centering
    \begin{subfigure}[b]{0.19\textwidth}
        \includegraphics[width=\textwidth]{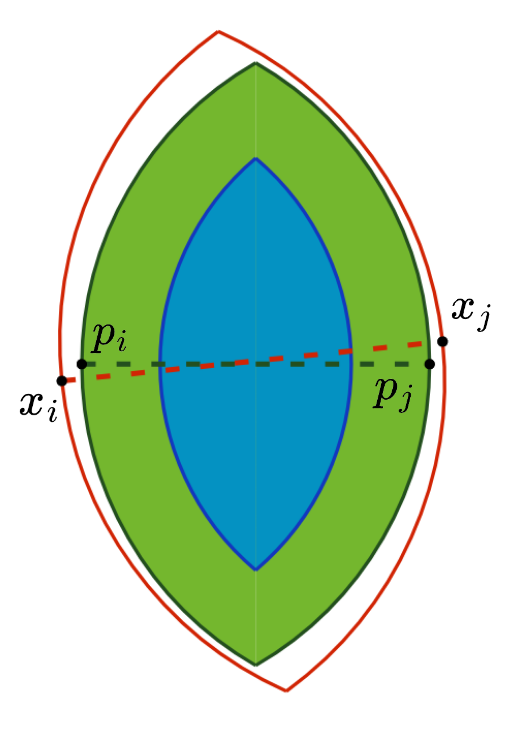}
    \end{subfigure}
    \begin{subfigure}[b]{0.19\textwidth}
        \includegraphics[width=\textwidth]{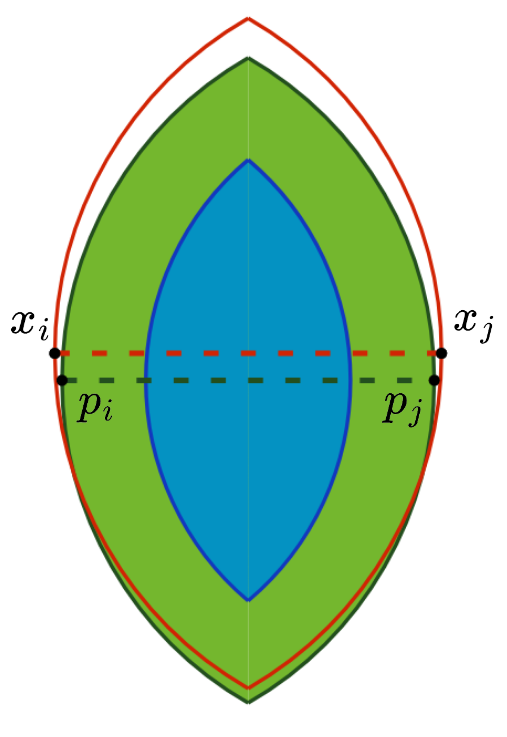}
    \end{subfigure}
    \begin{subfigure}[b]{0.19\textwidth}
        \includegraphics[width=\textwidth]{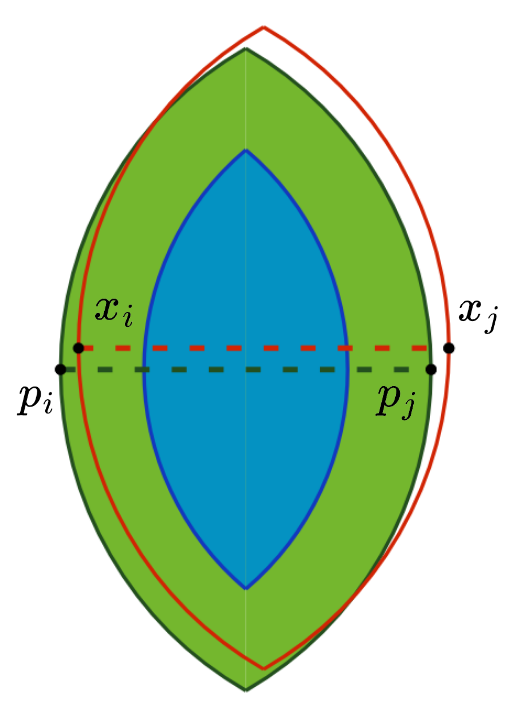}
    \end{subfigure}
    \begin{subfigure}[b]{0.19\textwidth}
        \includegraphics[width=\textwidth]{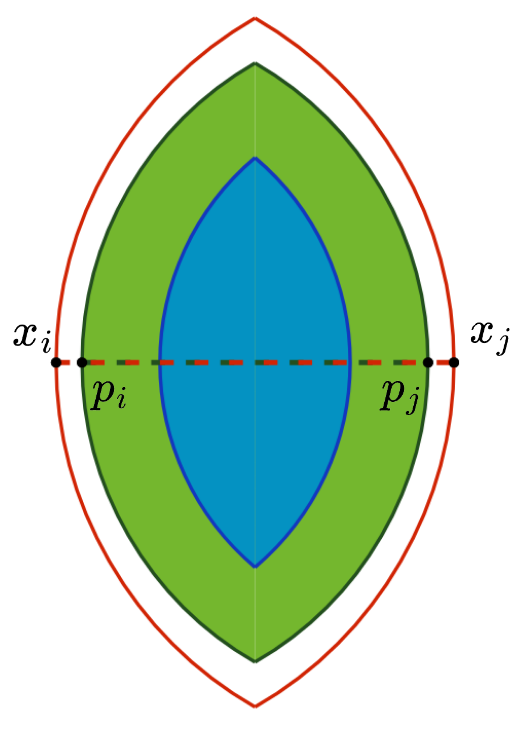}
    \end{subfigure}
    \begin{subfigure}[b]{0.19\textwidth}
        \includegraphics[width=\textwidth]{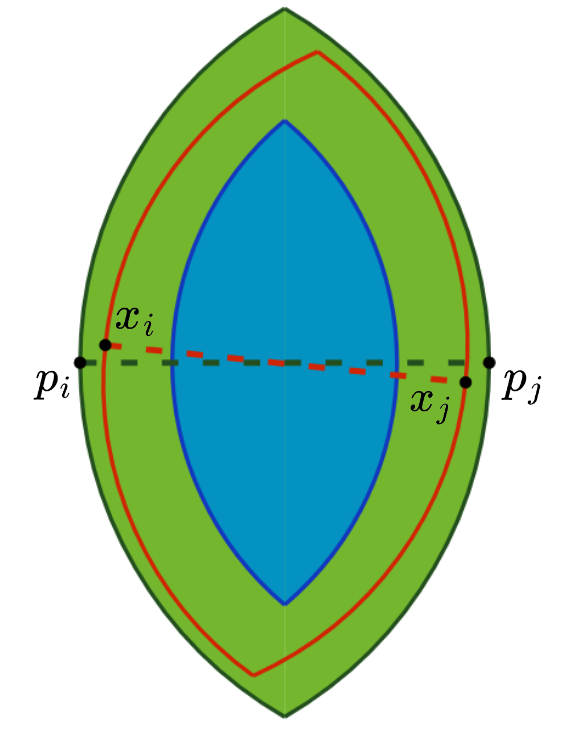}
    \end{subfigure}
    \caption[A few examples illustrating the similarity between lunes.]{A few examples illustrating the similarity between the $\text{lune}\left(x_{i},x_{j}\right)$ and the $\text{lune}\left(p_{i},p_{j}\right)$, where $x_{i}$ and $x_{j}$ are both fairly close to $p_{i}$ and $p_{j}$, respectively, relative to $d(x_i,x_j)$ but otherwise unconstrained. Lune($x_i,x_j$) is shown in red, lune($p_i,p_j$) is shown in green, and G-lune($p_i,p_j$) is shown in blue. Note how much smaller the generalized lune is compared to the RNG lunes.}
    \label{fig:luneVariationExamples}
\end{figure}
\fi

%
\begin{definition} 
    \textbf(Generalized Relative Neighborhood Graph (GRNG)): Two pivots $p_i,p_j \in \mathcal{P}$ have a GRNG link iff no pivots $p_k \in \mathcal{P}$ can be found inside the generalized lune defined by, 
    
    \begin{subnumcases}{}
        d\left(p_{k},p_{i}\right) < d\left(p_{i},p_{j}\right)-\left(2r_{i}+r_{j}\right) \\
        d\left(p_{k},p_{j}\right) < d\left(p_{i},p_{j}\right)-\left(r_{i}+2r_{j}\right).
    \end{subnumcases}
\end{definition}
Observe that GRNG($\mathcal{P}$) is just the RNG when $r_i=0$, $\forall i$, thus it is a generalization of it, Figure \ref{fig:GRNG_Examples}. Also, note that GRNG($\mathcal{P}$) is a superset of RNG($\mathcal{P}$) since lune($p_i,p_j$) is larger than the generalized-lune($p_i,p_j$), abbreviated as G-lune($p_i,p_j$). This implies that the larger $r_i$  and $r_j$ are, the denser the graph is, until it is effectively the complete graph. This places a constraint on how large $r_i$ and $r_j$ can be. Furthermore, it is easy to show that GRNG($\mathcal{P}$) is a connected graph. In practice, all pivots share the same uniform radius, \emph{i.e.}, $r_i=r, \forall i$. The single parameter $r$ is the minimum for which the union of all pivot domains cover $\mathcal{S}$. Thus, the number of pivots $M$ and $r$ are inversely related. In what follows $d(Q,p_i),i=1,2,\ldots,M$ is computed.

\begin{figure}[b!]

    %
    \begin{minipage}{0.19\textwidth}
     \centering
     \includegraphics[width=\textwidth]{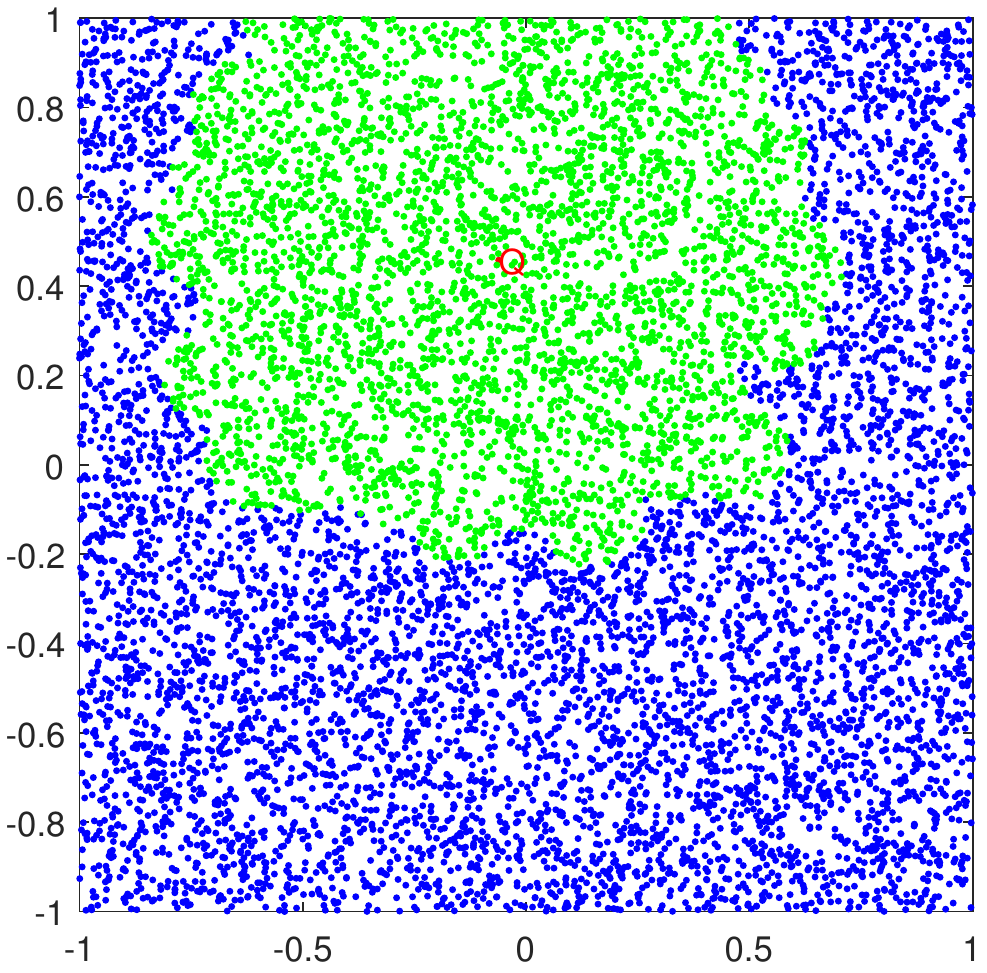}
     \subcaption{Stage \RNU{1}}
   \end{minipage}
   \begin{minipage}{0.19\textwidth}
     \centering
     \includegraphics[width=\textwidth]{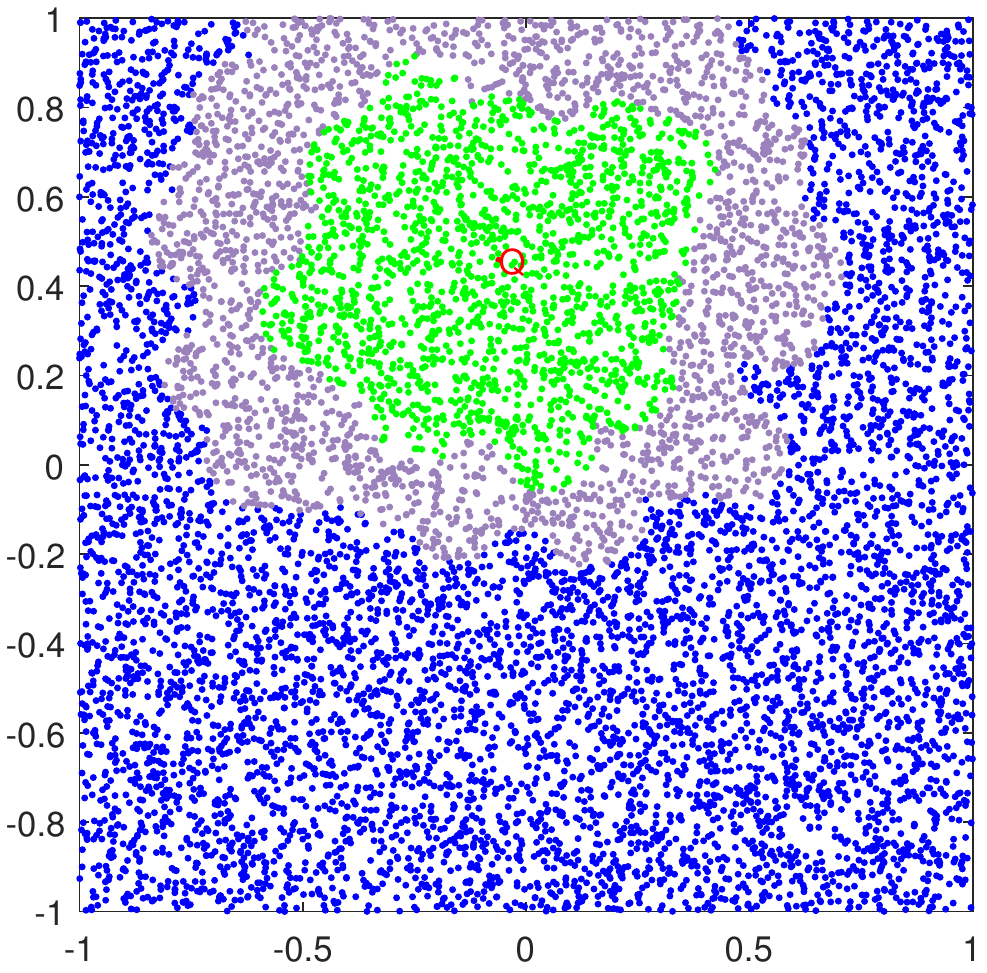}
     \subcaption{Stage \RNU{2}}
   \end{minipage}
   \begin{minipage}{0.19\textwidth}
     \centering
     \includegraphics[width=\textwidth]{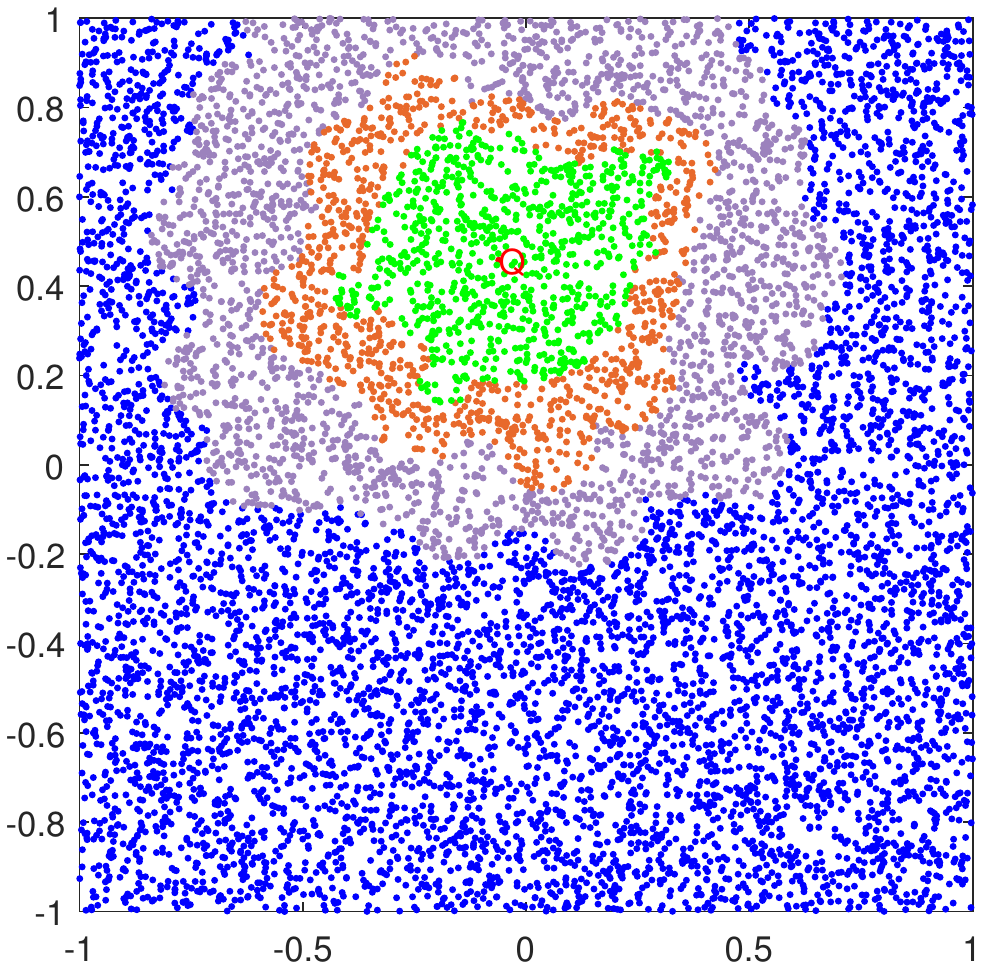}
     \subcaption{Stage \RNU{3}}
   \end{minipage}
   \begin{minipage}{0.19\textwidth}
     \centering
     \includegraphics[width=\textwidth]{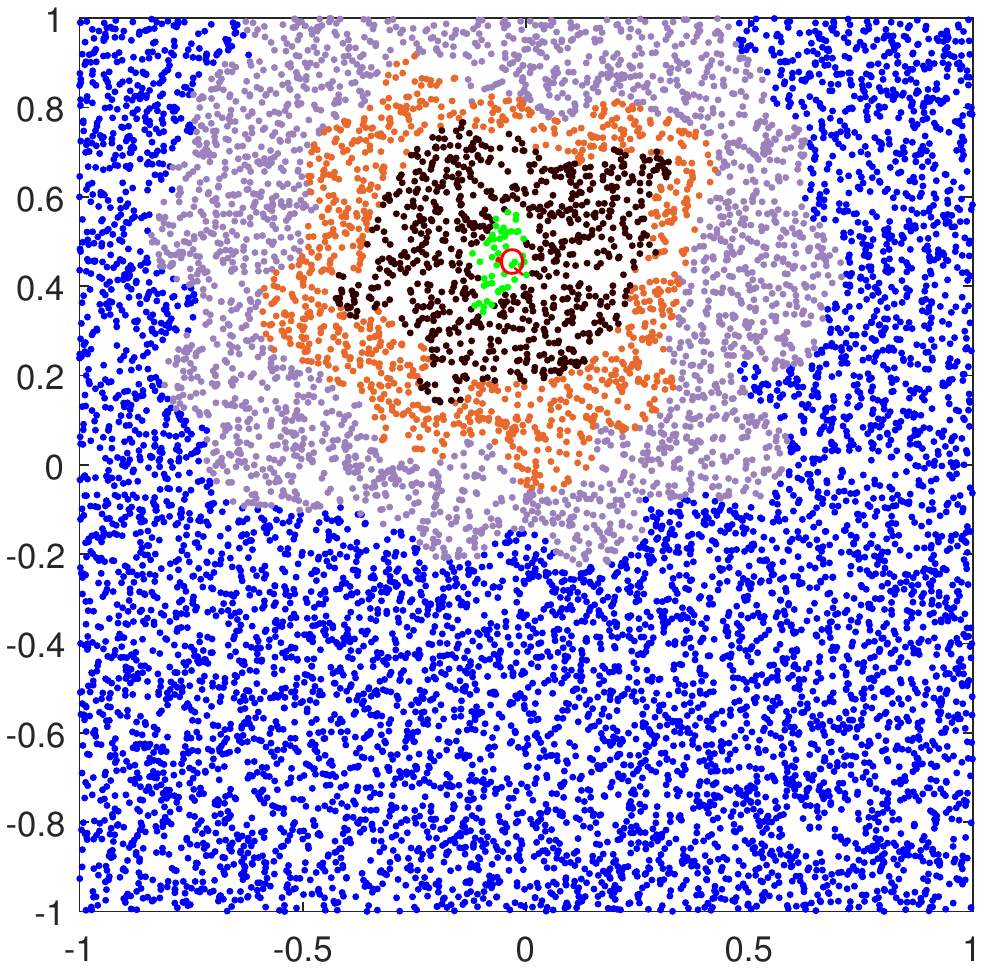}
     \subcaption{Stage \RNU{4}}
   \end{minipage}
   \begin{minipage}{0.19\textwidth}
     \centering
     \includegraphics[width=\textwidth]{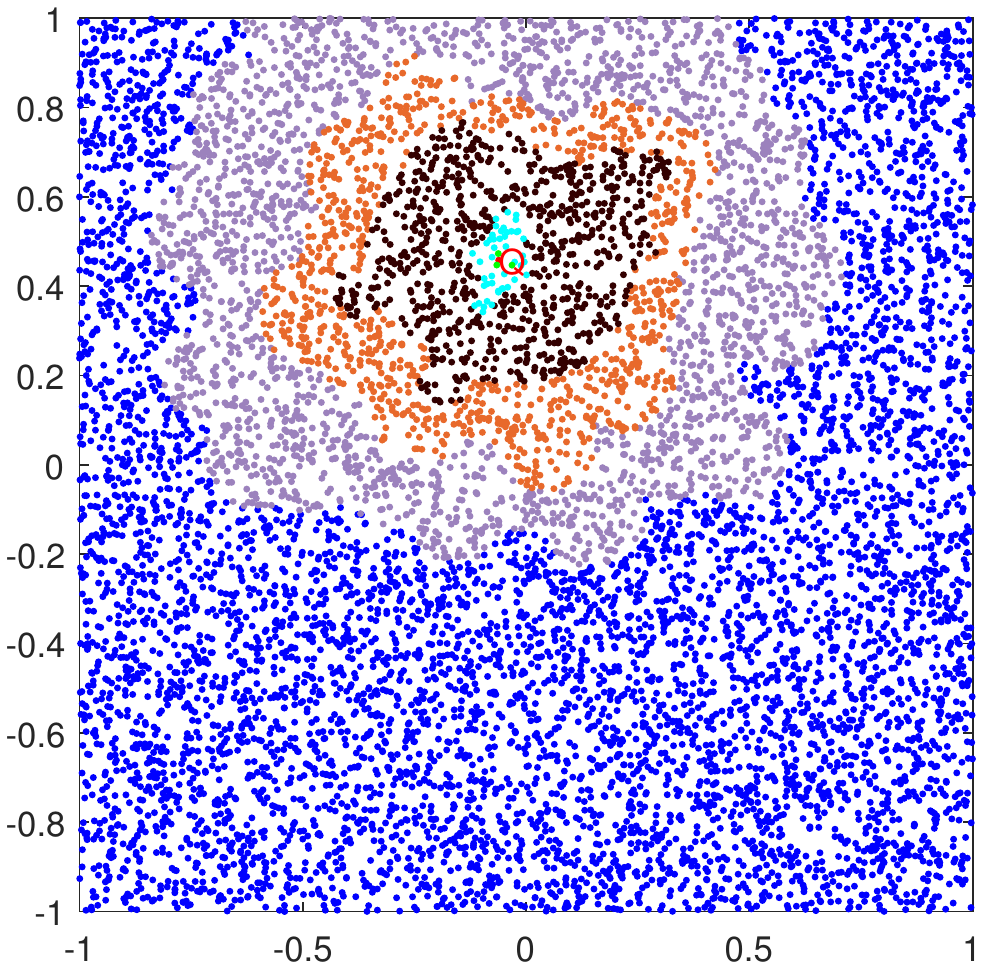}
     \subcaption{Stage \RNU{5}}
   \end{minipage}
   
    \begin{minipage}{0.16\textwidth}
     \centering
     \includegraphics[width=\textwidth]{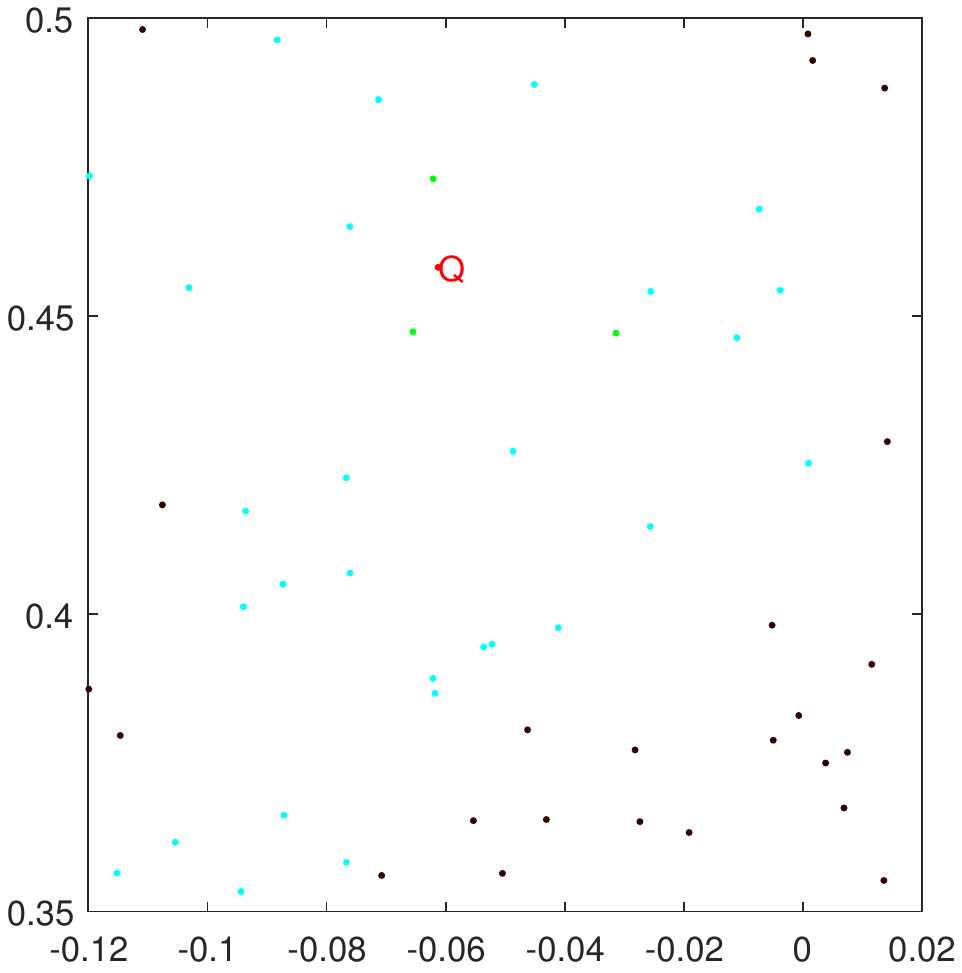}
     \subcaption{Stage \RNU{5}}
   \end{minipage}
   \begin{minipage}{0.16\textwidth}
     \centering
     \includegraphics[width=\textwidth]{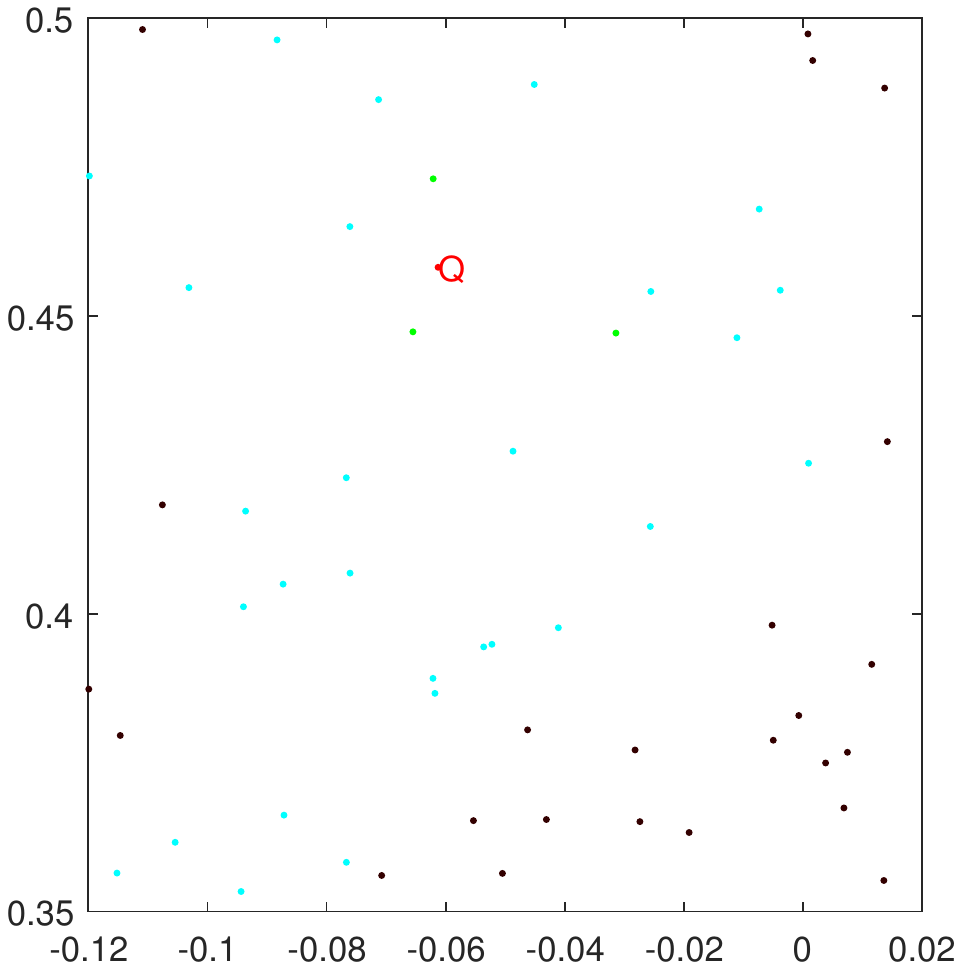}
     \subcaption{Stage \RNU{6}}
   \end{minipage}
   \begin{minipage}{0.21\textwidth}
     \centering
     \includegraphics[width=\textwidth]{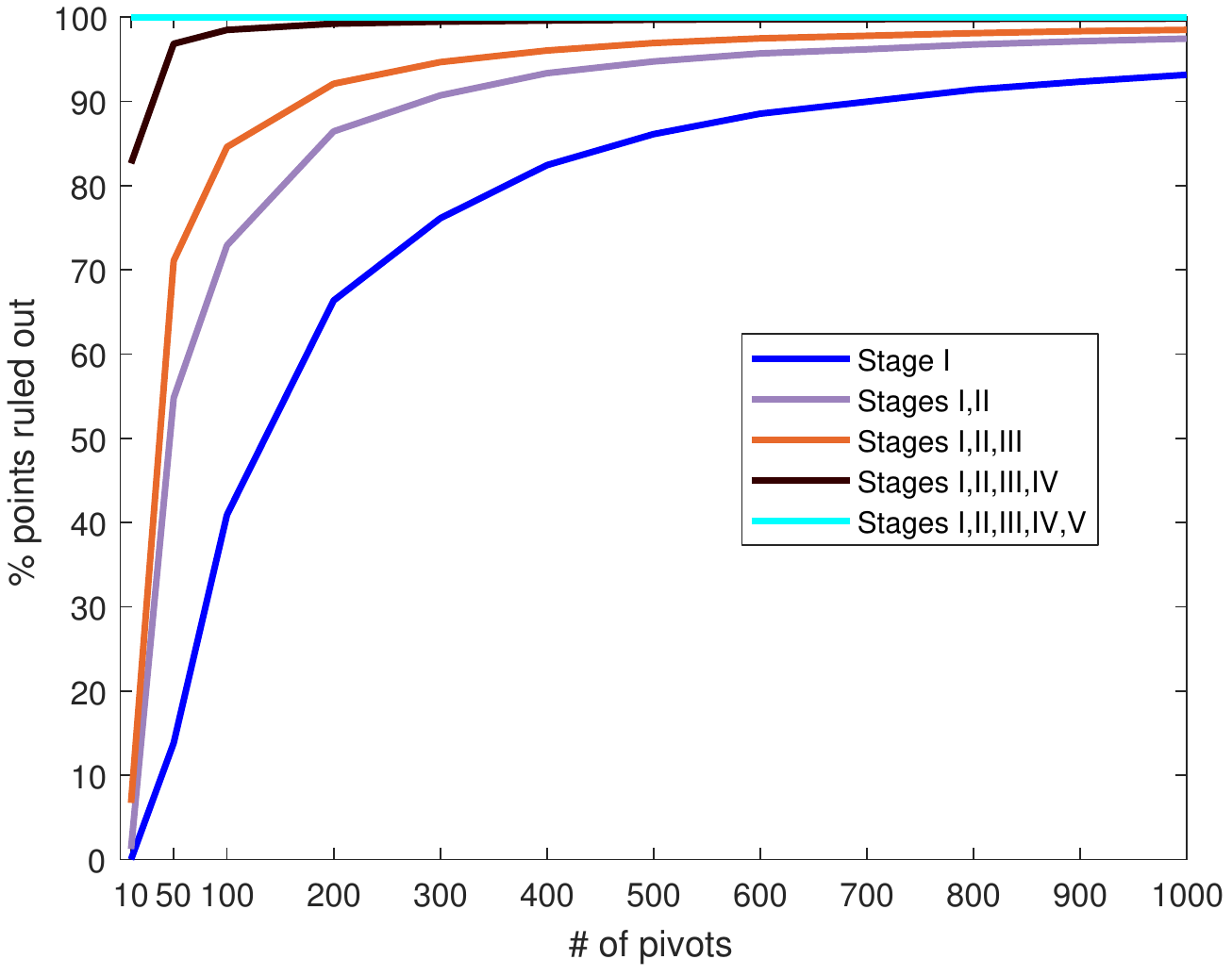}
     \subcaption{Stage \RNU{5}}
     \label{fig:RNGSavingsVisualized:RNGStage5SavingsCurve}
   \end{minipage}
   \begin{minipage}{0.21\textwidth}
     \centering
     \includegraphics[width=\textwidth]{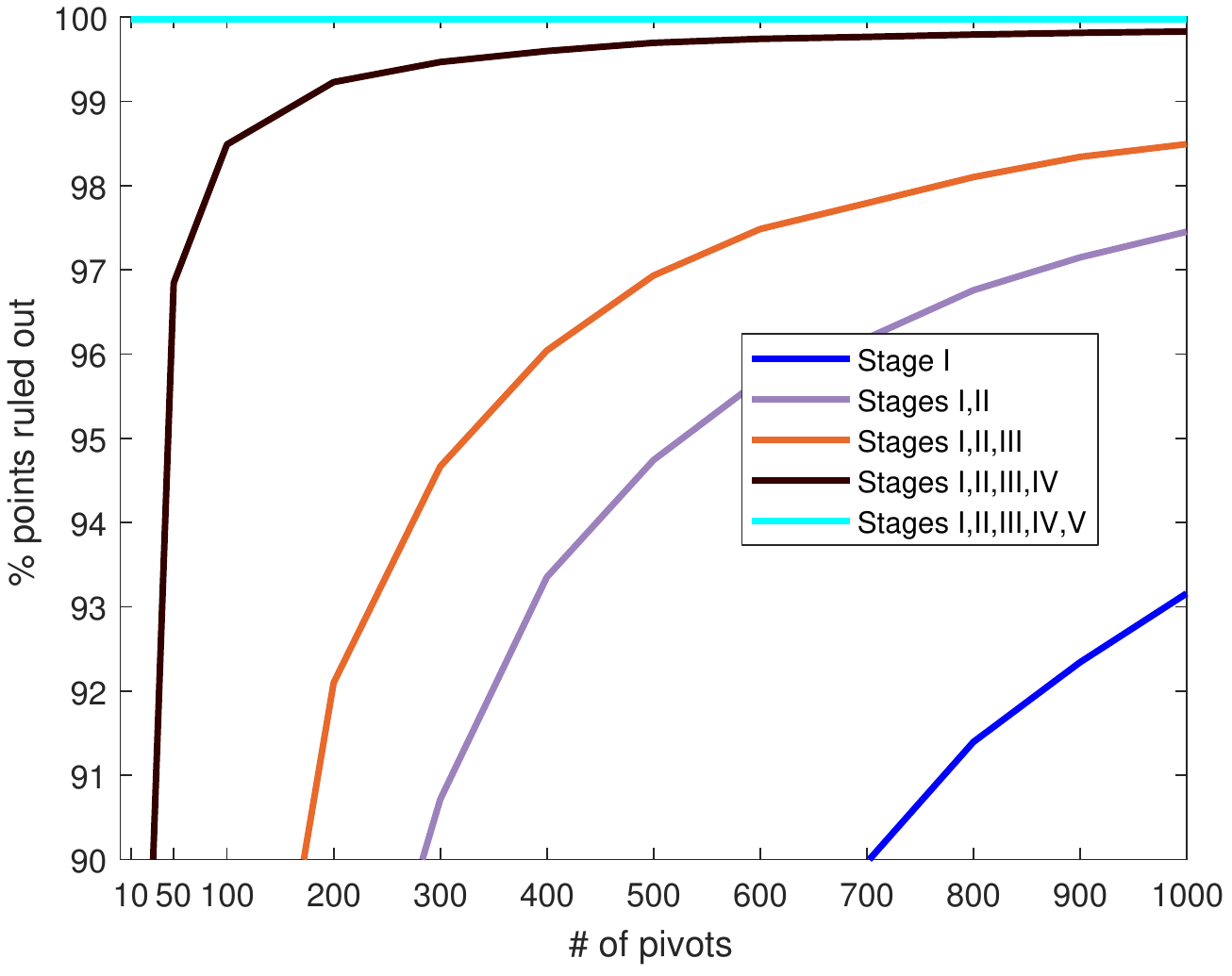}
     \subcaption{Stage \RNU{5}}
     \label{fig:RNGSavingsVisualized:RNGStage5SavingsCurveZoomed}
   \end{minipage}
   \begin{minipage}{0.21\textwidth}
     \centering
     \includegraphics[width=\textwidth]{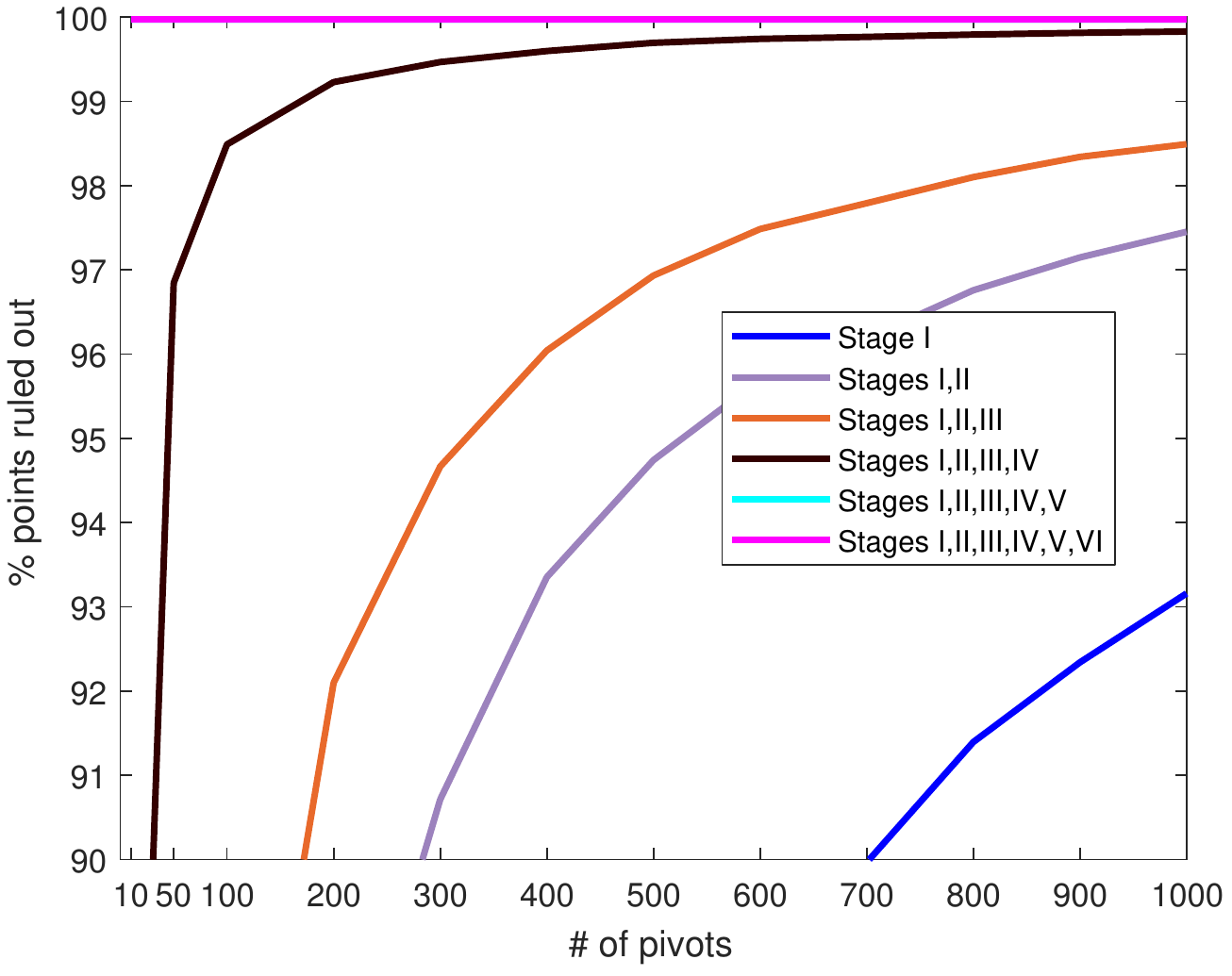}   
     \subcaption{Stage \RNU{6}}  
     \label{fig:RNGSavingsVisualized:RNGStage6SavingsCurveZoomed}
   \end{minipage}
   \vspace{-0.25cm}

   \caption{\footnotesize The savings achieved by Stages \RNU{1}-\RNU{6} for a GRNG-RNG Hierarchy with $M=200$ pivots on a dataset of $N=10,000$ uniformly distributed points in $[-1,1]^2$ where the green area shows remaining exemplars after each stage. (f),(g),(i), and (j) are zoomed in.}
   \label{fig:RNGSavingsVisualized}
\end{figure}

%
\noindent{\bf Stage \RNU{1}: Pivot-Pivot Interaction: } The most important implication of the GRNG($\mathcal{P}$) via Theorem \ref{thm:RNG_S1}, is that a lack of a GRNG link between $p_i$ and $p_j$ invalidates all potential links between their constituents. Stage \RNU{1} therefore begins by locating the pivot parents of $Q$ in $\mathcal{P}$, Equation \ref{eq:pivotMembership}. If $Q$ has no parents, $Q$ is added to the set of pivots $\mathcal{P}$ and GRNG($\mathcal{P}$) is updated. Otherwise, $Q$ can only have RNG links with the common GRNG neighbors of \emph{all} of $Q$'s parents. See Figure \ref{fig:RNGSavingsVisualized}.

%

%
\noindent{\bf Stage \RNU{2}: Query-Pivot Interaction: } Stage \RNU{1} removes entire pivot domains from interacting with $Q$, namely, those exemplars in the domain of pivots that do not have GRNG links to \emph{all} parents of $Q$. Note, however, that the GRNG lune is significantly reduced in size due to the increased radii, in comparison with RNG, \emph{i.e.}, by $2r_i + r_j, r_i + 2r_j$ on each side. This stage enlarges the G-lune by considering $Q$ itself as a virtual parent pivot with $r_Q=0$.

\begin{proposition} \label{prop:RNG_S2}
    If $p_k$ is in the G-lune of $(p_i,r_i)$ and $(Q,r_Q=0)$, i.e.,
    
    \begin{subnumcases}{\label{eq:RNG_S2_equations}}
        d\left(Q,p_{k}\right)< d\left(Q,p_{i}\right)-r_{i} \label{eq:RNG_S2_equations_1} \\ 
        d\left(p_{i},p_{k}\right)< d\left(Q,p_{i}\right)-2r_{i}. \label{eq:RNG_S2_equations_2} 
    \end{subnumcases}
    Then, $p_k$ is also in the RNG lune($Q,x_i$) $\forall x_i \in \mathcal{D}(p_i,r_i)$, thereby invalidating it, i.e., $\max\left(d\left(p_{k},Q\right),d\left(p_{k},x_{i}\right)\right)<d\left(Q,x_{i}\right)$. 
\end{proposition}
\iffull
\begin{proof}
    Apply Theorem~\ref{thm:RNG_S1} with $p_{i}=Q$, $p_{j}=p_{j}$ and $p_{k}=p_{k}$ with radii $r_{i}=r_{Q}=0$. The conditions of the theorem is then
    
    \begin{subnumcases}{\label{eq:RNG_S2_equations_proof}}
        d\left(p_{k},Q\right)< d\left(Q,p_{j}\right)-\left(2r_{Q}+r_{j}\right)\\
        d\left(p_{k},p_{j}\right)< d\left(Q,p_{j}\right)-\left(r_{Q}+2r_{j}\right),
    \end{subnumcases}
    which are Equations~\ref{eq:RNG_S2_equations} and thus holds by assumption. The consequence of the theorem is then $\max\left(d\left(p_{k},x_{i}\right),d\left(p_{k},x_{j}\right)\right) < d\left(x_{i},x_{j}\right)$, for any $x_{i}$ and $x_{j}$ in the pivot domains of $Q$ and $p_{j}$, respectively. Since the only member of the pivot $Q$ is $Q$, then $\max\left(d\left(p_{k},Q\right),d\left(p_{k},x_{j}\right)\right) < d\left(Q,x_{j}\right)$.
    \qed
\end{proof}
\fi
Note that since $Q$ is not really a pivot, we cannot simply lookup $GRNG$ neighbors of it. Rather, Equations \ref{eq:RNG_S2_equations} must be explicitly checked for all pivots $p_i$ that survive the elimination round of Stage \RNU{1}. Thus, additional entire pivot domains are eliminated, Figure \ref{fig:RNGSavingsVisualized}.

%
\noindent{\bf Stage \RNU{3}: Pivot-Exemplar Interaction: } This stage is symmetric with Stage \RNU{2} by enlarging the G-lune, but instead of using $Q$ as a virtual pivot, an exemplar is used a a virtual, zero-radius pivot. These exemplar are constituents $x_j$ of surviving pivots $p_j$.

%
\begin{proposition} \label{prop:RNG_S3}
    If a pivot $p_k$ falls in the G-lune of a parent $(p_i,r_i)$ of $Q$  and $(x_j,r_j=0)$, i.e.,
    
    \begin{subnumcases}{\label{eq:exemplarVirtualPivot}}
        d\left(p_{k},p_{i}\right)< d\left(p_{i},x_{j}\right)-2r_{i}\\
        d\left(p_{k},x_{j}\right)< d\left(p_{i},x_{j}\right)-r_{i},
    \end{subnumcases}
    then $\max\left(d\left(p_{k},Q\right),d\left(p_{k},x_{j}\right)\right)<d\left(Q,x_{j}\right)$ and $Q$ cannot have a RNG link with $x_j$.
\end{proposition}
\iffull
\begin{proof}
    Apply Theorem~\ref{thm:RNG_S1} with $x_j$ as $p_j$ with radii $r_{j}=0$. Then, the condition of Theorem~\ref{thm:RNG_S1} is
    
    \begin{subnumcases}{\label{eq:RNG_S3_equations_proof}}
        d\left(p_{k},p_{i}\right)< d\left(p_{i},x_{j}\right)-\left(2r_{i}+0\right)\\
        d\left(p_{k},x_{j}\right)< d\left(p_{i},x_{j}\right)-\left(r_{i}+0\right),
    \end{subnumcases}
    which is the premise of the proposition. Then by Theorem~\ref{thm:RNG_S1}, using $Q$ as an exemplar in the pivot domain of $p_{i}$, and $x_j$ as the sole exemplar in the pivot domain of $x_{j}$ gives $\max\left(d\left(p_{k},Q\right),d\left(p_{k},x_{j}\right)\right)<d\left(Q,x_{j}\right)$.
    \qed
\end{proof}
\fi
In Stage~\RNU{3}, then, for all parents of $Q$, ($p_i,r_i$), and each exemplar $x_j$ of the remaining pivots $p_j$, Equations \ref{eq:exemplarVirtualPivot} are checked which if valid rule out the exemplar $x_j$. Note that once a $p_k$ is found that eliminates $x_j$, the process stops, so it is judicious to pick $p_k$ in order of distance to $p_i$ as closer pivots are more likely to fall in the G-lune of $p_i$ and $x_j$, Figure \ref{fig:RNGSavingsVisualized}.

%
\noindent{\bf Stage \RNU{4}: Pivot-Mediated Exemplar-Exemplar Interactions:} The aim of the next three stages is to prevent brute-force examination of all exemplars $x_k$ potentially invalidating RNG link($Q$,$x_i$) by falling in lune($Q,x_i$). In Stage~\RNU{4} only pivots are checked, \emph{i.e.}, whether pivot $p_k$ satisfies 

%
\begin{equation} \label{eq:pivotPrioritization}
    \max\left(d\left(p_{k},Q\right),d\left(p_{k},x_{i}\right)\right)<d\left(Q,x_{i}\right), \quad k=1,2,...,M\,.
\end{equation}
Observe that only $p_k$ for which $d(p_k,Q) < d(Q,x_i)$ need to be considered, and for those $d(p_k,x_i) < d(Q,x_i)$ is checked. Note that if one $p_k$ satisfies this, link($Q,x_j$) is invalidated and the process is stopped, Figure \ref{fig:RNGSavingsVisualized}.

%
\noindent{\bf Stage \RNU{5}: Exemplar-Mediated Exemplar-Exemplar Interactions:} In this stage, all the exemplars $x_k$ which may invalidate the potential RNG link between $Q$ and $x_i$ are explored by checking

%
\begin{equation} \label{eq:stage5BruteForceCheck}
    \max\left(d\left(Q,x_{k}\right),d\left(x_{i},x_{k}\right)\right)<d\left(Q,x_{i}\right).
\end{equation}
Observe that since the process stops if one $x_k$ falls in the lune, so it is judicious to begin with a select group of $x_k$ that would more likely fall in the lune($Q,x_i$). First, the closest neighbors of $x_i$ can be found by consulting the RNG neighbors of $x_i$ and neighbors of neighbors, and so on until $d(x_i,x_k)$ exceeds $d(Q,x_i)$. Second, since some distances $d(Q,x_k)$ have been computed and cached for other purposes, these can be rank-ordered and these $x_k$ can be explored until $d(Q,x_k)$ exceeds $d(Q,x_j)$, Figure \ref{fig:RNGSavingsVisualized}.

%
\noindent{\bf Stage \RNU{6}: RNG Link Verification:}
If the potential RNG link($Q,x_i$) is not invalidated by the select group of exemplars $x_k$, the entire remaining set of $x_k$ must exhaustively be considered to complete the verification. Note, however, that exemplars $x_k$ in pivot domain $p_k$ can be excluded from this consideration and without the costly computation of $d(Q,x_k)$ if the entire pivot domain is fully outside the lune($Q,x_i$):

%
\begin{proposition} \label{prop:RNG_S6_dmax}
    No exemplar $x_k$ of pivot domain $p_k$ can fall in lune($Q$,$x_i$) if 
    
    \begin{equation}
        \max(d\left(Q,p_{k}\right) - \delta_{\max}(p_k),d\left(x_{i},p_{k}\right) - \delta_{\max}(p_k)) \geq d(Q,x_i),
    \end{equation}
    \textrm{where} $\delta_{\max}(p_k)=\max_{\forall x_{k},d(x_{k},p_{k})\leq r_{k}}d\left(p_{k},x_{k}\right)$ is the maximum distance of exemplar $x_k \in \mathcal{D}(p_k,r_k)$ from $p_k$.
\end{proposition}
\iffull
\begin{proof}
    Let $x_k$ be in the pivot domain of $p_k$. Then
    
    \begin{subnumcases}{}
        \scalebox{0.88}{%
        $\displaystyle d\left(Q,x_{k}\right) \geq d\left(Q,p_{k}\right) - d(p_k,x_k) \geq d\left(Q,p_{k}\right) - \max_{\forall x_{k},d(x_{k},p_{k})\leq r_{k}}d\left(p_{k},x_{k}\right) \geq d(Q,x_j)$ 
        } \\
        \scalebox{0.88}{%
        $\displaystyle d\left(x_{j},x_{k}\right) \geq d\left(x_{j},p_{k}\right) - d(p_k,x_k) \geq d(x_j,p_k) - \max_{\forall x_{k},d(x_{k},p_{k})\leq r_{k}}d\left(p_{k},x_{k}\right) \geq d(Q,x_j)$,
        }
    \end{subnumcases}
    \noindent or $\max\left(d\left(Q,x_{k}\right),d\left(x_{j},x_{k}\right)\right) \geq d\left(Q,x_{j}\right)$, which puts $x_k$ outside the $\text{lune}(Q,x_j)$.
    \qed
\end{proof}
\fi
For the remaining pivot domains, the computation of $d(Q,x_k)$ can still be avoided for some exemplar $x_k$:

%
\begin{proposition} \label{prop:RNG_S6_assisted}
    Any exemplar $x_k$ in the pivot domain of $p_k$ for which 
    
    \begin{equation}
        \max( d\left(Q,p_{k}\right) - d\left(x_k,p_{k}\right), d\left(x_{i},p_{k}\right) - d\left(x_k,p_{k}\right) ) \geq d(Q,x_i) 
    \end{equation}
    falls outside lune($Q,x_i$). 
\end{proposition}
\iffull
\begin{proof}
    \begin{subnumcases}{\label{eq:RNG_S6_assisted_proof}}
        d\left(Q,x_{k}\right) \geq d\left(Q,p_{k}\right) - d\left(x_k,p_{k}\right) \geq d(Q,x_j) \\
        d\left(x_{j},x_{k}\right) \geq d\left(x_{j},p_{k}\right) - d\left(x_k,p_{k}\right) \geq d(Q,x_j),
    \end{subnumcases}
    \noindent or $\max\left(d\left(Q,x_{k}\right),d\left(x_{j},x_{k}\right)\right) \geq d\left(Q,x_{j}\right)$, which puts $x_k$ outside the $\text{lune}(Q,x_j)$.
    \qed
\end{proof}
\fi
Any exemplar $x_k$ which is not ruled out by Proposition \ref{prop:RNG_S6_dmax} and \ref{prop:RNG_S6_assisted} must now be explicitly considered. If none are in the lune($Q,x_i$), then link($Q,x_i$) is validated. 

%
\begin{algorithm}[t!]
\DontPrintSemicolon
\SetKwFunction{cov}{cov}
\KwIn{ Query $Q$, pivots $\mathcal{P}$, radius $r_i$ for $p_i \in \mathcal{P}$, GRNG($\mathcal{P}$), children $C(p_i)$ for $p_i \in \mathcal{P}$, max child distance $\delta_{\max}(p_i)$ for $p_i \in \mathcal{P}$, exemplar $\mathcal{X}$, RNG($X$), parents $P(x_i)$ for $x_i \in \mathcal{X}$, GRNG neighbors GRNG$(x_i)$ for $x_i \in \mathcal{X}$.}
\KwOut{ RNG neighbors of $Q$, Parents of $Q$, GRNG neighbors of $Q$.}
\Begin{
    \textbf{Stage 1:} Find parents $P(Q)=\{p_i \in \mathcal{P}: d(Q,p_i) \leq r_i \}$. Collect potential GRNG neighbors of $Q$ as $\mathcal{A}(\mathcal{P})=\bigcup_{p_i \in P(Q)} \text{GRNG}(p_i)$. If $|P(Q)|=0$, $\mathcal{A}(\mathcal{P}) = \mathcal{P}$.
    
    \textbf{Stage 2:} Find GRNG($Q$) by validating all potential neighbors $p_j \in \mathcal{A}(\mathcal{P})$. If no $p_k \in \mathcal{P}$ satisfies both $d(Q,p_k) < d(Q,p_j) - r_j$ and $d(p_j,p_k) < d(Q,p_j) - 2r_j$, then $p_j$ is added to GRNG($Q$).
    
    \textbf{Stage 3:} Collect potential RNG neighbors of $Q$ as $\mathcal{A}(\mathcal{X}) = \bigcup_{p_i \in P(Q)} C(p_i)$. Remove $x_j \in \mathcal{A}(\mathcal{X})$ if any $p_j \in P(x_j)$ is not in GRNG($Q$). Remove $x_j \in \mathcal{A}(\mathcal{X})$ if any $p_i \in P(Q)$ is not in GRNG($x_j$).
    
    \textbf{Stage 4:} Consider invalidation of link($Q,x_j$) for $x_j \in \mathcal{A}(\mathcal{X})$ by checking $p_k \in GRNG(Q)$ for interference. If any $p_k$ satisfies both $d(Q,p_k) < d(Q,x_j)$ and $d(x_j,p_k) < d(Q,x_j)$, then $x_j$ is removed from $\mathcal{A}(\mathcal{X})$.
    
    \textbf{Stage 5:} Consider invalidation of link($Q,x_j$) for $x_j \in \mathcal{A}(\mathcal{X})$ by checking $x_k \in \mathcal{A}(\mathcal{X})$ for interference. If any $x_k$ satisfies both $d(Q,x_k) < d(Q,x_j)$ and $d(x_j,x_k) < d(Q,x_j)$, then $x_j$ is removed from $\mathcal{A}(\mathcal{X})$.
    
    \textbf{Stage 6:} Consider invalidation of link($Q,x_j)$ for $x_j \in \mathcal{A}(\mathcal{X})$ by performing exhaustive check for interference. Use $\delta_{\max}(p_k)$ and $C(p_k)$ for $p_k \in \mathcal{P}$ with Propositions \ref{prop:RNG_S6_dmax} and \ref{prop:RNG_S6_assisted} to narrow down the set of potentially interfering points $x_k$. If no $x_k$ satisfies both $d(Q,x_k) < d(Q,x_j)$ and $d(x_j,x_k) < d(Q,x_j)$, then $x_j$ is added to RNG$(Q)$.
}
\caption{RNG Localization for Query $Q$ in GRNG-RNG Hierarchy.}
\label{algo:2L_localization}
\end{algorithm}

%
\noindent{\bf Stage \RNU{7}: Existing RNG Link Validation:}
The above six stages locate $Q$ in the RNG and identify its RNG neighbors. This is sufficient for a RNG search query. However, if the dataset $\mathcal{S}$ is to be augmented with $Q$, a final check must be made as to which existing RNG links would be removed by the presence of $Q$. While this is a brute force $O(\alpha N)$ operation, it is important to avoid computing $d(Q,x_i)$ for all $x_i \in \mathcal{S}$. Observe that $Q$ does not threaten links that are "too far" from it. This notion can be implemented if two parameters are maintained, one for exemplars and one for pivots:
%

\begin{equation} \label{eq:RNGUpdate_Stage7_dmax_pi} \small
    \bar{\mu}_{\max}\left(x_{i}\right)=\underset{x_{j} \in \textrm{RNG}\,(x_{i})}{\max}d\left(x_{i},x_{j}\right), 
    \mu_{\max}\left(p_{i}\right)=\underset{d\left(x_{i},p_{i}\right)\leq r_{i}}{\max}\left[\bar{\mu}_{\max}\left(x_{i}\right) + d(x_i, p_i)\right].
\end{equation}

%
\begin{proposition} \label{prop:RNG_S7_umax}
    A query $Q$ does not invalidate RNG links at $x_i$ if $d(Q,x_i) \geq \bar{\mu}_{\max}(x_i)$. A query $Q$ does not invalidate any RNG link of any exemplars $x_i \in \mathcal{D}(p_i,r_i)$, if $d(Q,p_i) > \mu_{\max}(p_i)$.
\end{proposition}
\iffull
\begin{proof}
    The query $Q$ lies outside $\text{lune}(x_i,x_j)$ because
    
    \begin{equation}
        \max\left(d\left(Q,x_{i}\right),d\left(Q,x_{j}\right)\right)\geq d\left(Q,x_{i}\right)\geq \bar{\mu}_{\max}\left(x_{i}\right)\geq d\left(x_{i},x_{j}\right).
    \end{equation}
    Consider and arbitrary exemplar $x_i$ in the pivot domain of $p_i$ and show that $d(Q,x_i) \geq \bar{\mu}_{\max}\left(x_{i}\right)$ so:
    \begin{equation}
        \scalebox{0.76}{%
        $d(Q,x_i) \geq d(Q,p_i) - d(p_i,x_i) \geq \mu_{\max}\left(p_{i}\right) - d(p_i,x_i) \geq \bar{\mu}_{\max}\left(x_{i}\right) + d(x_i,p_i) - d(p_i,x_i) \geq \bar{\mu}_{\max}\left(x_{i}\right)$.
        }
    \end{equation}
    \qed
\end{proof}
\fi
This proposition suggests a three-step procedure: (\emph{\RNL{1}}) remove entire pivot domains if $d(Q,p_i) \geq \mu_{\max}(p_i)$; (\emph{\RNL{2}}) remove all exemplars in the remaining pivot domains for which $d(Q,x_i) \geq \bar{\mu}_{\max}(x_i)$; (\emph{\RNL{3}}) check the RNG condition explicitly for the remaining $x_i$ and any $x_j$ it links to. This completes the incremental update of $\mathcal{S}$ to $\mathcal{S} \cup \{Q\}$.

\subsubsection{Experimental Results}
\begin{figure}[t!]
    \centering
    \scriptsize{(a)}
    \begin{minipage}{0.205\textwidth}
        \includegraphics[width=\textwidth]{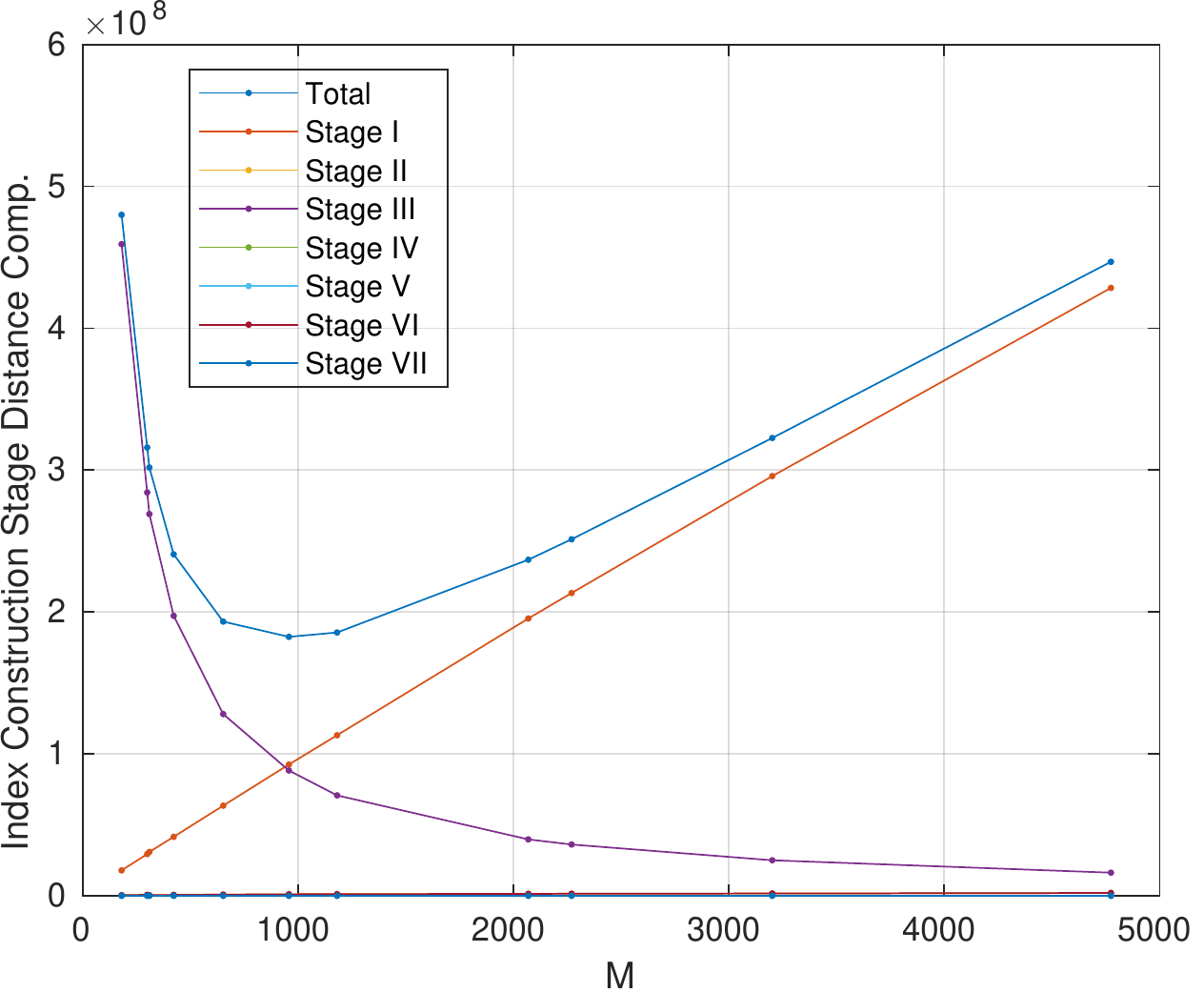}
    \end{minipage}
    \scriptsize{(b)}
    \begin{minipage}{0.205\textwidth}
        \includegraphics[width=\textwidth]{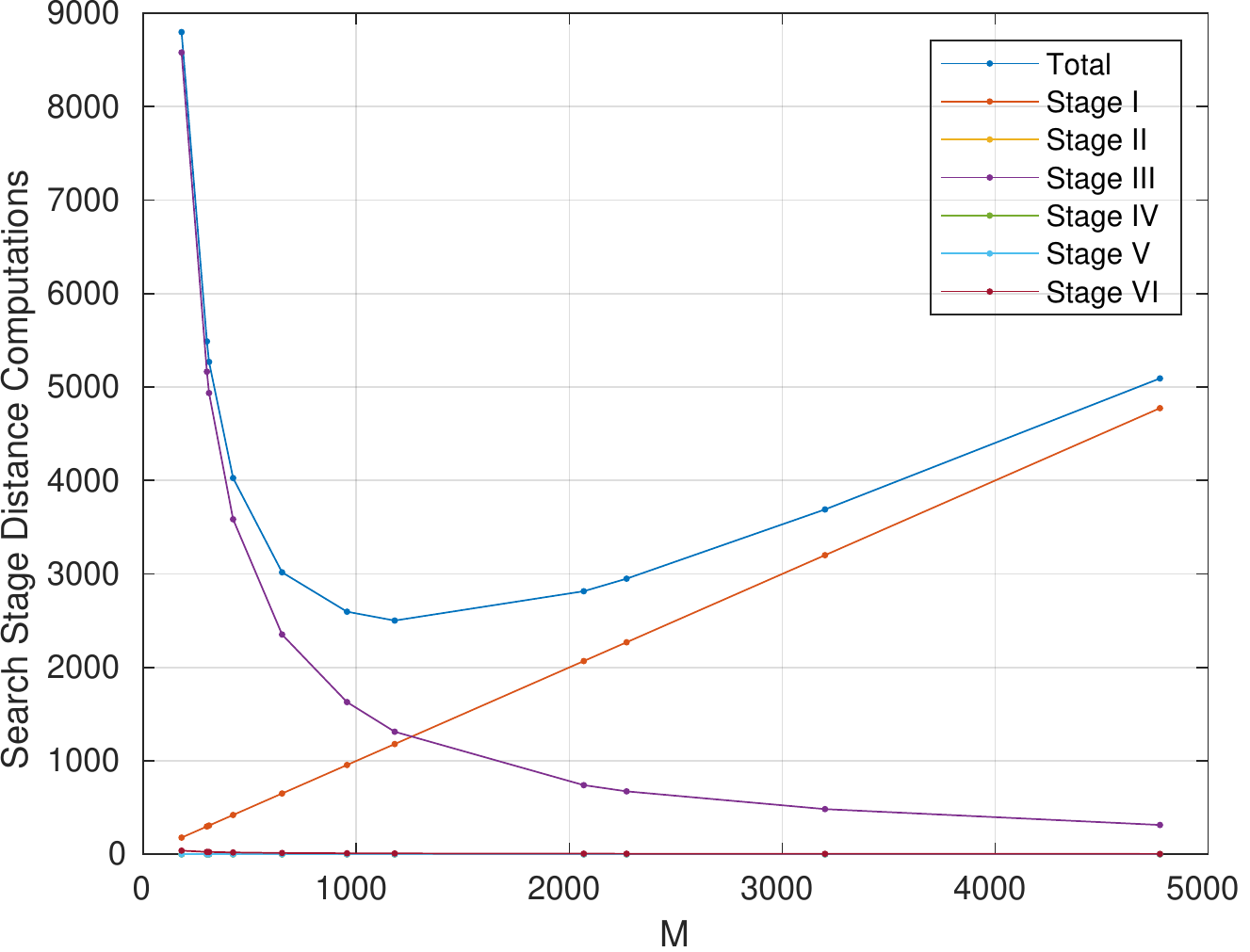}
    \end{minipage}
    \scriptsize{(c)}
    \begin{minipage}{0.205\textwidth}
        \centering
        \includegraphics[width=\textwidth]{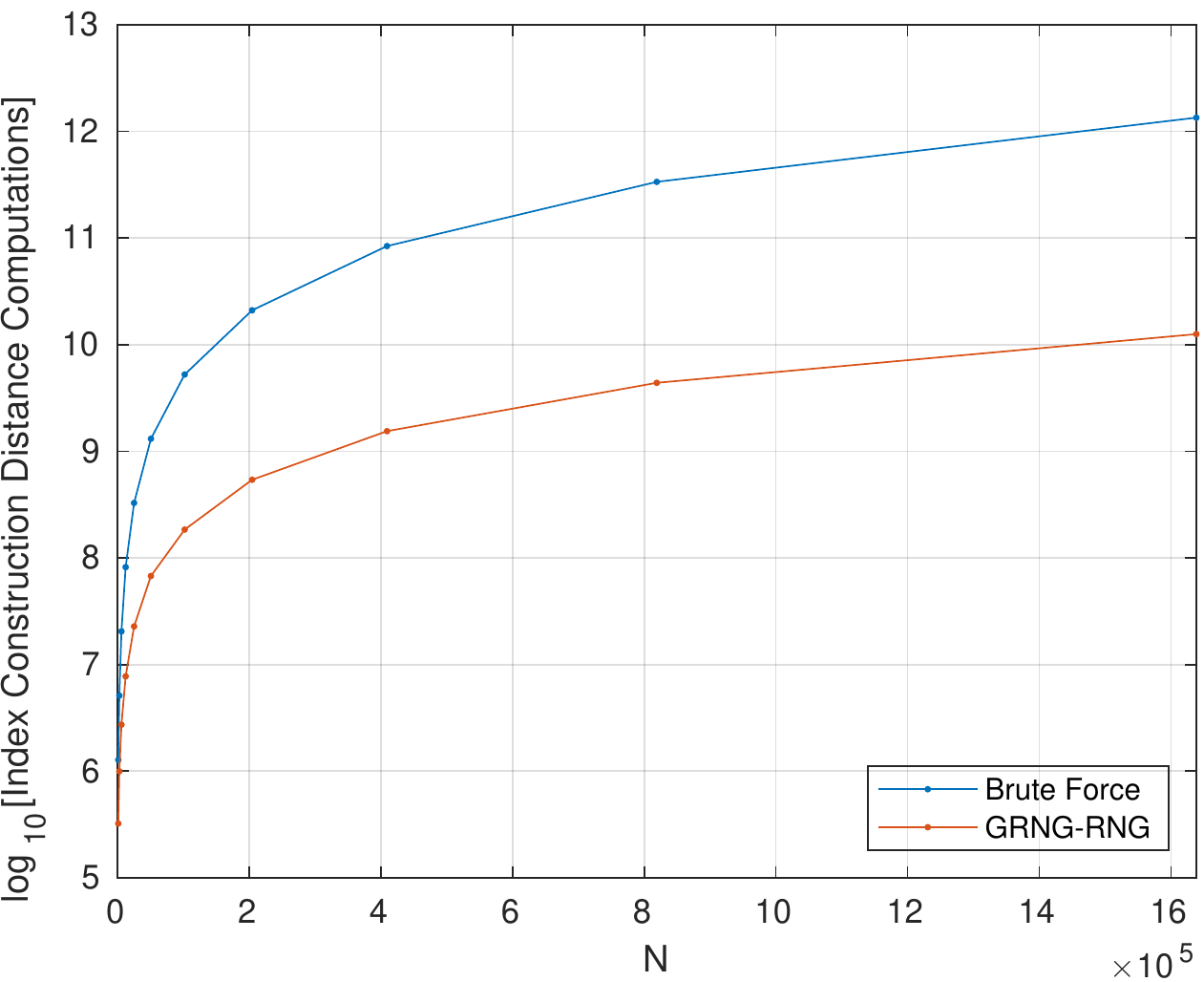}
    \end{minipage}
    \scriptsize{(d)}
    \begin{minipage}{0.205\textwidth}
        \centering
        \includegraphics[width=\textwidth]{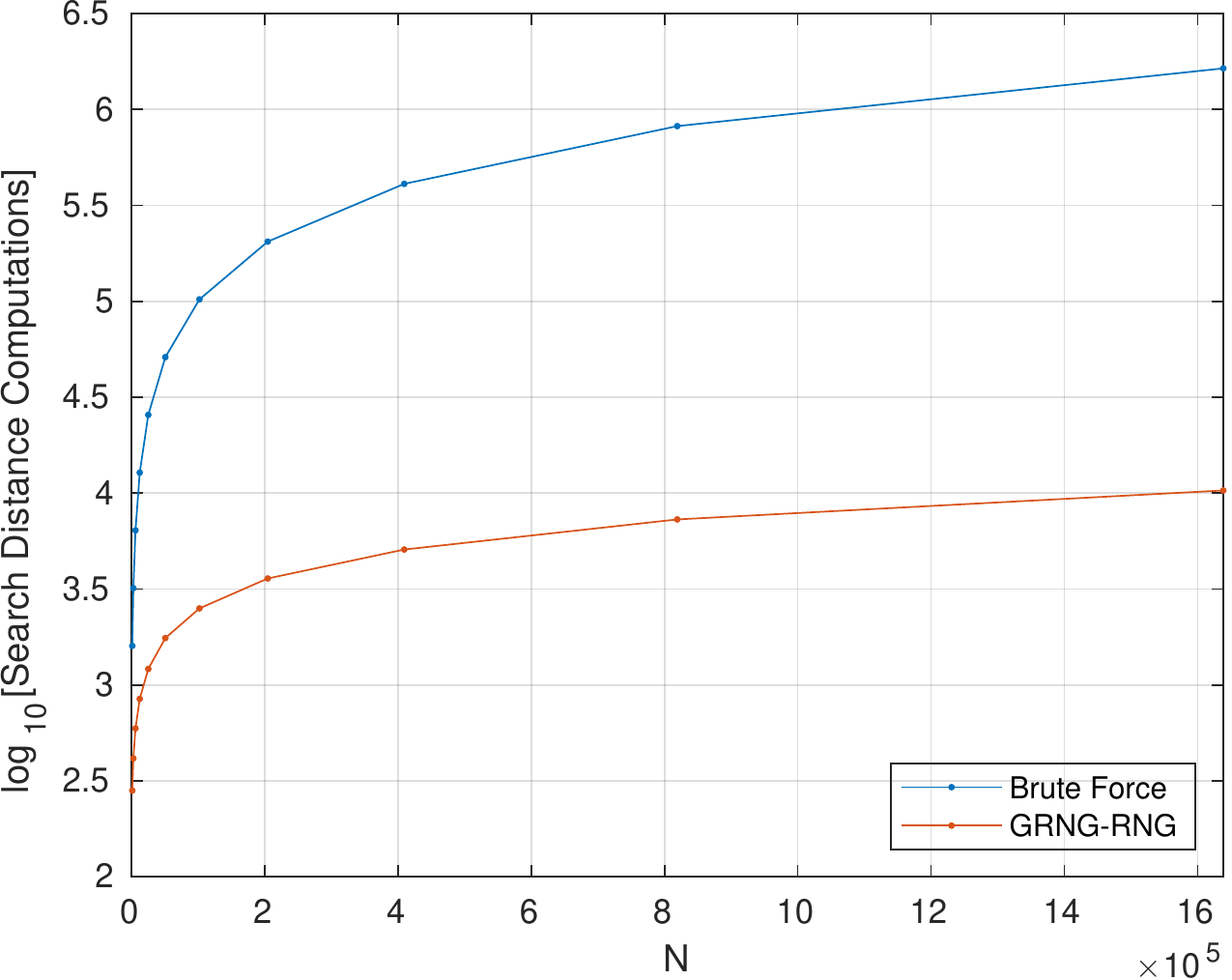}
    \end{minipage}
    \vspace{-0.25cm}

    \caption{\footnotesize Stage-by-stage distance computations for construction (a) and search (b) across different numbers of pivots $M$ for N=102,400 exemplars uniformly distributed in 2D. Comparison of our GRNG-RNG hierarchy for RNG construction (c) and search (d) to a Brute Force RNG algorithm that precomputes all distances.}
    \label{fig:2L_GRNG_RNG_Results}
\end{figure}

The improvements due to this two-layer GRNG-RNG configuration are examined in experiments by varying dimensions and number of exemplars. Figure \ref{fig:2L_GRNG_RNG_Results} examines the number of distance computations required for construction and search per stage as a function of the number of pivots. Observe that the first stage cost \CB increases exponentially \CB while the remaining stages experience an exponential drop. This is also observed for search distances per query. The total cost thus has an optimum for each. Since construction is offline while search is online, the number of pivots is optimized for the latter. Figure \ref{fig:2L_GRNG_RNG_Results}(c) examines the search costs for different dimensions. It is clear that search time rises exponentially with increasing dimension. Observe from Figure \ref{fig:2L_GRNG_RNG_Results}(b) that additional pivots would have enjoyed the exponential drop in all stages except for Stage \RNU{1} which involves GRNG Construction. If the cost of this stage as a function of $M$ can be lowered, the overall cost will be decreased dramatically. The next section proposes a two-layer scheme for constructing GRNG using a coarser GRNG in the same way the RNG construction was guided by a GRNG.

%


\section{Incremental Construction of the GRNG} \label{sec:build-grng}

The question naturally arises whether the construction of the GRNG of the pivot layer itself can benefit from a two-layer pivot-based indexing approach similar to the construction of the same for the RNG of the exemplars. Formally, let $\bar{\mathcal{P}}=\left\{(\bar{p}_{\bar{i}},\bar{r}_{\bar{i}}) | i=1,2,\ldots,\bar{M}\right\}$ denote pivots obtained from the previous section; refer to these as \emph{fine-scale pivots} to distinguish them from the \emph{coarse-scale pivots} $\mathcal{P}=\left\{(P_i,r_i) | i=1,2,\ldots,M\right\}$. The idea is for each coarse-scale pivot $p_i$ to represent a number of fine-scale pivots $\bar{p}_{\bar{i}}$. Define the \emph{Relative Pivot Domain} $\mathcal{D}(p_i,r_i)$ as the set of all fine-scale domain pivots $(\bar{p}_{\bar{i}},\bar{r}_{\bar{i}})$ whose entire exemplar domain is within a radius of $r_i$, \emph{i.e.}, $d(p_i,\bar{p}_{\bar{i}}) \leq r_i - \bar{r}_{\bar{i}}$. In this scenario, a query $Q$ is either a fine-scale pivot for now with $r_Q$ matching that of other fine-scale pivots, or it can be considered a fine-scale pivot with zero radius. The query computes $d(Q,p_i), i=1,2,\ldots,M$ and if $d(Q,p_i) < r_i - r_Q$, $p_i$ is a parent of $Q$. The question then arises as to what kind of graph structure for the coarse-scale pivots can efficiently locate a query in the GRNG of the fine-scale pivots. The following shows that the GRNG of coarse-scale pivots can accomplish this:

%
\begin{figure}[b!]

    %
    \begin{minipage}{0.19\textwidth}
     \centering
     \includegraphics[width=\textwidth]{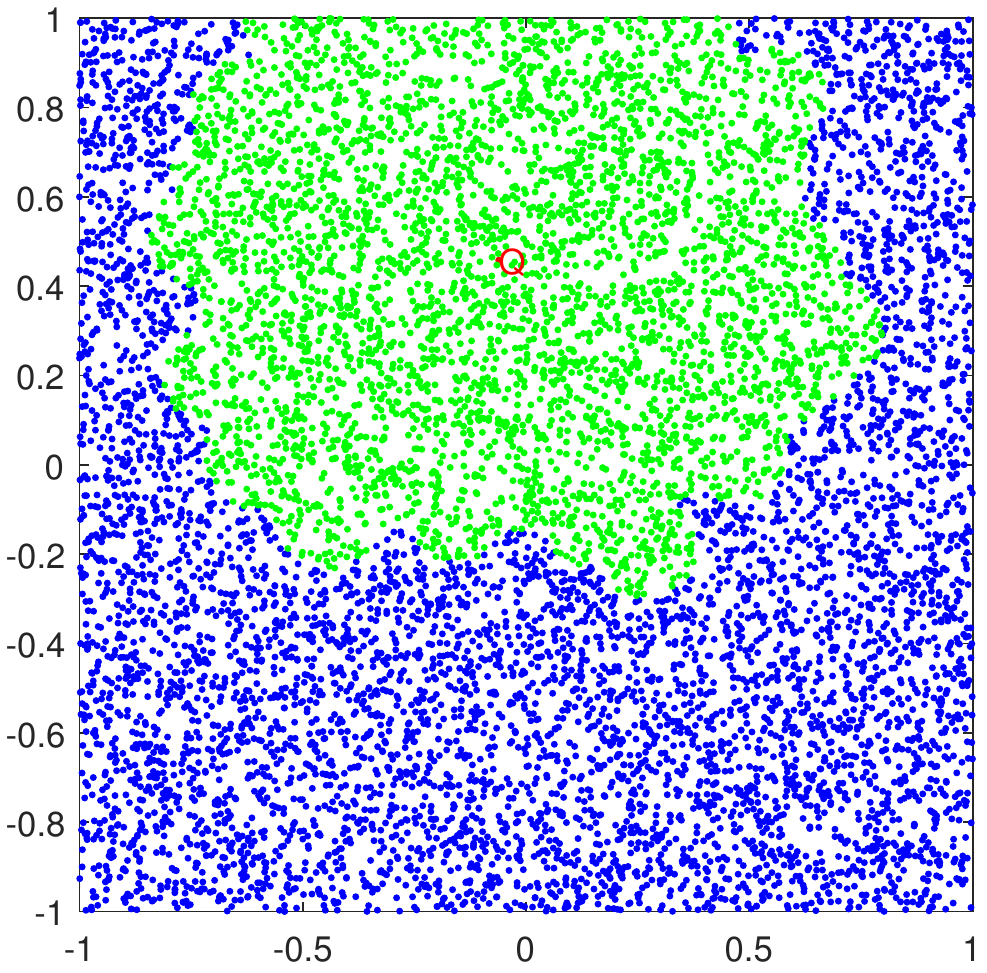}
     \subcaption{Stage \RNU{1}}
   \end{minipage}
   \begin{minipage}{0.19\textwidth}
     \centering
     \includegraphics[width=\textwidth]{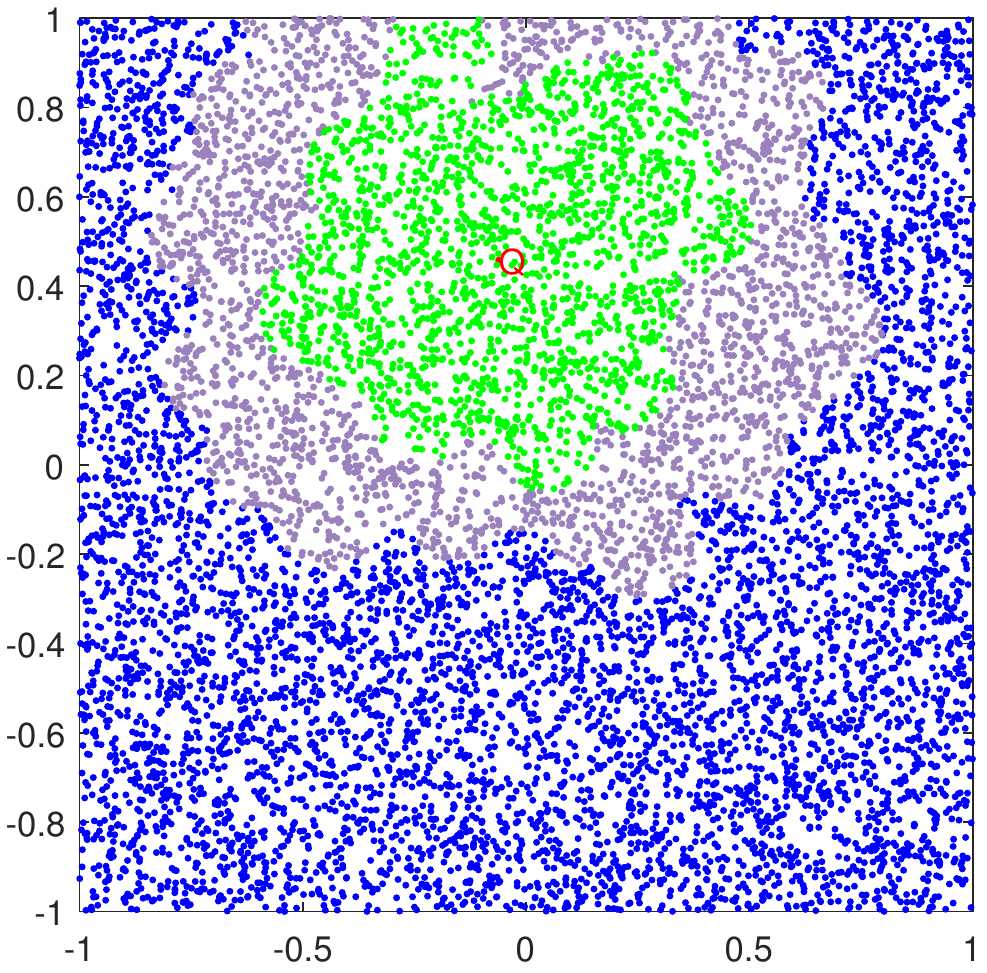}
     \subcaption{Stage \RNU{2}}
   \end{minipage}
   \begin{minipage}{0.19\textwidth}
     \centering
     \includegraphics[width=\textwidth]{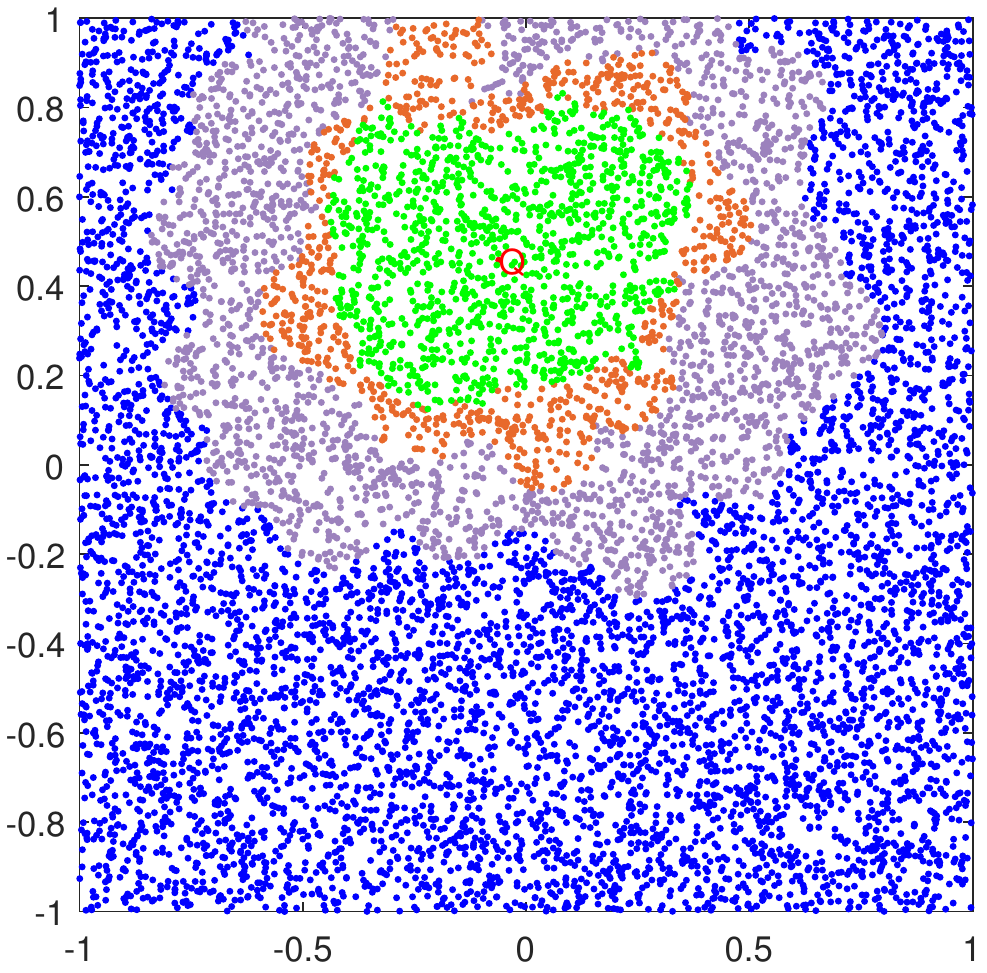}
     \subcaption{Stage \RNU{3}}
   \end{minipage}
   \begin{minipage}{0.19\textwidth}
     \centering
     \includegraphics[width=\textwidth]{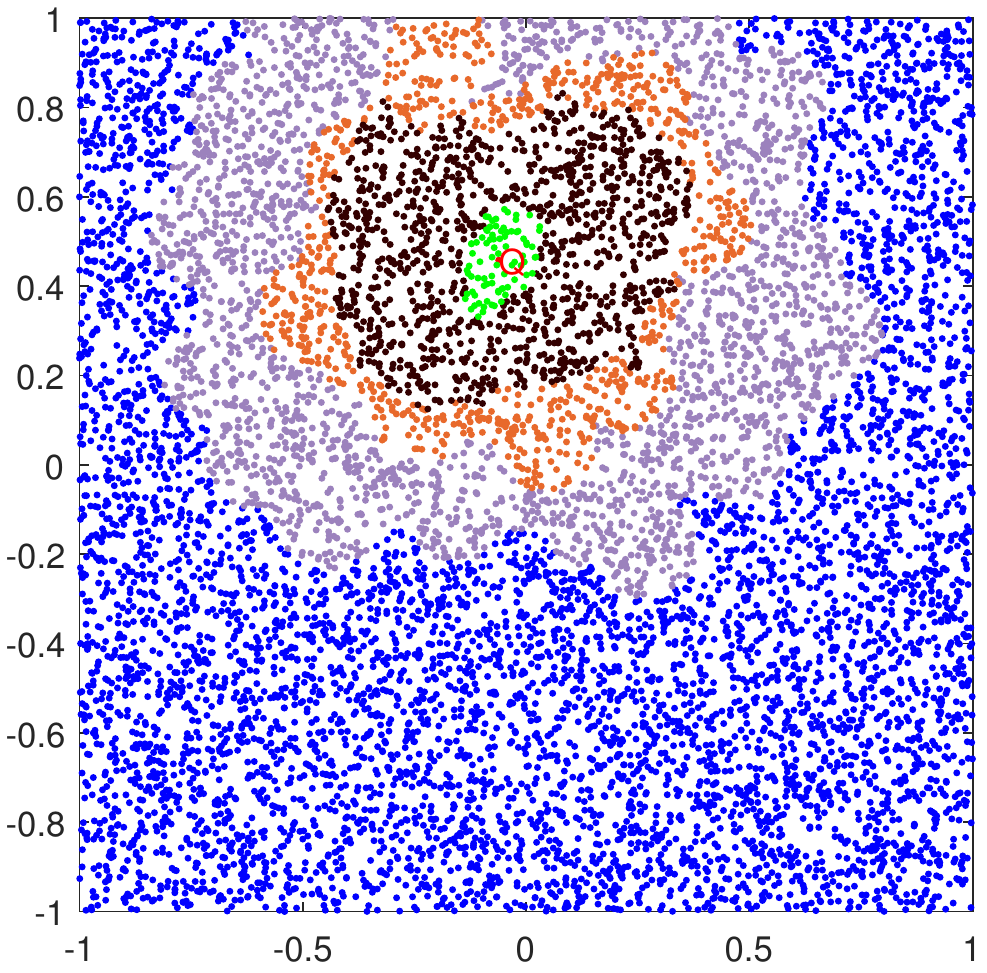}
     \subcaption{Stage \RNU{4}}
   \end{minipage}
   \begin{minipage}{0.19\textwidth}
     \centering
     \includegraphics[width=\textwidth]{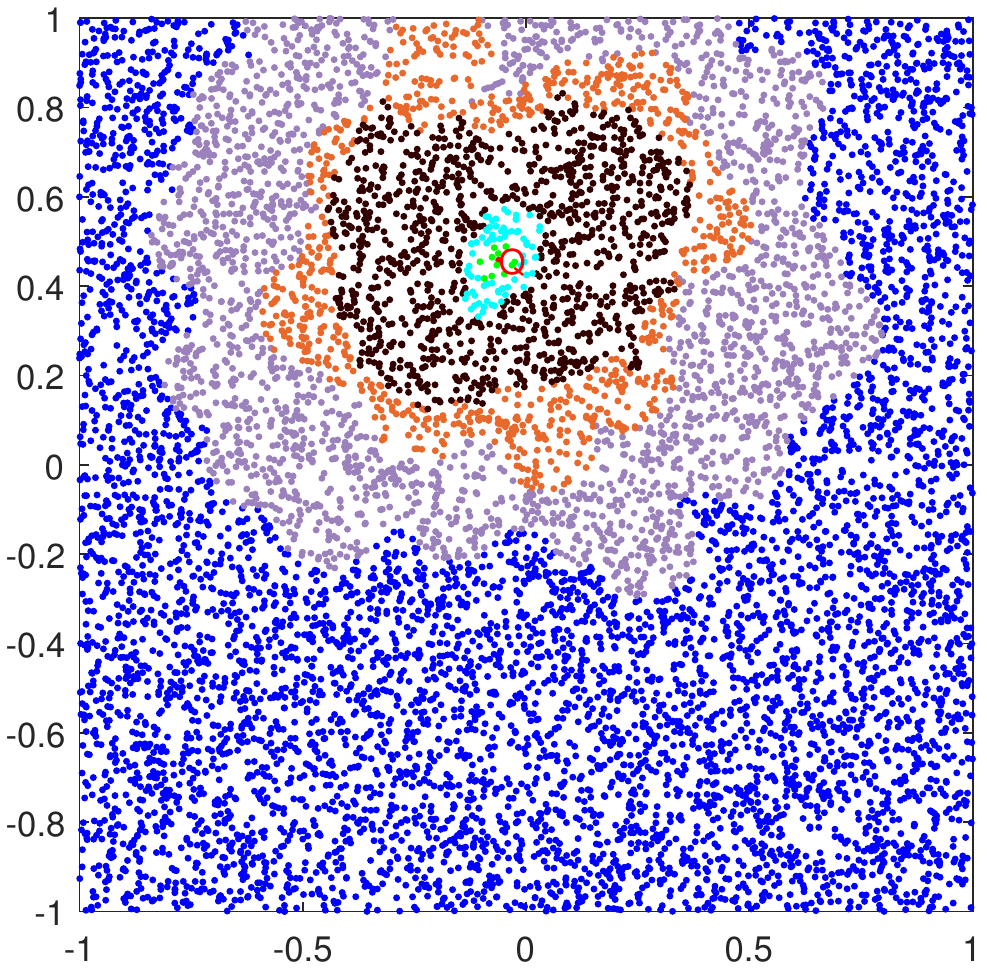}
     \subcaption{Stage \RNU{5}}
   \end{minipage}
   
    \begin{minipage}{0.16\textwidth}
     \centering
     \includegraphics[width=\textwidth]{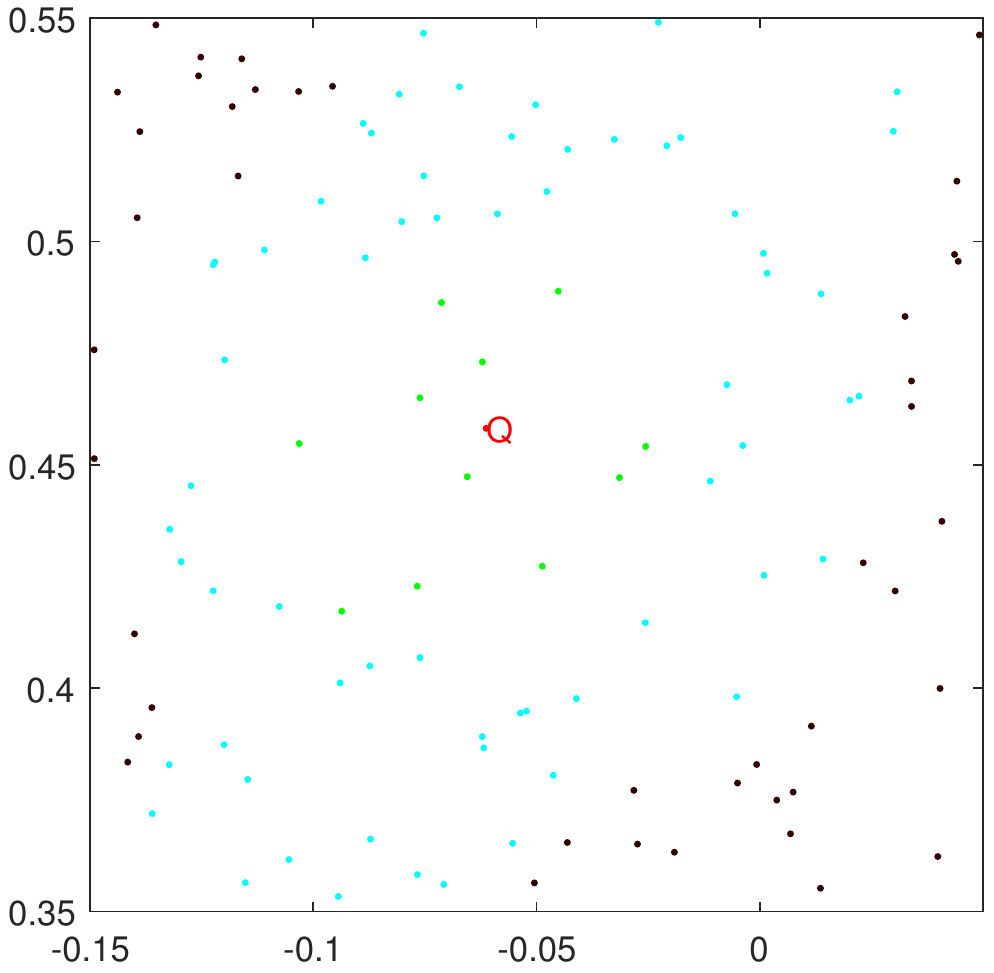}
     \subcaption{Stage \RNU{5}}
   \end{minipage}
   \begin{minipage}{0.16\textwidth}
     \centering
     \includegraphics[width=\textwidth]{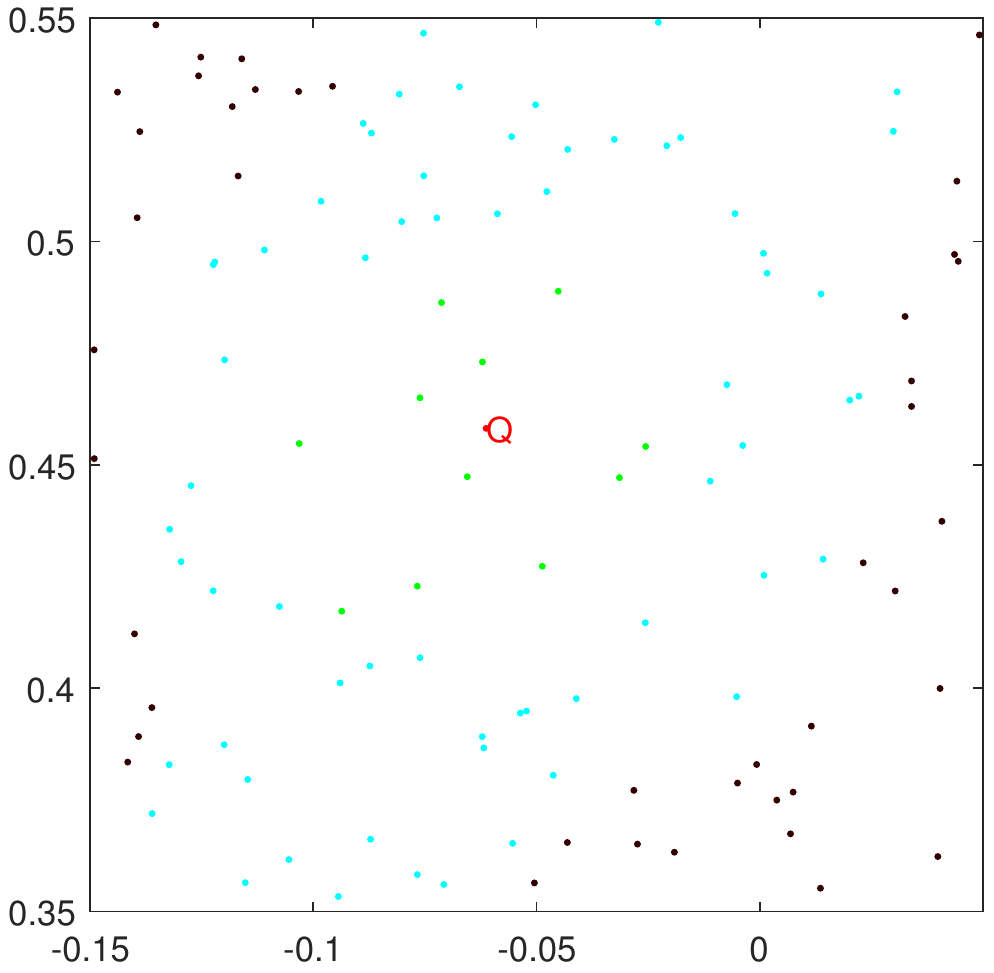}
     \subcaption{Stage \RNU{6}}
   \end{minipage}
   \begin{minipage}{0.21\textwidth}
     \centering
     \includegraphics[width=\textwidth]{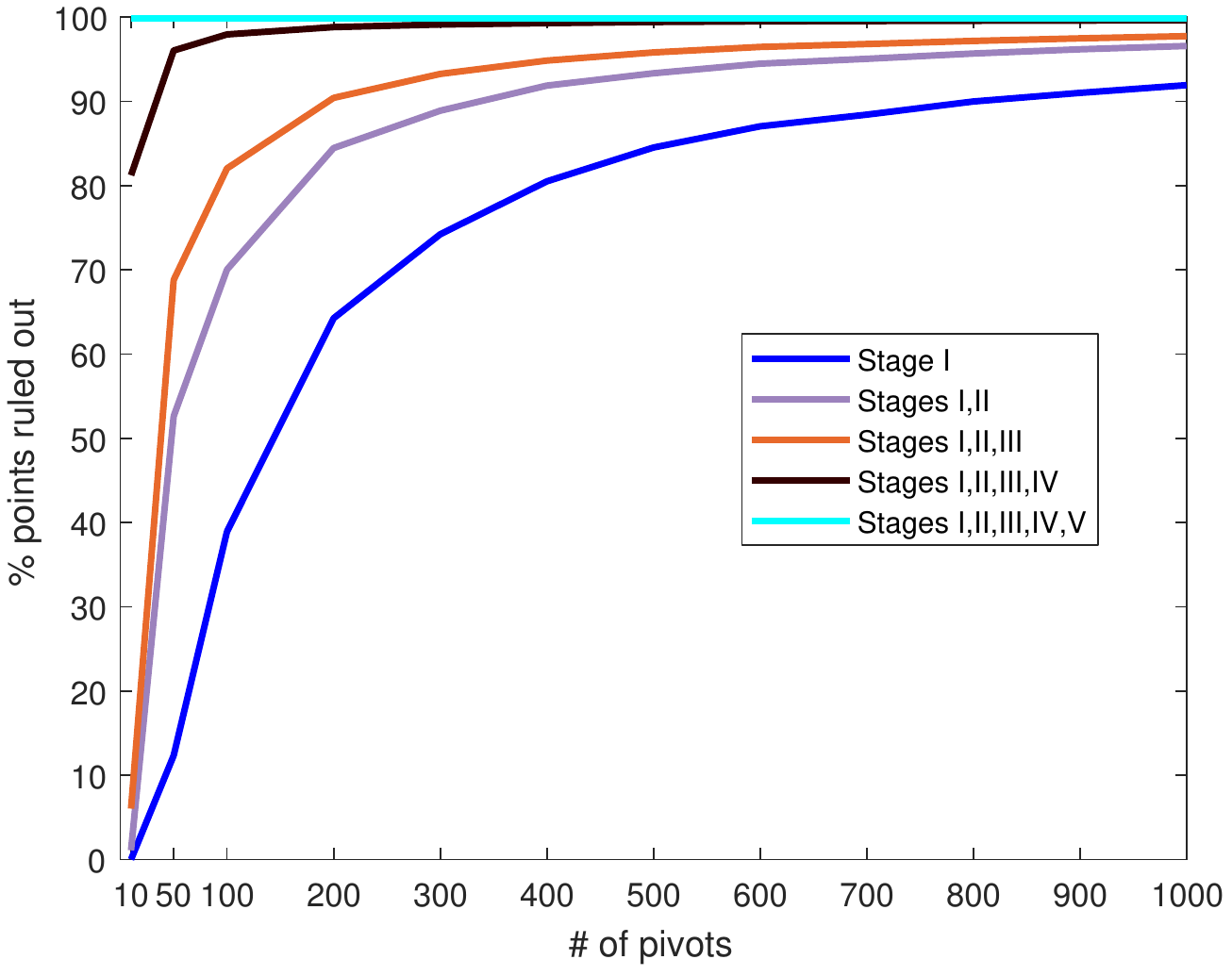}
     \subcaption{Stage \RNU{5}}
     \label{fig:GRNGSavingsVisualized:RNGStage5SavingsCurve}
   \end{minipage}
   \begin{minipage}{0.21\textwidth}
     \centering
     \includegraphics[width=\textwidth]{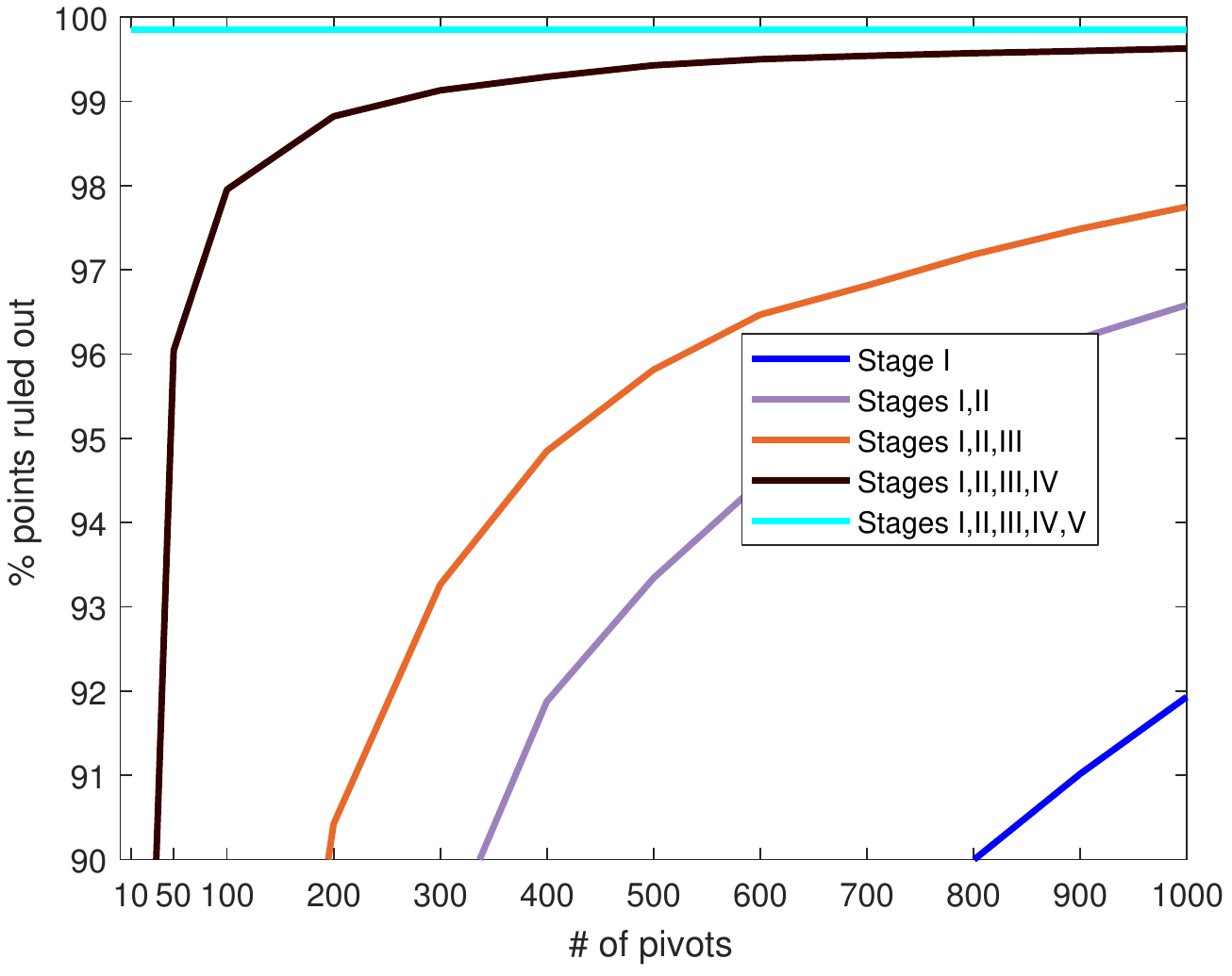}
     \subcaption{Stage \RNU{5}}
     \label{fig:GRNGSavingsVisualized:RNGStage5SavingsCurveZoomed}
   \end{minipage}
   \begin{minipage}{0.21\textwidth}
     \centering
     \includegraphics[width=\textwidth]{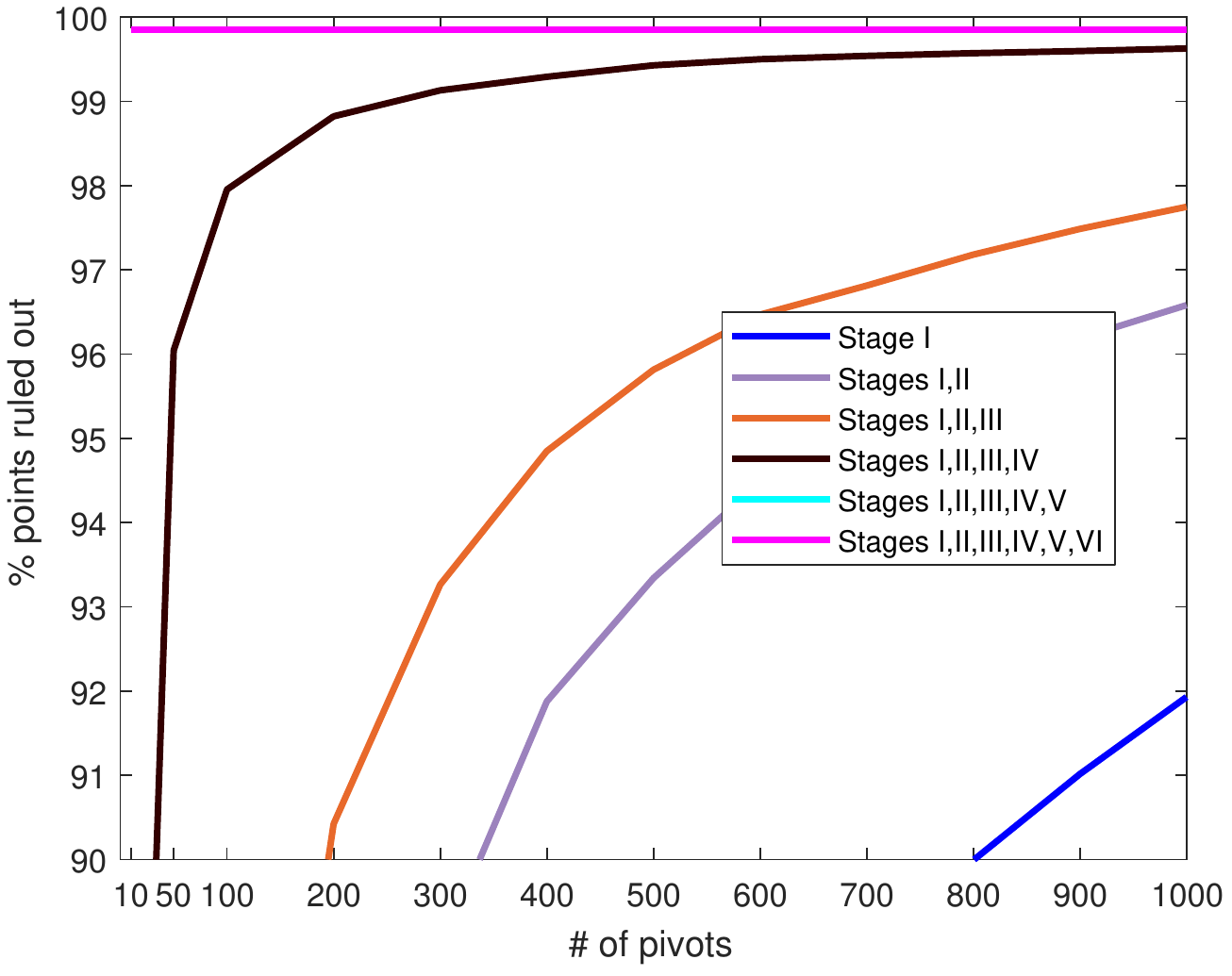}   
     \subcaption{Stage \RNU{6}}  
     \label{fig:GRNGSavingsVisualized:RNGStage6SavingsCurveZoomed}
   \end{minipage}
   \vspace{-0.25cm}

   \caption{\footnotesize The savings achieved by Stages \RNU{1}-\RNU{6} for a GRNG-GRNG Hierarchy of $M=200$ pivots with radius $r=0.005$ on a dataset of 10,000 uniformly distributed points in $[-1,1]^2$ where the green area shows remaining exemplars after each stage. (f),(g),(i), and (j) are zoomed in.}
   \label{fig:GRNGSavingsVisualized}
\end{figure}

%
\noindent{\bf Stage \RNU{1}: ``Coarse-Scale Pivot'' - ``Coarse-Scale Pivot'' Interactions:}

\begin{theorem} \label{thm:GRNG_S1}
    Consider two fine-scale pivots $(\bar{p}_{\bar{i}},\bar{r}_{\bar{i}}) \in \mathcal{D}(p_i,r_i)$ and  $(\bar{p}_{\bar{j}},\bar{r}_{\bar{j}}) \in \mathcal{D}(p_j,r_j)$. Then, if ($p_i,r_i$) and ($p_j,r_j$) do not share a GRNG link, ($\bar{p}_{\bar{i}},\bar{r}_{\bar{i}}$) and ($\bar{p}_{\bar{j}},\bar{r}_{\bar{j}}$) cannot have a GRNG link either.
\end{theorem}
\iffull
\begin{proof}
    Since $\bar{p}_{\bar{i}}$ and $\bar{p}_{\bar{j}}$ are in the relative-pivot domain of their coarse-scale pivots $p_i$ and $p_j$, respectively,
    
    \begin{subnumcases}{\label{eq:theorem:BerkGraphUpdate:stage1:preconditionsCombined}}
    d\left(\bar{p}_{\bar{i}},p_{i}\right) \leq r_{i}-\bar{r}_{\bar{i}} \label{eq:theorem:BerkGraphUpdate:stage1:precondition1}\\
    d\left(\bar{p}_{\bar{j}},p_{j}\right) \leq r_{j}-\bar{r}_{\bar{j}} \label{eq:theorem:BerkGraphUpdate:stage1:precondition2}.
    \end{subnumcases}
    That $p_i$ and $p_j$ do not have a GRNG link implies that there exists a coarse-level pivot $p_k$ such that
    
    \begin{subnumcases}{\label{eq:theorem:preconditionsCombined:hierarchicalBerkGraph}}
        d\left(p_{k},p_{i}\right) < d\left(p_{i},p_{j}\right)-\left(2r_{i}+r_{j}\right) \label{eq:theorem:precondition:hierarchicalBerkGraph:1}\\
        d\left(p_{k},p_{j}\right) < d\left(p_{i},p_{j}\right)-\left(r_{i}+2r_{j}\right) \label{eq:theorem:precondition:hierarchicalBerkGraph:2},
    \end{subnumcases}
    \textit{i.e.,} $p_k$ is in the G-lune of $p_i$ and $p_j$. It is now shown that $p_k$ is also in the G-lune of fine-scale pivots $\bar{p}_{\bar{i}}$ and $\bar{p}_{\bar{j}}$, and since all coarse-scale pivots are also fine-scale pivots, a GRNG link does not exist between $\bar{p}_{\bar{i}}$ and $\bar{p}_{\bar{j}}$.
    
    Observe first that $d\left(p_{k},\bar{p}_{\bar{i}}\right)$ can be related to $d\left(p_k,p_i\right)$ by a double application of the triangle inequality, and similarly for the $d\left(p_{k},\bar{p}_{\bar{j}}\right)$,
    
    \begin{subnumcases}{\label{theorem:hierarchicalBerkGraph:Stage1:cond1}}
        \scalebox{0.65}{%
        $d\left(p_{k},\bar{p}_{\bar{i}}\right) \leq d\left(p_{k},p_i\right) + d\left(p_{i},\bar{p}_{\bar{i}}\right) \leq d\left(p_{k},p_i\right) + \left(r_{i} - \bar{r}_{\bar{i}}\right) < d(p_i,p_j) - (2r_i + r_j) + (r_i - \bar{r}_{\bar{i}}) = d(p_i,p_j) - r_i - r_j - \bar{r}_{\bar{i}}$
        } \\
        \scalebox{0.65}{%
        $d\left(p_{k},\bar{p}_{\bar{j}}\right) \leq d\left(p_{k},p_j\right) + d\left(p_{j},\bar{p}_{\bar{j}}\right) \leq d\left(p_{k},p_j\right) + \left(r_{j} - \bar{r}_{\bar{j}}\right) < d(p_i,p_j) - (r_i + 2r_j) + (r_j - \bar{r}_{\bar{j}}) = d(p_i,p_j) - r_i - r_j - \bar{r}_{\bar{j}}$.
        }
    \end{subnumcases}
    Similarly, $d(p_i,p_j)$ can be related to $d\left(\bar{p}_{\bar{i}},\bar{p}_{\bar{j}}\right)$ by an application of the triangle inequality
    
    \begin{equation}
    \label{theorem:hierarchicalBerkGraph:Stage1:cond2}
    d\left(p_{i},p_{j}\right) \leq d\left(p_i,\bar{p}_{\bar{i}}\right) + d\left(\bar{p}_{\bar{i}},\bar{p}_{\bar{j}}\right) + d\left(\bar{p}_{\bar{j}},p_j\right) \leq d\left(\bar{p}_{\bar{i}},\bar{p}_{\bar{j}}\right) + (r_i - \bar{r}_{\bar{i}}) + (r_j - \bar{r}_{\bar{j}}).
    \end{equation}
    Combining Equations~\ref{theorem:hierarchicalBerkGraph:Stage1:cond1} and~\ref{theorem:hierarchicalBerkGraph:Stage1:cond2}
    
    \begin{subnumcases}{}
        \scalebox{0.92}{%
        $d\left(p_{k},\bar{p}_{\bar{i}}\right) < d\left(\bar{p}_{\bar{i}},\bar{p}_{\bar{j}}\right) + (r_i - \bar{r}_{\bar{i}}) + (r_j - \bar{r}_{\bar{j}}) - r_i - r_j - \bar{r}_{\bar{i}} = d\left(\bar{p}_{\bar{i}},\bar{p}_{\bar{j}}\right) - (2\bar{r}_{\bar{i}} + \bar{r}_{\bar{j}})$
        } \\
        \scalebox{0.92}{%
        $d\left(p_{k},\bar{p}_{\bar{j}}\right) < d\left(\bar{p}_{\bar{i}},\bar{p}_{\bar{j}}\right) + (r_i - \bar{r}_{\bar{i}}) + (r_j - \bar{r}_{\bar{j}}) - r_i - r_j - \bar{r}_{\bar{j}} = d\left(\bar{p}_{\bar{i}},\bar{p}_{\bar{j}}\right) - (\bar{r}_{\bar{i}} + 2\bar{r}_{\bar{j}})$,
        }
    \end{subnumcases}
    showing that $p_k$ falls in the G-lune of $d\left(\bar{p}_{\bar{i}},\bar{p}_{\bar{j}}\right)$ and thus no GRNG link can exist between them.
    \qed
\end{proof}
\fi
This theorem, in analogy to Theorem \ref{thm:RNG_S1} of the previous section, allows for the efficient localization of a query $Q$ for search in stating that the fine-scale GRNG neighbors of $Q$ are only among children of coarse-scale GRNG neighbors of $Q$'s parents, thus, removing entire pivot domains of non-neighbors, see Figure \ref{fig:GRNGSavingsVisualized}.

%
\noindent{\bf Stage \RNU{2}: Query - “Coarse-Scale Pivot” Interactions:}
In this stage, $(Q,r_Q)$ is considered as a virtual pivot. 

%
\begin{proposition} \label{prop:GRNG_S2}
    The query $Q$ does not form GRNG links with any children $(\bar{p}_{\bar{i}},\bar{r}_{\bar{i}})$ of those coarse-scale pivots $(p_i,r_i)$ that do not form a GRNG link with $Q$ when considered as a virtual pivot with $r_Q=0$. 
\end{proposition}
\iffull
\begin{proof}
    Apply Theorem~\ref{thm:GRNG_S1} with $(Q,\bar{r}_Q)$ acting both as $(p_i,r_i)$ and also as $(\bar{p}_{\bar{i}},\bar{r}_{\bar{i}})$, \textit{i.e.,} $Q$ is a coarse-scale pivot with a single member, itself, since $r_i - \bar{r}_{\bar{i}} = \bar{r}_Q - \bar{r}_Q = 0$.
    \qed
\end{proof}
\fi
%

%
\noindent{\bf Stage \RNU{3}: “Coarse-Scale Pivot” – “Fine-Scale Pivot” Interactions:}
This stage is mirror symmetric to Stage \RNU{2}, except that instead of treating $Q$ as a virtual coarse-scale pivot, a specific fine-scale pivot $(\bar{p}_{\bar{j}},\bar{r}_{\bar{j}})$ is considered a virtual pivot. 

%
\begin{proposition}\label{prop:GRNG_S3}
    If $(\bar{p}_{\bar{j}},\bar{r}_{\bar{j}})$ does not form a coarse-scale GRNG link with a parent $(p_i,r_i)$ of $Q$, then $(\bar{p}_{\bar{j}},\bar{r}_{\bar{j}})$ does not form a fine-scale GRNG link with $(Q,r_Q)$.
\end{proposition}
The proof is simply an application of Theorem \eqref{thm:GRNG_S1} with $(\bar{p}_{\bar{j}},\bar{r}_{\bar{j}})$ considered as both a fine-scale and a coarse-scale pivot. This third stage rules out all the remaining fine-scale pivots which are not a GRNG neighbor of \textbf{all} $Q$'s parents, Figure \ref{fig:GRNGSavingsVisualized}.

%
\noindent{\bf Stage \RNU{4}: ``Coarse-Scale Pivot''--Mediated ``Fine-Scale Pivot'' Interactions:}
All the GRNG links between the remaining fine-scale pivots and $Q$ must now be investigated. In Stage \RNU{4} only coarse-scale pivots are considered as potential occupiers of the G-lune by probing

%
\begin{subnumcases}{\label{eq:GRNG-S4}}
    d\left(p_{k},Q\right) < d\left(Q,\bar{p}_{\bar{j}}\right) - \left(2\bar{r}_{Q}+\bar{r}_{\bar{j}}\right) \\
    d\left(p_{k},\bar{p}_{\bar{j}}\right) < d\left(Q,\bar{p}_{\bar{j}}\right) - \left(\bar{r}_{Q}+2\bar{r}_{\bar{j}}\right).
\end{subnumcases}
Since $d(Q,\bar{p}_{\bar{j}}) - (2r_Q + \bar{r}_{\bar{j}})$ is a known value, only pivots $p_k$ closer to $Q$ than this value need to be considered. Similarly, for $d(\bar{p}_{\bar{j}},\bar{r}_{\bar{j}}) \in \mathcal{D}(p_j,r_j)$, observe that $d(p_k,\bar{p}_{\bar{j}}) \geq d(p_k,p_j) - (r_j - \bar{r}_{\bar{j}})$, so that if $d(p_k,p_j) \geq d(Q,\bar{p}_{\bar{j}}) - (r_Q + 2\bar{r}_{\bar{j}}) + (r_j - \bar{r}_{\bar{j}})$, then Equation (\ref{eq:GRNG-S4}b) does not hold and there is no need to consider such $p_k$. Thus, very few $p_k$ are actually considered, Figure \ref{fig:GRNGSavingsVisualized}.

%
\begin{figure}[t!]
    \centering
    \begin{minipage}{0.70\textwidth}
    
        (a)
        \begin{minipage}{0.40\textwidth}
            \centering
            \includegraphics[width=\textwidth]{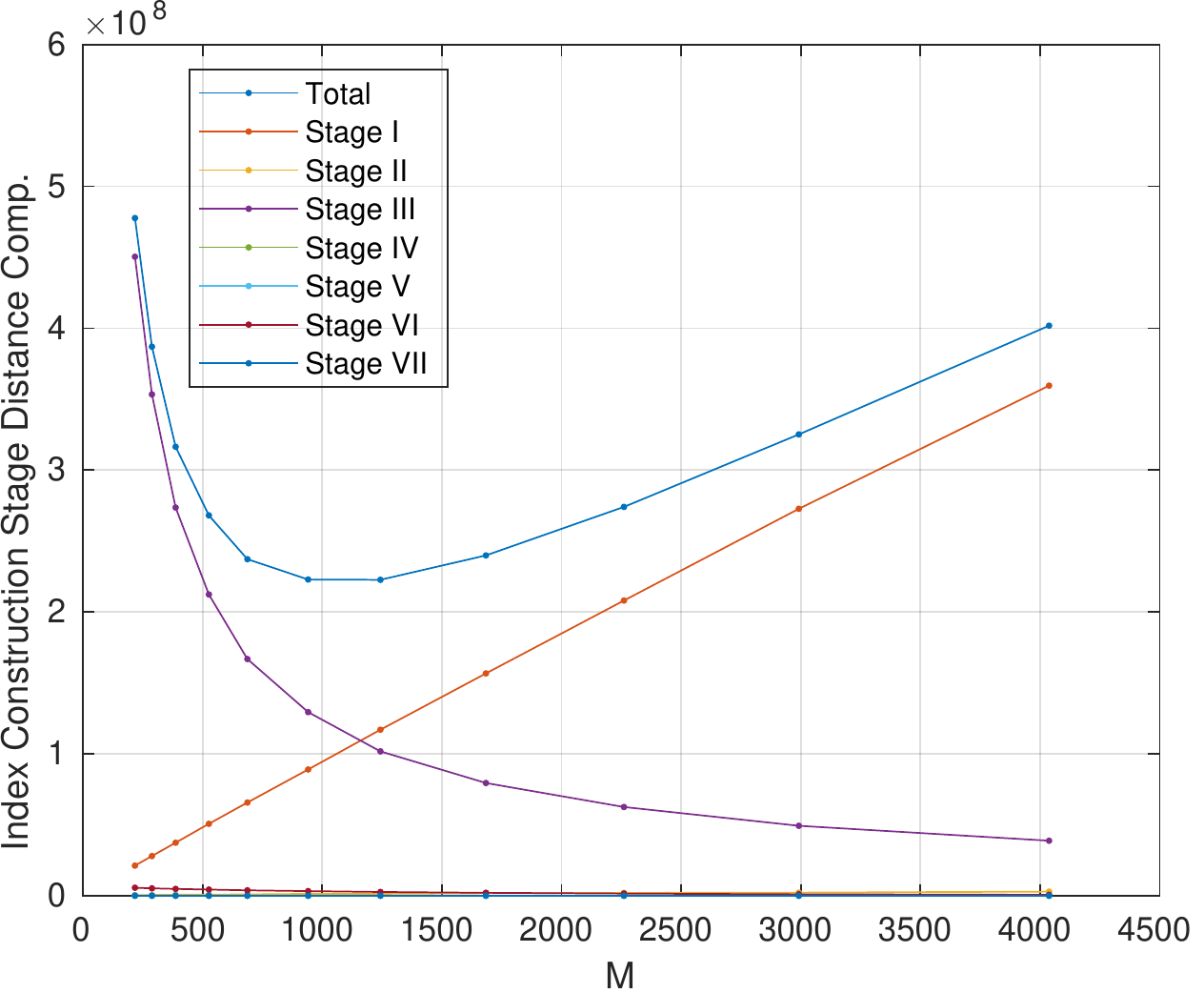}
        \end{minipage}
        (b)
        \begin{minipage}{0.40\textwidth}
            \centering
            \includegraphics[width=\textwidth]{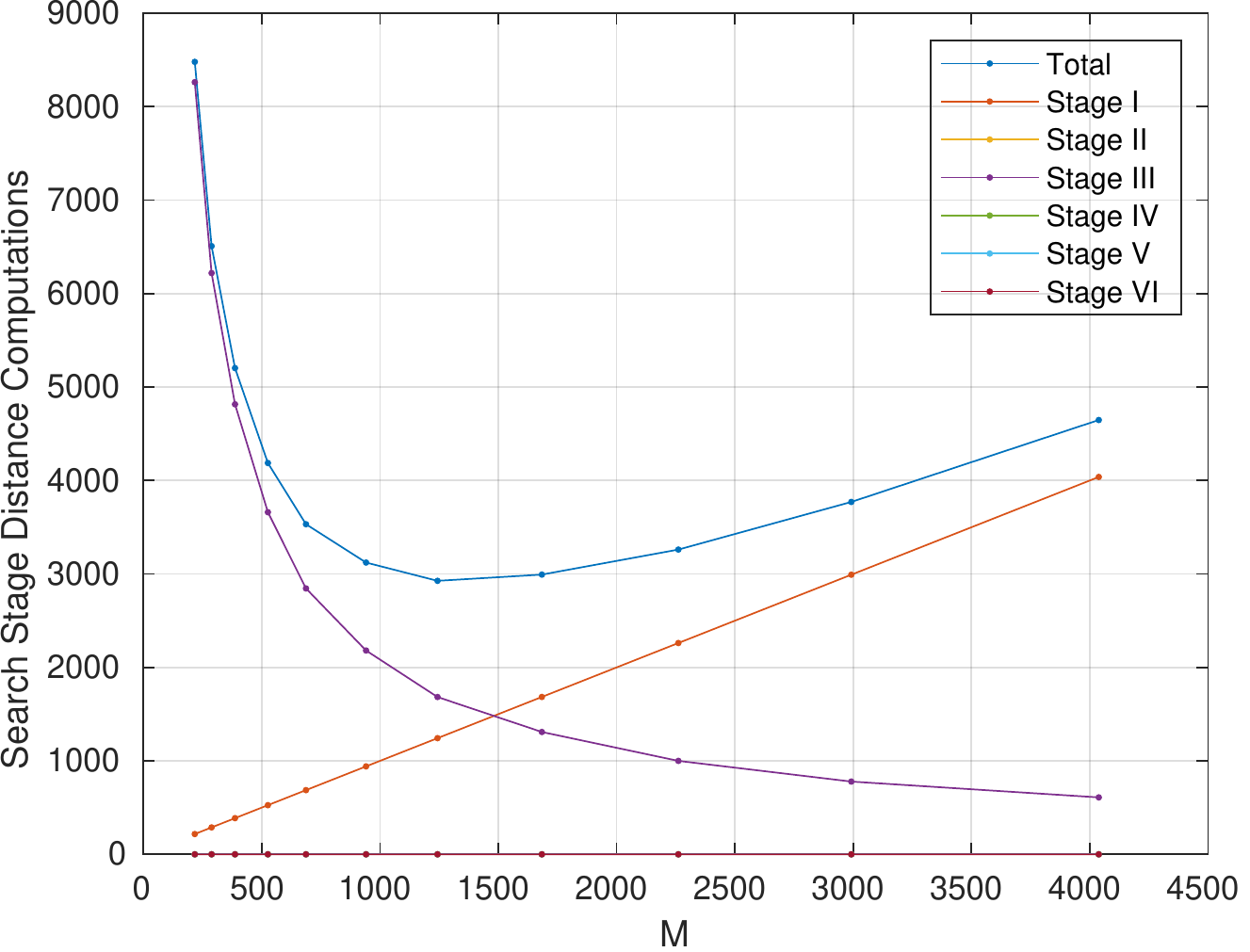}
        \end{minipage}
    
        (c)
        \begin{minipage}{0.40\textwidth}
            \centering
            \includegraphics[width=\textwidth]{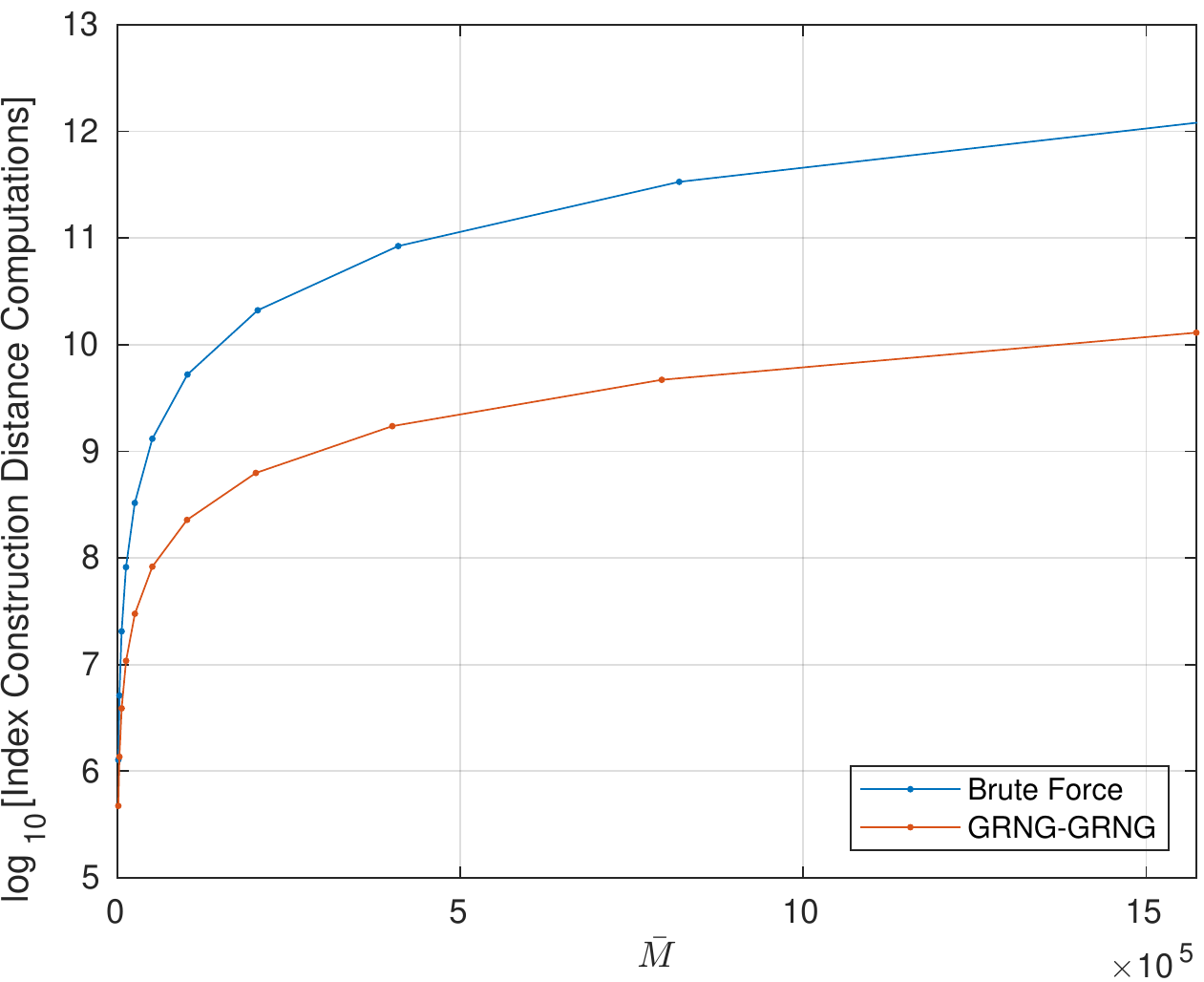}
        \end{minipage}
        (d)
        \begin{minipage}{0.40\textwidth}
            \centering
            \includegraphics[width=\textwidth]{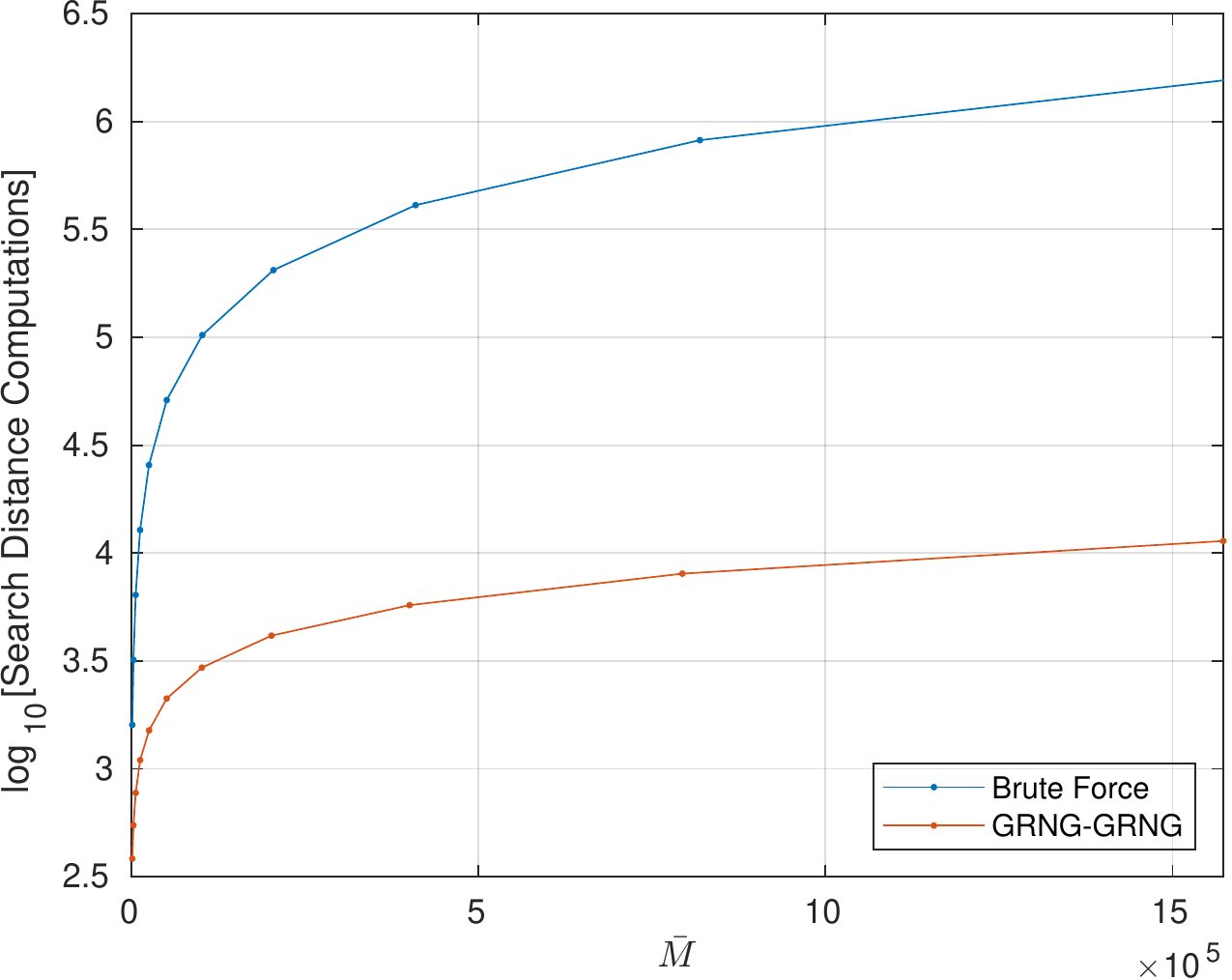}
        \end{minipage}
  \end{minipage}
  (e)
   \begin{minipage}{0.25\textwidth}
    \begin{subfigure}[b]{\textwidth}
        \includegraphics[width=\textwidth,height=0.25\textheight]{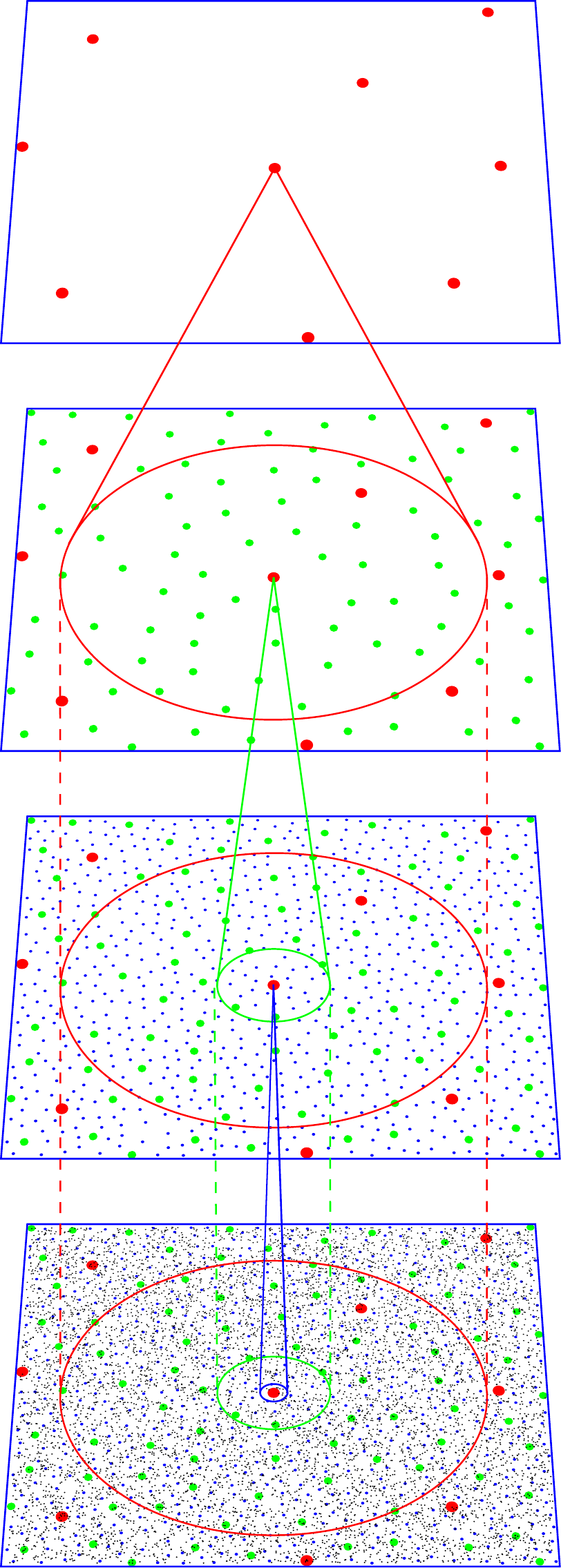}
        \vspace{-0.25cm}
    \end{subfigure}
    \end{minipage}
    
    \vspace{-0.25cm}

    \caption{\footnotesize Stage by stage analysis for GRNG-GRNG hierarchy for $\bar{M}=102,400$ uniformly distributed fine-scale pivots in 2D as a function of $M$, the number of coarse-scale pivots. The number of distance computations for construction (a) and search (b) show Stage \RNU{1} is increasing with $M$ while other stages exponentially decay with an optimum for each in total. The improvements of GRNG-GRNG with respect to brute-force as a function of $M$ for construction (c) and search (d) distances is significant.  (e) The monotonically increasing Stage \RNU{1} in (a-b) suggest using a multi-layer hierarchy.}
    \label{fig:grng_grng_efficiency}
\end{figure}

%
\noindent{\bf Stage \RNU{5}: ``Fine-Scale Pivot'' -- Mediated ``Fine-Scale Pivot'' Interactions:}
Those links between $Q$ and $\bar{p}_{\bar{j}}$ that survive the pivot test must now test against occupancy of G-lune($Q,\bar{p}_{\bar{j}}$) by exemplars $\bar{p}_{\bar{k}}$. In this stage, a select group of $\bar{p}_{\bar{k}}$, namely those close to $Q$ and $\bar{p}_{\bar{j}}$ which are more likely to be in G-lune($Q,\bar{p}_{\bar{j}}$) are considered, leaving the rest to Stage \RNU{6}. Specifically, these are the $k=25$ nearest neighbors of $Q$ and $\bar{p}_{\bar{j}}$, Figure \ref{fig:GRNGSavingsVisualized}.

%
\noindent{\bf Stage \RNU{6}: ``Fine-Scale Pivot'' ``Fine-Scale Pivot'' Interactions:}
Very few fine-scale pivots $\bar{p}_{\bar{j}}$ remain at this stage. These need to be validated with all other fine-scale pivots $\bar{p}_{\bar{k}}$. However, the following proposition prevents consideration of a majority of them. Define 
%
\begin{equation} \label{eq:GRNG_S6_dmax_definition}
    \delta_{\max}\left(p_k\right)=\underset{\forall \bar{p}_{\bar{k}},\,d\left(p_k,\bar{p}_{\bar{k}}\right) \leq (r_k - \bar{r}_{\bar{k}})}{\max} d(p_k, \bar{p}_{\bar{k}}).
\end{equation}

%
\begin{proposition} \label{prop:GRNG_S6_dmax}
    All fine-scale pivots $(\bar{p}_{\bar{k}},\bar{r}_{\bar{k}}) \in \mathcal{D}(p_k,r_k)$ satisfying
    
    \begin{subnumcases}{\label{eq:GRNG_S6_dmax}}
        d\left(Q,p_{k}\right) - \delta_{\max}(p_k) \geq d\left(Q,\bar{p}_{\bar{j}}\right) - \left(2\bar{r}_{Q}+\bar{r}_{\bar{j}}\right) \label{eq:BerkGraphUpdate_Stage6_ineq1}\\
        d\left(\bar{p}_{\bar{j}},p_{k}\right) - \delta_{\max}(p_k) \geq d\left(Q,\bar{p}_{\bar{j}}\right) - \left(2\bar{r}_{\bar{j}}+\bar{r}_{Q}\right)
    \end{subnumcases}
    fall outside the G-lune($Q,\bar{p}_{\bar{j}})$, for a query $(Q,\bar{r}_Q)$ and a fine-scale pivot $(\bar{p}_{\bar{j}},\bar{r}_{\bar{j}})$. 
\end{proposition}
\iffull
\begin{proof}
    Observe that
    
    \begin{subnumcases}{\label{eq:BerkGraphUpdate_Stage6_rule_proof}}
        \scalebox{0.74}{%
        $d\left(Q,\bar{p}_{\bar{k}}\right) \geq d\left(Q,p_k\right) -  d\left(\bar{p}_{\bar{k}},p_k\right) \geq d\left(Q,p_k\right) - \underset{\forall \bar{p}_{\bar{k}},\,d\left(p_k,\bar{p}_{\bar{k}}\right) \leq (r_k - \bar{r}_{\bar{k}})}{\max} d\left(\bar{p}_{\bar{k}},p_k\right) = d\left(Q,p_k\right) - \delta_{\max}(p_k)$,  \label{eq:BerkGraphUpdate_Stage6_ineq3}
        } \\
        \scalebox{0.74}{%
        $d\left(\bar{p}_{\bar{j}},\bar{p}_{\bar{k}}\right) \geq d\left(\bar{p}_{\bar{j}},p_k\right) -  d\left(\bar{p}_{\bar{k}},p_k\right) \geq d\left(\bar{p}_{\bar{j}},p_k\right) - \underset{\forall \bar{p}_{\bar{k}},\,d\left(p_k,\bar{p}_{\bar{k}}\right) \leq (r_k - \bar{r}_{\bar{k}})}{\max} d\left(\bar{p}_{\bar{k}},p_k\right) = d\left(\bar{p}_{\bar{j}},p_k\right) - \delta_{\max}(p_k)$. \label{eq:BerkGraphUpdate_Stage6_ineq4}
        }
    \end{subnumcases}
    Now, combining Equations~\ref{eq:GRNG_S6_dmax} and~\ref{eq:BerkGraphUpdate_Stage6_rule_proof},
    
    \begin{subnumcases}{}
        d\left(Q,\bar{p}_{\bar{k}}\right) \geq d\left(Q,\bar{p}_{\bar{j}}\right) - (2\bar{r}_Q + \bar{r}_{\bar{j}}) \\
        d\left(\bar{p}_{\bar{j}},\bar{p}_{\bar{k}}\right) \geq d\left(Q,\bar{p}_{\bar{j}}\right) - (\bar{r}_Q + 2\bar{r}_{\bar{j}}),
    \end{subnumcases}
    \textit{i.e.,} $\bar{p}_{\bar{k}}$ cannot be inside the G-lune$(Q,\bar{p}_{\bar{j}})$.
    \qed
\end{proof}
\fi
This proposition excludes entire pivot domains from the validation process. The following proposition further restricts the remaining sets.

%
\begin{proposition} \label{prop:GRNG_S6_assisted}
    All fine-scale pivots $(\bar{p}_{\bar{k}},\bar{r}_{\bar{k}}) \in \mathcal{D}(p_k,r_k)$ satisfying
    
    \begin{subnumcases}{\label{eq:GRNG_S6_assisted}}
        d\left(Q,p_k\right) - d\left(p_k,\bar{p}_{\bar{k}}\right) \geq d\left(Q,\bar{p}_{\bar{j}}\right) - (2\bar{r}_Q + \bar{r}_{\bar{j}}) \\
        d\left(\bar{p}_{\bar{j}},p_k\right) - d\left(p_k,\bar{p}_{\bar{k}}\right) \geq d\left(Q,\bar{p}_{\bar{j}}\right) - (\bar{r}_Q + 2\bar{r}_{\bar{j}}),
    \end{subnumcases}
    falls outside the GRNG-lune($Q,\bar{p}_{\bar{j}})$ for a query $(Q,\bar{r}_Q)$ and a fine-scale pivot $(\bar{p}_{\bar{j}},\bar{r}_{\bar{j}})$. 
\end{proposition}
\iffull
\begin{proof}
    Combine the first inequality in Equations~\ref{eq:BerkGraphUpdate_Stage6_rule_proof} with the premise to show Equations~\ref{eq:GRNG_S6_assisted}. 
    \qed
\end{proof}

\fi
After the majority of fine-scale pivots $(\bar{p}_{\bar{k}},\bar{r}_{\bar{k}})$ have been eliminated, the remaining ones must test the two GRNG conditions. For efficiency, if first condition $d(Q,\bar{p}_{\bar{k}}) < d(Q,\bar{p}_{\bar{j}} - (2\bar{r}_Q + \bar{r}_{\bar{j}})$ does not hold, the second condition $d(\bar{p}_{\bar{j}},\bar{p}_{\bar{k}}) < d(Q,\bar{p}_{\bar{j}} - (\bar{r}_Q + 2\bar{r}_{\bar{j}})$ need not be tested, Figure \ref{fig:GRNGSavingsVisualized}.

%
\begin{figure}[b!]
    \centering
   \scriptsize{(a)}
   \begin{minipage}{0.29\textwidth}
     \centering
     \includegraphics[width=\textwidth]{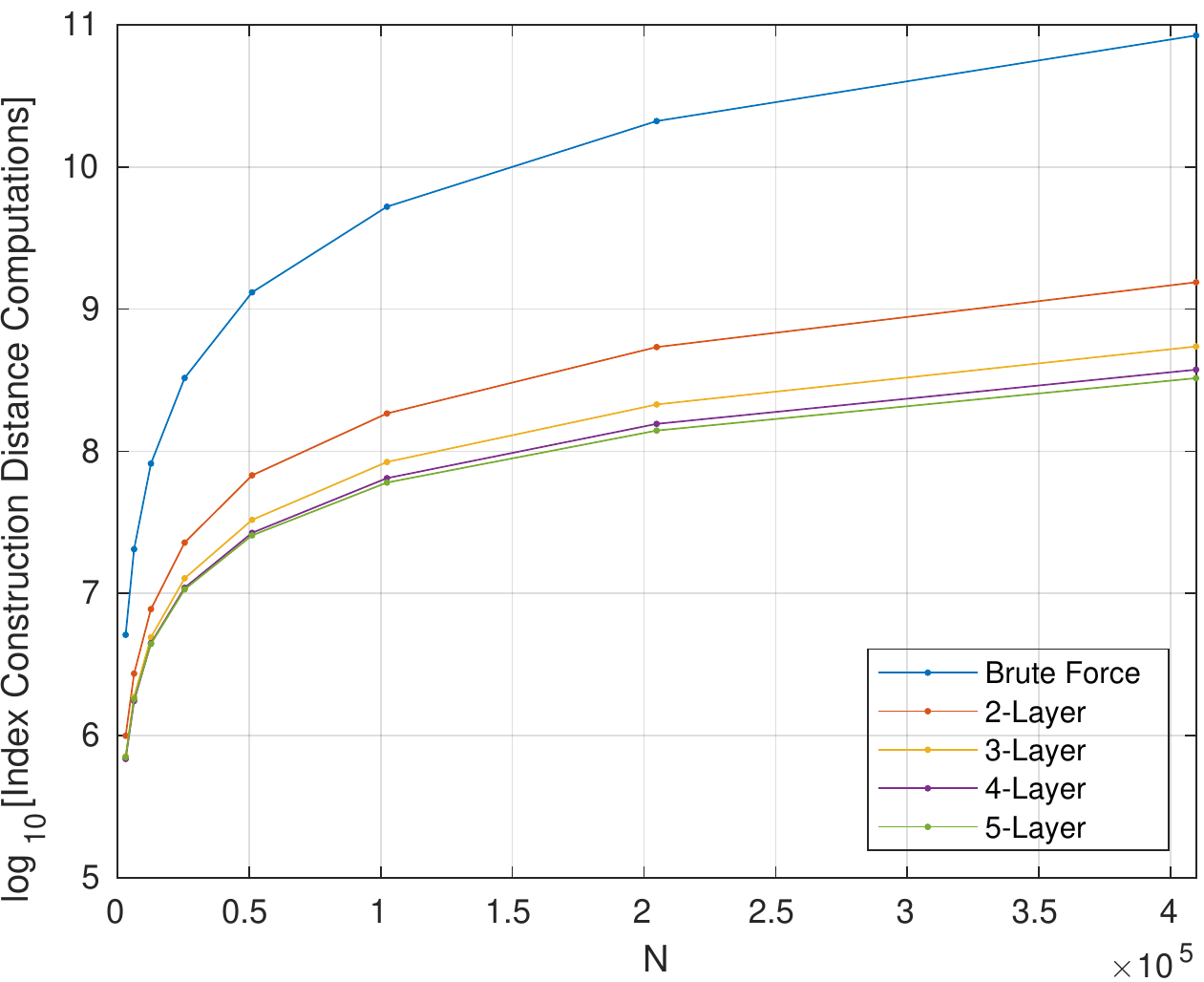}
   \end{minipage}
   \scriptsize{(b)}
   \begin{minipage}{0.29\textwidth}
     \centering
     \includegraphics[width=\textwidth]{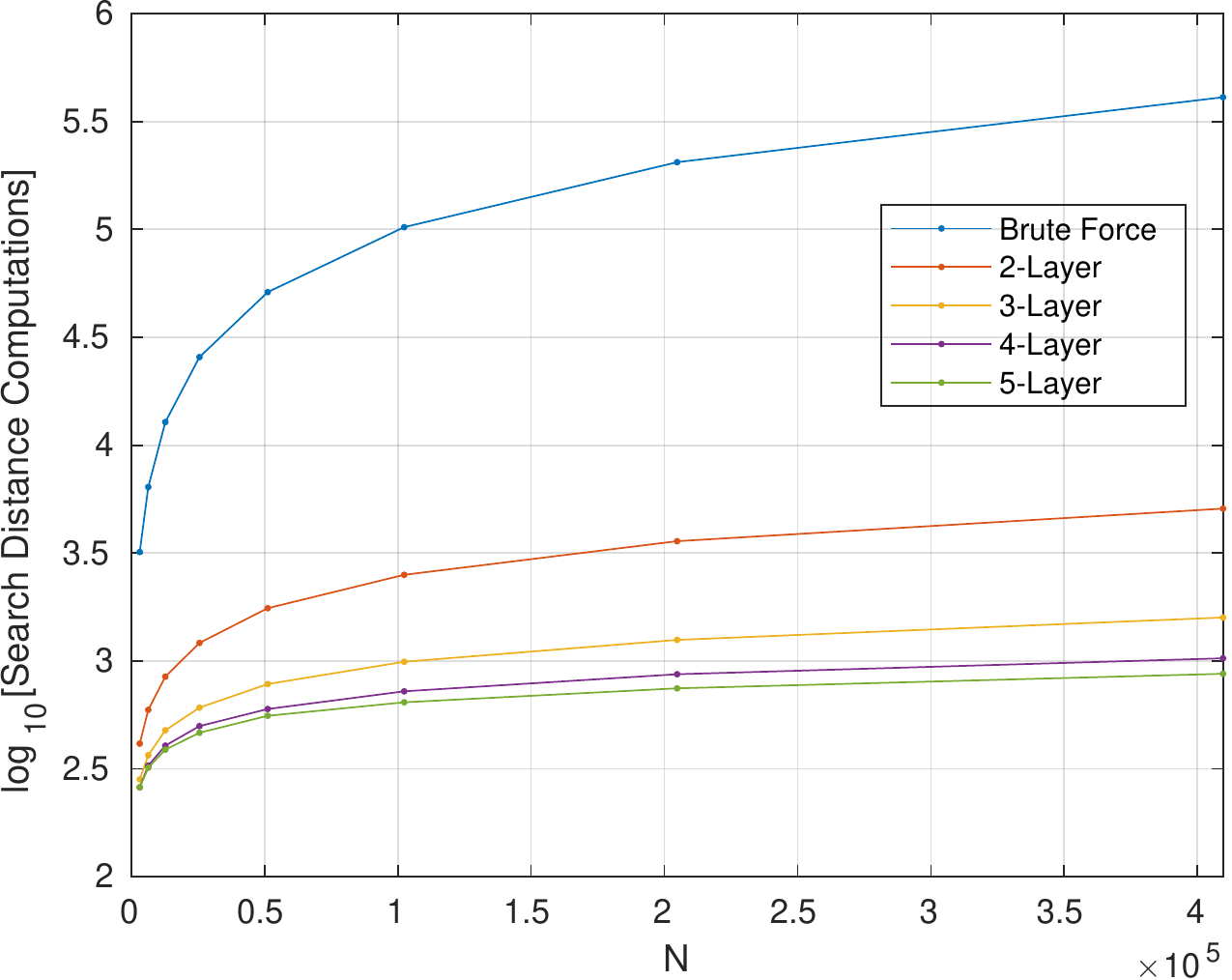}
   \end{minipage}
   \scriptsize{(c)}
   \begin{minipage}{0.29\textwidth}
     \centering
     \includegraphics[width=\textwidth]{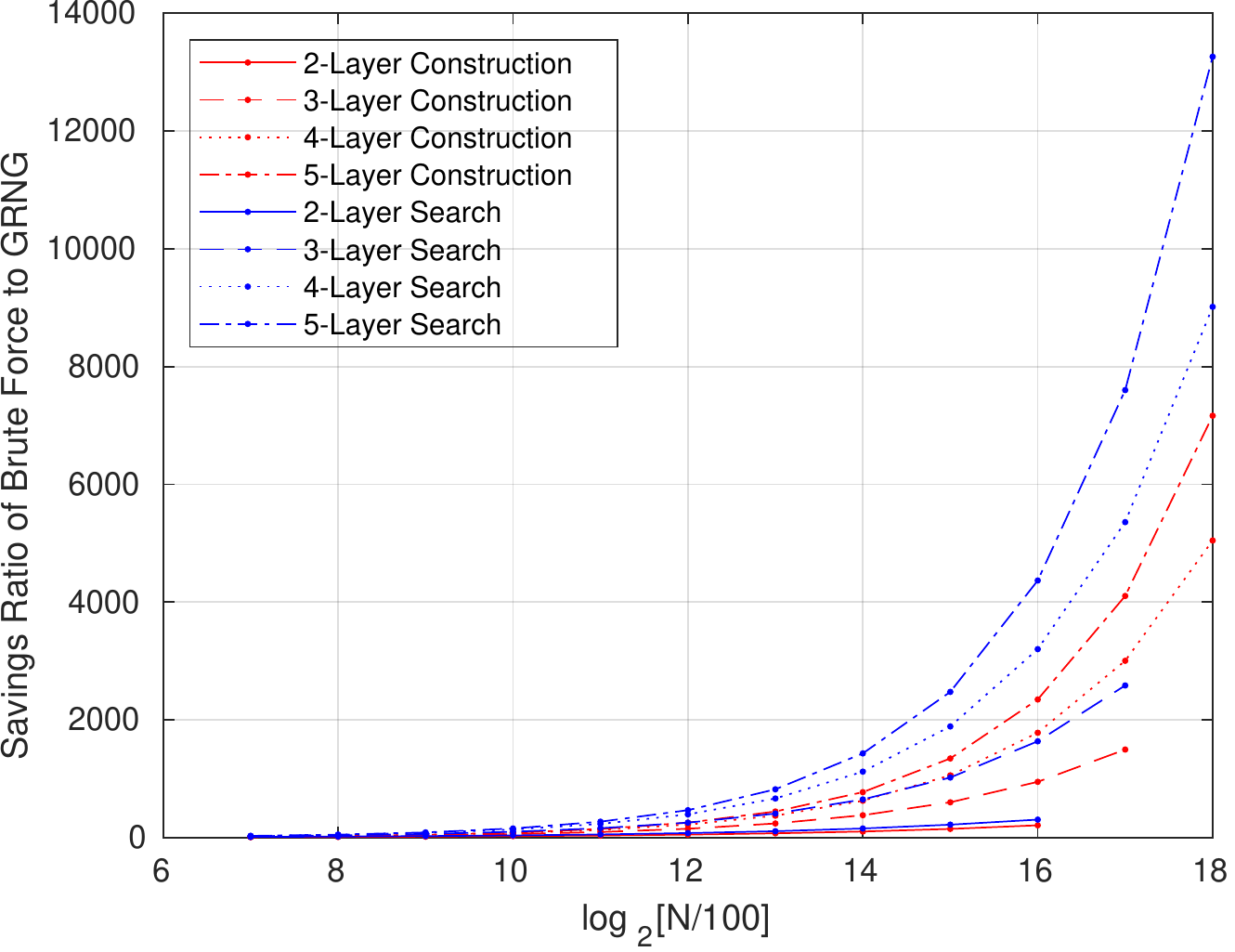}
    \end{minipage}
    \vspace{-0.25cm}
    
   \caption{\footnotesize Comparing the efficiency of multi-layer GRNG hierarchies on 2D uniformly distributed data to a Brute Force algorithm that would precompute all pairwise distances (a) for RNG construction or precompute all $N$ distances to dataset members (b) for search. (c) The ratio of distance computation savings across $N$ and number of layers.} 
   \label{fig:grng_savings_bruteForce}
\end{figure}

%
\noindent{\bf Stage \RNU{7}:``Coarse-Scale Pivot'' -- ``Fine-Scale Pivot'' Validations:}
The incremental construction requires checking which existing GRNG links may be invalidated by the addition of $Q$. Define first, 

%
\begin{subnumcases}{\label{eq:GRNG_S7_umax_both}}
    \bar{\mu}_{\max}\left(\bar{p}_{\bar{i}}\right)=\underset{\bar{p}_{\bar{j}},\,\textrm{GRNG}(\bar{p}_{\bar{i}})}{\max} \left[ d\left(\bar{p}_{\bar{i}},\bar{p}_{\bar{j}}\right)-(2\bar{r}_{\bar{i}}+\bar{r}_{\bar{j}})\right] \\
    \mu_{\max}\left(p_i\right)=\underset{\forall (\bar{p}_{\bar{i}},\bar{r}_{\bar{i}}) \in \mathcal{D}(p_i,r_i)}{\max} \left[\bar{\mu}_{\max}\left(\bar{p}_{\bar{i}}\right)+d(p_i,\bar{p}_{\bar{i}})\right].
\end{subnumcases}

%
\begin{proposition} \label{prop:GRNG_S7_umax}
    The insertion of $Q$ does not invalidate any GRNG links involving fine-scale pivot $\bar{p}_{\bar{i}}$ for which 
    
    \begin{equation} \label{eq:GRNG_S7_umax_fine}
        d(Q,\bar{p}_{\bar{i}}) \geq \bar{\mu}_{\max}(\bar{p}_{\bar{i}}).
    \end{equation}
    Furthermore, the insertion of $Q$ does not interfere with the GRNG link involving fine-scale pivots $(\bar{p}_{\bar{i}},\bar{r}_{\bar{i}}) \in \mathcal{D}(p_i,r_i)$ if 
    
    \begin{equation} \label{eq:GRNG_S7_umax_coarse}
        d(Q,p_i) \geq \mu_{\max}(p_i).
    \end{equation} 
\end{proposition}
\iffull
\begin{proof}
    It is clear that if $d\left(Q,\bar{p}_{\bar{i}}\right)\geq \bar{\mu}_{\max}\left(\bar{p}_{\bar{i}}\right)$, then $\forall \bar{p}_{\bar{j}}$
    
    \begin{equation}
    d\left(Q,\bar{p}_{\bar{i}}\right)\geq d\left(\bar{p}_{\bar{i}},\bar{p}_{\bar{j}}\right)-(2\bar{r}_{\bar{i}}+\bar{r}_{\bar{j}}),
    \end{equation}
    \noindent so that $Q$ is outside $\text{G-lune}\left(\bar{p}_{\bar{i}},\bar{p}_{\bar{j}}\right)$. By the triangle inequality,
    
    \begin{align} 
        d\left(Q,\bar{p}_{\bar{i}}\right) &\geq d\left(Q,p_i\right) - d\left(p_i,\bar{p}_{\bar{i}}\right) \nonumber \\ 
         &\geq \mu_{\max}\left(p_{i}\right) - d\left(p_i,\bar{p}_{\bar{i}}\right) \nonumber \\
         &\geq \underset{\forall \bar{p}_{\bar{k}},\,d(p_i,\bar{p}_{\bar{k}}) \leq r_i - \bar{r}_{\bar{k}}}{\max}(\bar{\mu}_{\max}\left(\bar{p}_{\bar{k}}\right)+d(p_i,\bar{p}_{\bar{k}})) - d\left(p_i,\bar{p}_{\bar{i}}\right) \nonumber \\
         &\geq \bar{\mu}_{\max}\left(\bar{p}_{\bar{i}}\right)+d(p_i,\bar{p}_{\bar{i}}) - d\left(p_i,\bar{p}_{\bar{i}}\right) \nonumber \\
         &\geq \bar{\mu}_{\max}\left(\bar{p}_{\bar{i}}\right).
         \label{eq:BerkGraphUpdate_Stage7_queryFineScalePivotInequalityCheck}
    \end{align}
    
    Thus, $Q$ cannot interfere with any GRNG link formed from $\bar{p}_{\bar{i}}$.
    \qed
\end{proof}
\fi
The proposition suggests a three-step approach to examining existing links: (\emph{\RNL{1}}) Remove all coarse-scale pivot domains $p_i$ satisfying Equation \ref{eq:GRNG_S7_umax_coarse}; (\emph{\RNL{2}}) Remove all fine-scale pivot domains $(\bar{p}_{\bar{i}},\bar{r}_{\bar{i}})$ satisfying Equation 
\ref{eq:GRNG_S7_umax_fine}; (\emph{\RNL{3}}) For any remaining fine-scale pivot $(\bar{p}_{\bar{i}},\bar{r}_{\bar{i}})$ connecting with $(\bar{p}_{\bar{j}},\bar{r}_{\bar{j}})$, if $Q$ is in the G-lune($\bar{p}_{\bar{i}},\bar{p}_{\bar{j}}$), then the link needs to be removed.

%
\begin{figure}[t!]
    \centering

    \begin{subfigure}[b]{0.19\textwidth}
        \includegraphics[width=\textwidth]{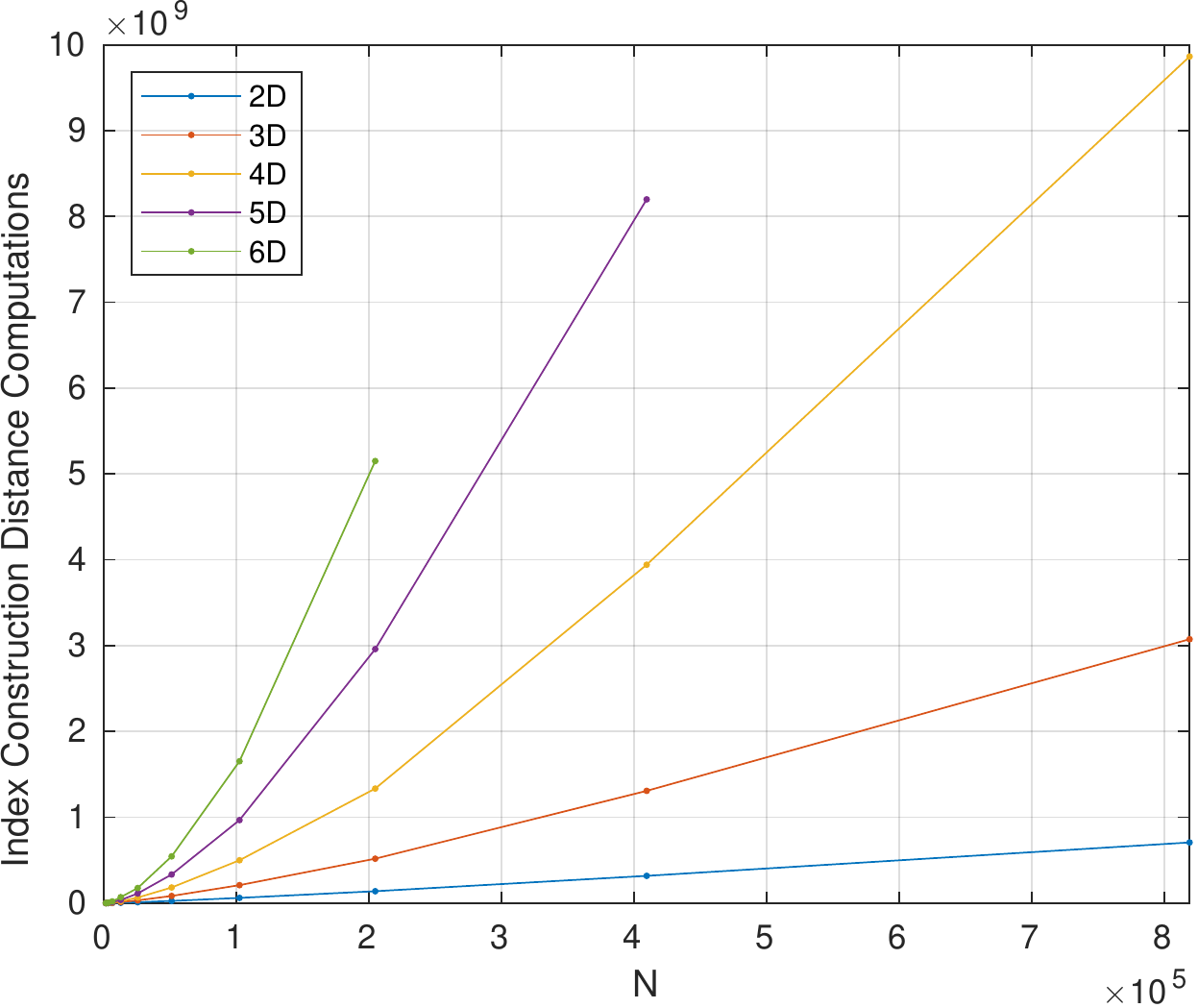}
        \caption{}
        \label{fig:results_synthetic:uniform_search}
    \end{subfigure}   
    \begin{subfigure}[b]{0.19\textwidth}
        \includegraphics[width=\textwidth]{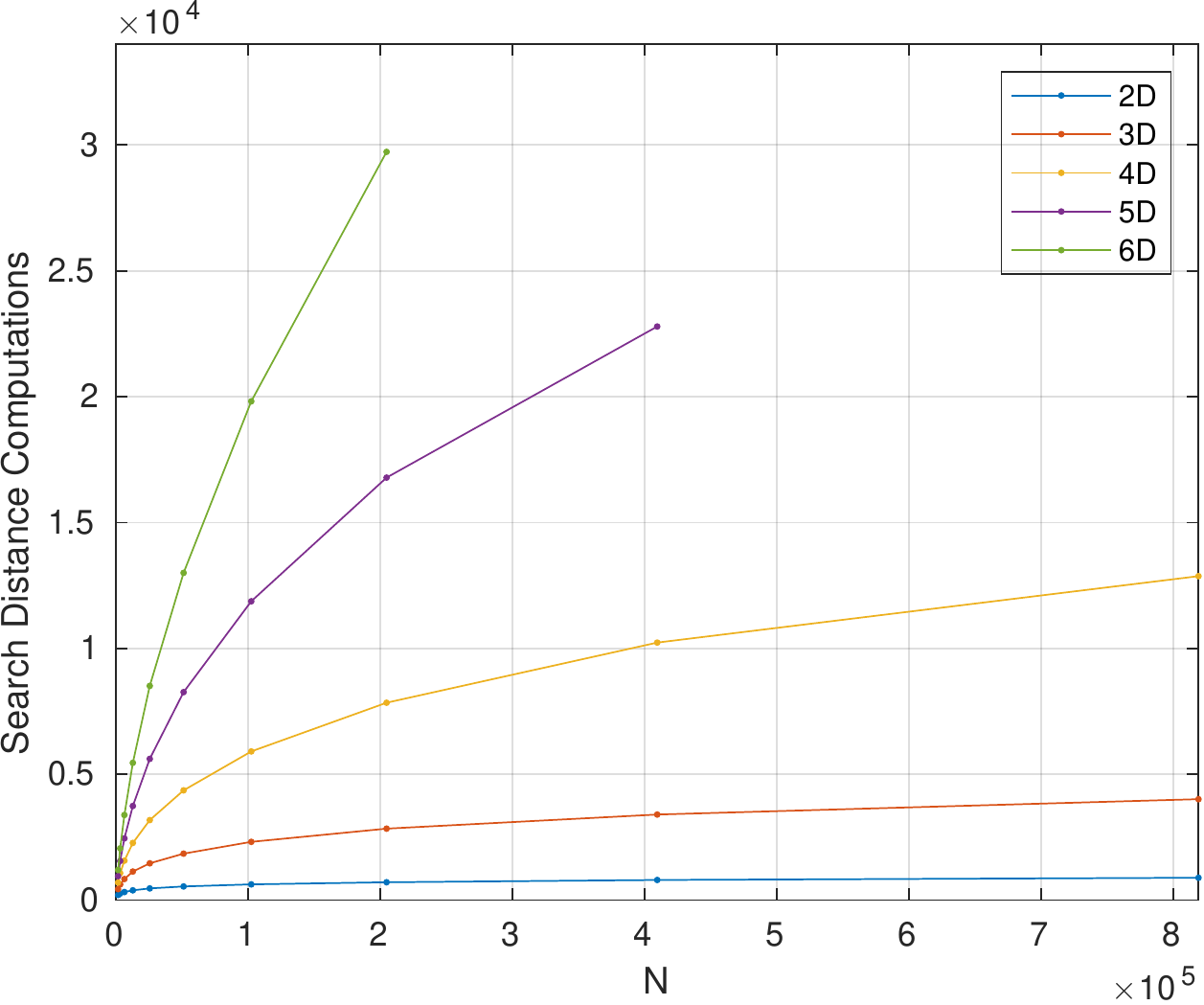}
        \caption{}
        \label{fig:results_synthetic:uniform_index}
    \end{subfigure}  
    \begin{subfigure}[b]{0.19\textwidth}
        \includegraphics[width=\textwidth]{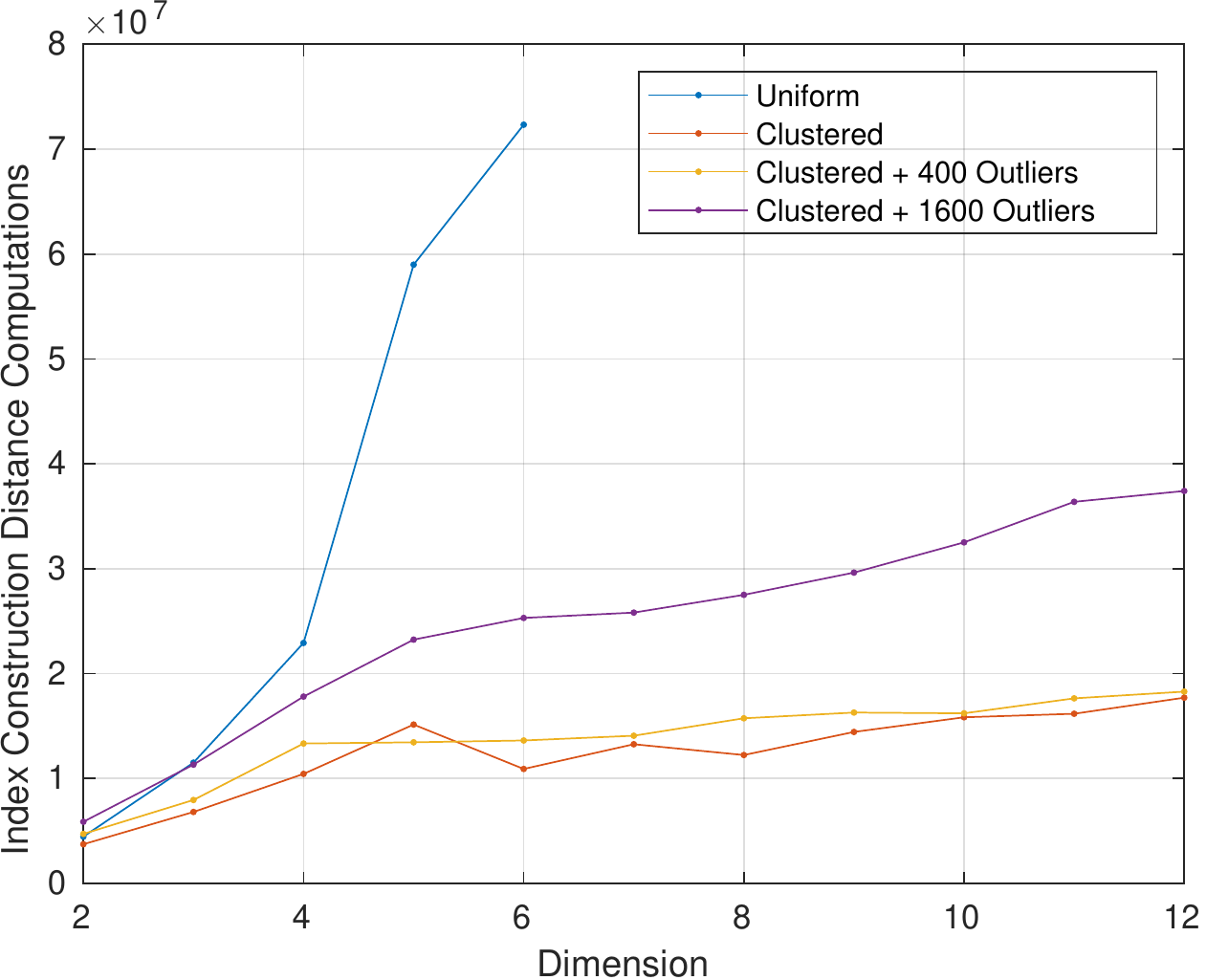}
        \caption{}
        \label{fig:results_synthetic:cluster_index}
    \end{subfigure}
    \begin{subfigure}[b]{0.19\textwidth}
        \includegraphics[width=\textwidth]{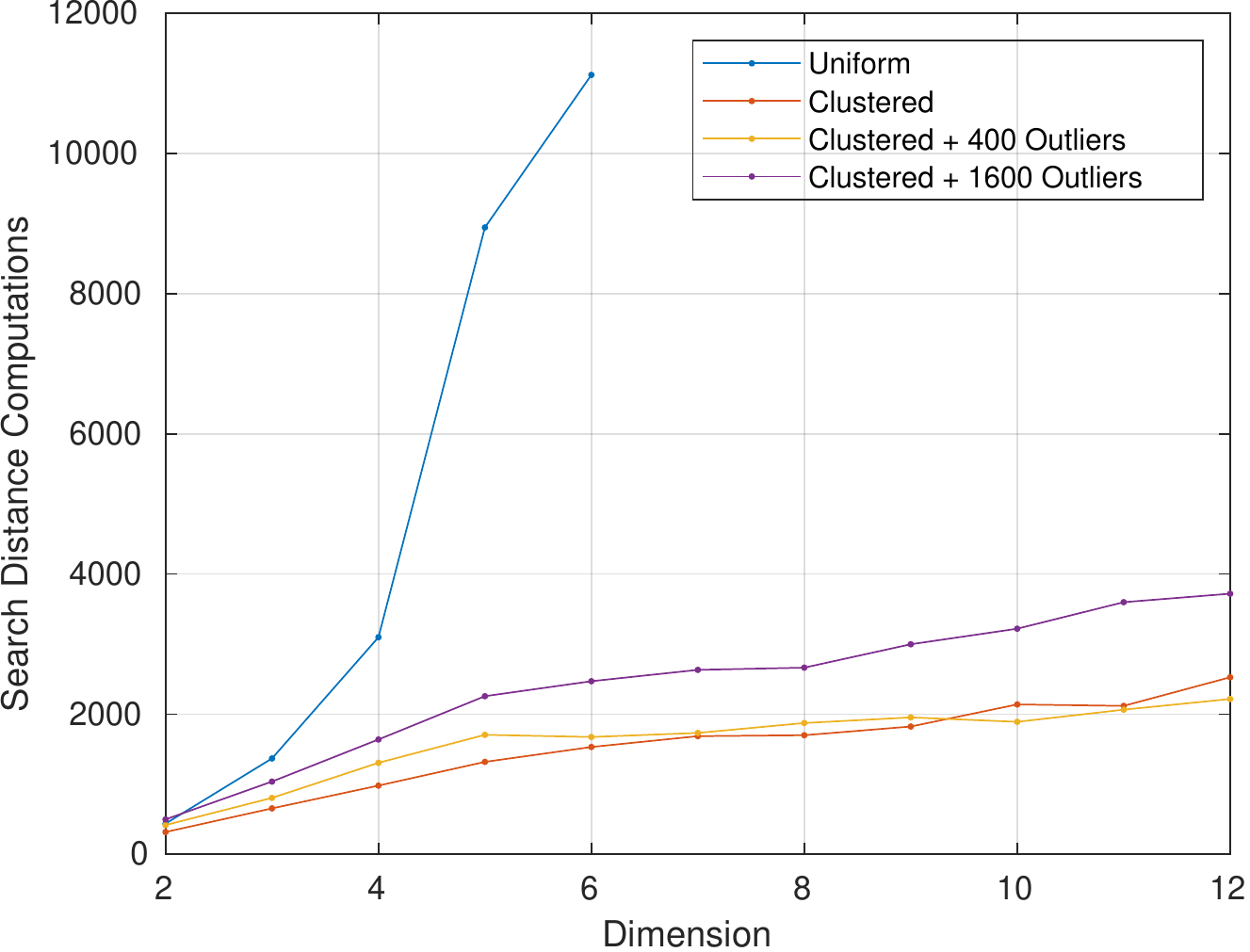}
        \caption{}
        \label{fig:results_synthetic:cluster_search}
    \end{subfigure}
    \begin{subfigure}[b]{0.19\textwidth}
        \includegraphics[width=\textwidth]{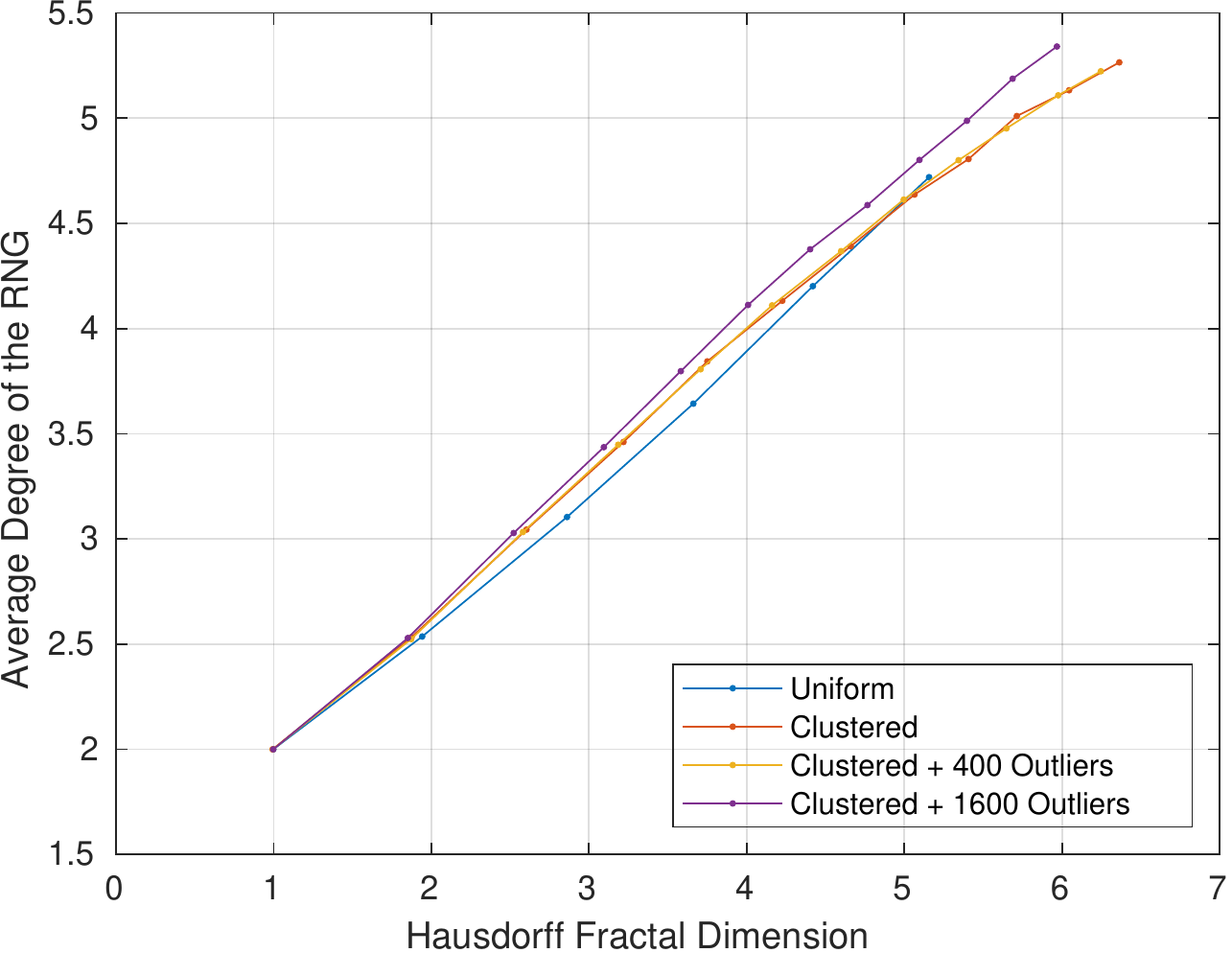}
        \caption{}
        \label{fig:results_synthetic:HausdorffVsAverageDegreeRNGCombined}
    \end{subfigure} 
    \vspace{-0.25cm}
    \caption{\footnotesize The distance computations for index construction (a) and search (b) increases as a function of number of exemplars and dimensions. However, with clustered data (c)(d), even with outliers, both construction costs and search distances increase much less rapidly. (e) The average degree of the RNG is related to the intrinsic dimensionality. }
    \label{fig:results_synthetic}
\end{figure}
%

\iffull

    \begin{table}[]
    \centering
    \caption{Statistics for Optimal 2-Layer GRNG-RNG Hierarchies on Uniformly Distributed Data.}
    
    \begin{minipage}{\textwidth}
        \scalebox{0.85}{%
\begin{tabular}{|c|c|c|c|c|c|}
\hline
\multicolumn{6}{|c|}{\textbf{Search Distance Computations}} \\
\hline
\textbf{N} & \textbf{2D} & \textbf{3D} & \textbf{4D} & \textbf{5D}           & \textbf{6D}           \\ \hline \hline
1,600      & 281.77      & 463.03      & 696.97      & 950.93      & 1,196.06    \\ \hline
3,200      & 414.40      & 701.99      & 1,089.12    & 1,552.94    & 2,055.56    \\ \hline
6,400      & 594.15      & 1,042.35    & 1,614.03    & 2,452.24    & 3,379.85    \\ \hline
12,800     & 846.60      & 1,511.29    & 2,453.67    & 3,773.75    & 5,454.52    \\ \hline
25,600     & 1,213.10    & 2,187.77    & 3,634.22    & 5,677.06    & 8,489.30    \\ \hline
51,200     & 1,757.14    & 3,136.99    & 5,348.49    & 8,603.36    & 13,014.10   \\ \hline
102,400    & 2,508.88    & 4,527.96    & 7,808.32    & 12,859.70   & 20,154.40   \\ \hline
204,800    & 3,591.85    & 6,601.45    & 11,373.70   & 18,972.20   & 30,015.00   \\ \hline
409,600    & 5,083.95    & 9,388.31    & 16,636.20   & 28,045.50   &             \\ \hline
819,200    & 7,297.52    & 13,491.90   & 23,792.50   & 41,247.20   &             \\ \hline
1,638,400  & 10,331.30   & 19,432.80   & 34,413.90   &             &             \\ \hline
\end{tabular}
        }
        \scalebox{0.85}{%
\begin{tabular}{|c|c|c|c|c|c|}
\hline
\multicolumn{6}{|c|}{\textbf{Search Time (ms)}} \\
\hline
\textbf{N} & \textbf{2D} & \textbf{3D} & \textbf{4D} & \textbf{5D}           & \textbf{6D}           \\ \hline \hline
1,600      & 0.274       & 0.427       & 0.978       & 2.080       & 3.649       \\ \hline
3,200      & 0.400       & 0.723       & 2.300       & 4.752       & 5.987       \\ \hline
6,400      & 0.653       & 1.314       & 3.506       & 8.705       & 19.454      \\ \hline
12,800     & 0.867       & 1.974       & 4.250       & 15.986      & 42.176      \\ \hline
25,600     & 1.477       & 3.334       & 7.012       & 25.474      & 77.990      \\ \hline
51,200     & 1.644       & 5.190       & 12.655      & 47.144     & 124.103     \\ \hline
102,400    & 3.375       & 6.682       & 17.684      & 67.754      & 226.059     \\ \hline
204,800    & 5.638       & 8.909       & 28.389      & 91.606      & 290.796     \\ \hline
409,600    & 6.304       & 12.289      & 30.600      & 118.376     &             \\ \hline
819,200    & 11.481      & 18.770      & 52.273      & 141.868     &             \\ \hline
1,638,400  & 14.739      & 30.528      & 53.540      &             &             \\ \hline
\end{tabular}
        }
    \end{minipage}
    
    \vspace{0.25cm}
    
    \begin{minipage}{\textwidth}
        \centering
        \scalebox{0.85}{%
    \begin{tabular}{|c|c|c|c|c|c|c|}
    \hline
    \multicolumn{7}{|c|}{\textbf{Index Construction Distance Computations}} \\
    \hline
    \textbf{N} & \textbf{N(N-1)/2} & \textbf{2D}    & \textbf{3D}    & \textbf{4D}    & \textbf{5D}    & \textbf{6D}   \\ 
    \hline \hline
1,600      & 1,279,200         & 323,362        & 576,421        & 871,615        & 1,265,981      & 1,424,701     \\ \hline
3,200      & 5,118,400         & 998,165        & 1,746,168      & 2,980,553      & 4,380,391      & 5,389,184     \\ \hline
6,400      & 20,476,800        & 2,731,860      & 5,268,490      & 9,124,159      & 13,977,441     & 20,194,814    \\ \hline
12,800     & 81,913,600        & 7,756,808      & 15,056,026     & 27,667,618     & 46,749,155     & 69,953,472    \\ \hline
25,600     & 327,667,200       & 22,781,408     & 42,295,510     & 76,583,845     & 141,335,210    & 234,636,962   \\ \hline
51,200     & 1,310,694,400     & 67,708,441     & 121,429,407    & 225,220,284    & 417,456,903    & 723,257,748   \\ \hline
102,400    & 5,242,828,800     & 184,344,339    & 346,602,393    & 644,199,988    & 1,205,946,173  & 2,193,952,436 \\ \hline
204,800    & 20,971,417,600    & 540,102,922    & 982,784,171    & 1,804,674,193  & 3,399,421,853  & 6,398,877,358 \\ \hline
409,600    & 83,885,875,200    & 1,543,057,134  & 2,792,989,300  & 5,108,689,230  & 9,550,546,084  &               \\ \hline
819,200    & 335,543,910,400   & 4,381,197,495  & 7,939,742,585  & 14,528,254,634 & 26,692,822,584 &               \\ \hline
1,638,400  & 1,342,176,460,800 & 12,537,531,662 & 22,602,120,265 & 41,198,331,167 &                &               \\ \hline
\end{tabular}
        }
    \end{minipage}
    
    \vspace{0.25cm}
    
    \begin{minipage}{\textwidth}
        \scalebox{0.75}{%
\begin{tabular}{|c|c|c|c|c|c|}
\hline
\multicolumn{6}{|c|}{\textbf{Index Construction Time (hr)}} \\
\hline
\textbf{N} & \textbf{2D} & \textbf{3D} & \textbf{4D} & \textbf{5D} & \textbf{6D} \\ 
\hline \hline
1,600      & 3.931E-03   & 1.906E-04   & 6.376E-04   & 1.665E-03   & 7.062E-03   \\ \hline
3,200      & 3.707E-04   & 6.220E-04   & 2.416E-03   & 6.863E-03   & 1.566E-02   \\ \hline
6,400      & 1.336E-03   & 2.151E-03   & 7.185E-03   & 5.037E-02   & 9.357E-02   \\ \hline
12,800     & 1.955E-03   & 6.007E-03   & 2.009E-02   & 0.104       & 0.353       \\ \hline
25,600     & 5.727E-03   & 2.572E-02   & 5.196E-02   & 0.381       & 1.434       \\ \hline
51,200     & 1.312E-02   & 0.242       & 0.228       & 1.214       & 4.176       \\ \hline
102,400    & 4.814E-02   & 0.509       & 0.633       & 3.557       & 13.573      \\ \hline
204,800    & 0.153       & 0.737       & 1.792       & 9.514       & 39.091      \\ \hline
409,600    & 0.382       & 1.703       & 4.038       & 23.313      &             \\ \hline
819,200    & 1.718       & 3.139       & 11.183      & 54.078      &             \\ \hline
1,638,400  & 2.821       & 7.641       & 25.147      &             &             \\ \hline
\end{tabular}
        }
        \scalebox{0.75}{%
\begin{tabular}{|c|c|c|c|c|c|}
\hline
\multicolumn{6}{|c|}{\textbf{Memory Usage (GB)}} \\
\hline
1,600      & 5.428E-03   & 6.954E-03   & 1.033E-02   & 1.785E-02   & 2.692E-02   \\ \hline
3,200      & 8.476E-03   & 1.219E-02   & 2.136E-02   & 3.832E-02   & 6.510E-02   \\ \hline
6,400      & 1.360E-02   & 2.262E-02   & 4.289E-02   & 7.547E-02   & 0.159       \\ \hline
12,800     & 2.436E-02   & 4.267E-02   & 8.587E-02   & 0.174       & 0.345       \\ \hline
25,600     & 4.607E-02   & 8.223E-02   & 0.179       & 0.401       & 0.862       \\ \hline
51,200     & 8.268E-02   & 0.156       & 0.370       & 0.8815       & 1.981       \\ \hline
102,400    & 0.157       & 0.312       & 0.778       & 1.949       & 4.587       \\ \hline
204,800    & 0.317       & 0.625       & 1.602       & 4.161       & 10.251      \\ \hline
409,600    & 0.633       & 1.258       & 3.299       & 8.947       &             \\ \hline
819,200    & 1.258       & 2.531       & 6.856       & 18.824      &             \\ \hline
1,638,400  & 2.531       & 5.181       & 14.194      &             &             \\ \hline
\end{tabular}

        }
    \end{minipage}

\end{table}
    
    \begin{table}[]
    \centering
    \caption{Statistics for Optimal 3-Layer GRNG-GRNG-RNG Hierarchies on Uniformly Distributed Data.}
    
    \begin{minipage}{\textwidth}
        \scalebox{0.85}{%
            \begin{tabular}{|c|c|c|c|c|c|}
\hline
\multicolumn{6}{|c|}{Search Distance Computations} \\
\hline
\textbf{N} & \textbf{2D} & \textbf{3D}           & \textbf{4D}           & \textbf{5D}           & \textbf{6D}           \\ \hline \hline
1,600      & 212.14      & 435.44      & 698.10      & 952.02      & 1,219.46    \\ \hline
3,200      & 282.71      & 622.46      & 1,073.40    & 1,557.88    & 2,074.42    \\ \hline
6,400      & 366.40      & 836.98      & 1,566.64    & 2,451.58    & 3,382.67    \\ \hline
12,800     & 477.65      & 1,139.28    & 2,270.06    & 3,739.39    & 5,450.51    \\ \hline
25,600     & 608.46      & 1,513.72    & 3,178.08    & 5,611.01    & 8,510.70    \\ \hline
51,200     & 782.17      & 1,980.12    & 4,360.25    & 8,262.83    & 13,001.20   \\ \hline
102,400    & 992.31      & 2,575.46    & 5,906.43    & 11,871.50   & 19,817.10   \\ \hline
204,800    & 1,252.96    & 3,317.01    & 7,841.76    & 16,810.30   & 29,527.10   \\ \hline
409,600    & 1,591.09    & 4,240.76    & 10,296.60   & 23,215.30   & 43,297.60   \\ \hline
819,200    & 1,990.32    & 5,406.54    & 13,521.00   & 31,769.40   &             \\ \hline
1,638,400  & 2,514.94    & 6,927.56    &             &             &             \\ \hline
3,276,800  & 3,201.18    & 8,794.71    &             &             &             \\ \hline
6,553,600  & 3,999.04    & 11,263.40   &             &             &             \\ \hline
13,107,200 & 5,067.76    &             &             &             &             \\ \hline
\end{tabular}
        }
        \scalebox{0.85}{%
            \begin{tabular}{|c|c|c|c|c|c|}
\hline
\multicolumn{6}{|c|}{Search Time (ms)} \\
\hline
\textbf{N} & \textbf{2D} & \textbf{3D}           & \textbf{4D}           & \textbf{5D}           & \textbf{6D}           \\ \hline \hline
1,600      & 0.289       & 0.926       & 1.718       & 5.164       & 9.070       \\ \hline
3,200      & 0.357       & 1.199       & 3.165       & 8.746       & 18.928      \\ \hline
6,400      & 0.436       & 1.682       & 6.410       & 20.792      & 34.202      \\ \hline
12,800     & 0.571       & 2.372       & 9.189       & 27.869      & 73.279      \\ \hline
25,600     & 0.945       & 3.319       & 15.835      & 53.142      & 145.327     \\ \hline
51,200     & 1.288       & 4.122       & 22.940      & 74.519      & 151.275     \\ \hline
102,400    & 1.608       & 6.780       & 33.555      & 109.313     & 231.166     \\ \hline
204,800    & 1.607       & 8.223       & 44.257      & 144.389     & 294.095     \\ \hline
409,600    & 2.119       & 8.821       & 69.808      & 189.334     & 504.418     \\ \hline
819,200    & 3.524       & 11.882      & 69.172      & 273.151     &             \\ \hline
1,638,400  & 2.865       & 11.776      &             &             &             \\ \hline
3,276,800  & 3.690       & 16.5714     &             &             &             \\ \hline
6,553,600  & 5.908       & 20.0029     &             &             &             \\ \hline
13,107,200 & 5.873       &             &             &             &             \\ \hline
\end{tabular}
        }
    \end{minipage}
    
    \vspace{0.25cm}
    
    \begin{minipage}{\textwidth}
        \centering
        \scalebox{0.85}{%
            \begin{tabular}{|c|c|c|c|c|c|c|}
\hline
\multicolumn{7}{|c|}{Index Construction Distance Computations} \\
\hline
\textbf{N} & \textbf{N(N-1)/2}  & \textbf{2D}    & \textbf{3D}           & \textbf{4D}           & \textbf{5D}           & \textbf{6D}           \\ \hline \hline
1,600      & 1,279,200          & 265,623        & 516,628        & 760,225       & 1,029,621      & 1,221,605      \\ \hline
3,200      & 5,118,400          & 715,370        & 1,529,411      & 2,545,681     & 3,502,217      & 4,533,864      \\ \hline
6,400      & 20,476,800         & 1,875,159      & 4,237,192      & 7,647,949     & 11,573,618     & 16,048,613     \\ \hline
12,800     & 81,913,600         & 4,914,263      & 11,882,697     & 22,838,137    & 36,807,770     & 54,388,897     \\ \hline
25,600     & 327,667,200        & 12,777,657     & 32,292,006     & 65,019,984    & 112,659,381    & 176,000,348    \\ \hline
51,200     & 1,310,694,400      & 32,927,002     & 84,975,099     & 182,367,161   & 333,940,276    & 545,109,215    \\ \hline
102,400    & 5,242,828,800      & 84,017,423     & 222,786,598    & 499,604,440   & 967,623,862    & 1,653,233,653  \\ \hline
204,800    & 20,971,417,600     & 213,276,161    & 578,454,113    & 1,334,902,118 & 2,724,468,870  & 4,878,923,837  \\ \hline
409,600    & 83,885,875,200     & 545,584,549    & 1,490,406,238  & 3,515,273,754 & 7,508,607,499  & 14,083,342,711 \\ \hline
819,200    & 335,543,910,400    & 1,377,413,839  & 3,818,355,231  & 9,212,801,209 & 20,370,378,025 &                \\ \hline
1,638,400  & 1,342,176,460,800  & 3,497,518,219  & 9,781,461,413  &               &                &                \\ \hline
3,276,800  & 5,368,707,481,600  & 8,902,672,288  & 25,018,301,725 &               &                &                \\ \hline
6,553,600  & 21,474,833,203,200 & 22,597,217,032 & 64,078,700,856 &               &                &                \\ \hline
13,107,200 & 85,899,339,366,400 & 57,317,868,141 &                &               &                &                \\ \hline
\end{tabular}
        }
    \end{minipage}
    
    \vspace{0.25cm}
    
    \begin{minipage}{\textwidth}
        \scalebox{0.75}{%
            \begin{tabular}{|c|c|c|c|c|c|}
\hline
\multicolumn{6}{|c|}{Index Construction Time (hr)} \\
\hline
\textbf{N} & \textbf{2D} & \textbf{3D} & \textbf{4D} & \textbf{5D} & \textbf{6D} \\ \hline  \hline
1,600      & 1.370E-04   & 4.368E-04   & 8.014E-04   & 2.883E-03   & 4.154E-03             \\ \hline
3,200      & 3.161E-04   & 1.171E-03   & 2.966E-03   & 1.314E-02   & 1.711E-02             \\ \hline
6,400      & 7.230E-04   & 2.794E-03   & 1.024E-02   & 3.564E-02   & 6.159E-02             \\ \hline
12,800     & 1.752E-03   & 7.762E-03   & 3.132E-02   & 9.312E-02   & 0.266                 \\ \hline
25,600     & 1.723E-02   & 2.074E-02   & 0.101       & 0.399       & 1.768                 \\ \hline
51,200     & 1.508E-02   & 5.164E-02   & 0.296       & 0.928       & 2.158                 \\ \hline
102,400    & 3.599E-02   & 0.164       & 0.868       & 2.607       & 8.542                 \\ \hline
204,800    & 6.814E-02   & 0.399       & 2.293       & 7.878       & 15.067                \\ \hline
409,600    & 0.178       & 0.999       & 7.774       & 19.018      & 50.068                \\ \hline
819,200    & 0.638       & 2.708       & 13.508      & 48.048      & \multicolumn{1}{l|}{} \\ \hline
1,638,400  & 1.074       & 4.365       &             &             &                       \\ \hline
3,276,800  & 2.416       & 12.095      &             &             &                       \\ \hline
6,553,600  & 8.062       & 26.993      &             &             &                       \\ \hline
13,107,200 & 16.734      &             &             &             &                       \\ \hline
\end{tabular}
        }
        \scalebox{0.75}{%
            \begin{tabular}{|c|c|c|c|c|c|}
\hline
\multicolumn{6}{|c|}{Memory Usage (GB)} \\
\hline
\textbf{N} & \textbf{2D} & \textbf{3D} & \textbf{4D} & \textbf{5D} & \textbf{6D} \\ \hline \hline
1,600      & 6.199E-03   & 8.556E-03   & 1.305E-02   & 2.096E-02   & 2.599E-02   \\ \hline
3,200      & 9.457E-03   & 1.443E-02   & 2.731E-02   & 4.394E-02   & 7.261E-02   \\ \hline
6,400      & 1.579E-02   & 2.674E-02   & 4.927E-02   & 9.863E-02   & 0.154       \\ \hline
12,800     & 2.802E-02   & 4.917E-02   & 0.115       & 0.220       & 0.386       \\ \hline
25,600     & 5.230E-02   & 0.101       & 0.238       & 0.481       & 0.928       \\ \hline
51,200     & 9.154E-02   & 0.186       & 0.469       & 1.044       & 2.165       \\ \hline
102,400    & 0.177       & 0.365       & 0.966       & 2.299       & 5.000       \\ \hline
204,800    & 0.345       & 0.714       & 1.964       & 4.882       & 11.097      \\ \hline
409,600    & 0.680       & 1.400       & 3.951       & 10.228      & 24.772      \\ \hline
819,200    & 1.344       & 2.747       & 7.954       & 21.361      &             \\ \hline
1,638,400  & 2.940       & 5.678       &             &             &             \\ \hline
3,276,800  & 5.851       & 10.669      &             &             &             \\ \hline
6,553,600  & 11.665      & 22.260      &             &             &             \\ \hline
13,107,200 & 23.370      &             &             &             &             \\ \hline
\end{tabular}
        }
    \end{minipage}

\end{table}
    
    \begin{table}[]
    \centering
    \caption{Statistics for Optimal GRNG Hierarchies on Uniformly Distributed Data.}
    
    \begin{minipage}{\textwidth}
        \scalebox{0.85}{%
            \begin{tabular}{|c|c|c|c|c|c|}
\hline
\multicolumn{6}{|c|}{Search Distance Computations} \\
\hline
\textbf{N} & \textbf{2D} & \textbf{3D}           & \textbf{4D}           & \textbf{5D}           & \textbf{6D}           \\ \hline \hline
1,600      & 203.07      & 435.44      & 696.97      & 950.93      & 1,196.06    \\ \hline
3,200      & 260.03      & 622.46      & 1,073.40    & 1,552.94    & 2,055.56    \\ \hline
6,400      & 320.48      & 836.98      & 1,566.64    & 2,451.58    & 3,379.85    \\ \hline
12,800     & 388.54      & 1,135.04    & 2,270.06    & 3,739.39    & 5,454.52    \\ \hline
25,600     & 464.61      & 1,461.50    & 3,178.08    & 5,611.01    & 8,510.70    \\ \hline
51,200     & 541.92      & 1,845.55    & 4,360.25    & 8,262.83    & 13,001.20   \\ \hline
102,400    & 624.96      & 2,314.21    & 5,906.43    & 11,871.50   & 19,817.10   \\ \hline
204,800    & 709.05      & 2,838.78    & 7,841.76    & 16,785.40   & 29,728.20   \\ \hline
409,600    & 799.18      & 3,399.34    & 10,230.80   & 22,788.50   &             \\ \hline
819,200    & 888.07      & 4,009.96    & 12,870.40   &             &             \\ \hline
1,638,400  & 982.24      & 4,702.36    & 15,859.40   &             &             \\ \hline
3,276,800  & 1,072.47    & 5,457.28    &             &             &             \\ \hline
6,553,600  & 1,168.89    & 6,224.37    &             &             &             \\ \hline
13,107,200 & 1,264.24    &             &             &             &             \\ \hline
26,214,400 & 1,359.76    &             &             &             &             \\ \hline
\end{tabular}
        }
        \scalebox{0.85}{%
            \begin{tabular}{|c|c|c|c|c|c|}
\hline
\multicolumn{6}{|c|}{Search Time (ms)} \\
\hline
\textbf{N} & \textbf{2D} & \textbf{3D} & \textbf{4D} & \textbf{5D} & \textbf{6D}           \\ \hline \hline
1,600      & 0.416       & 0.926       & 0.978       & 2.080       & 3.649       \\ \hline
3,200      & 0.550       & 1.199       & 3.165       & 4.752       & 5.987       \\ \hline
6,400      & 1.326       & 1.682       & 6.410       & 20.792      & 19.454      \\ \hline
12,800     & 1.008       & 3.961       & 9.189       & 27.869      & 42.176      \\ \hline
25,600     & 1.510       & 5.804       & 15.835      & 53.142      & 145.327     \\ \hline
51,200     & 1.711       & 7.246       & 22.940      & 74.519      & 151.275     \\ \hline
102,400    & 3.488       & 7.627       & 33.555      & 109.313     & 231.166     \\ \hline
204,800    & 3.837       & 10.204      & 44.257      & 275.756     & 429.461     \\ \hline
409,600    & 4.607       & 20.330      & 87.536      & 357.445     &             \\ \hline
819,200    & 4.155       & 17.256      & 106.133     &             &             \\ \hline
1,638,400  & 4.111       & 20.921      & 132.551     &             &             \\ \hline
3,276,800  & 5.026       & 20.941      &             &             &             \\ \hline
6,553,600  & 5.229       & 33.643      &             &             &             \\ \hline
13,107,200 & 6.217       &             &             &             &             \\ \hline
26,214,400 & 7.024       &             &             &             &             \\ \hline
\end{tabular}
        }
    \end{minipage}
    
    \vspace{0.25cm}
    
    \begin{minipage}{\textwidth}
        \centering
        \scalebox{0.62}{%
\begin{tabular}{|c|c|c|c|c|c|}
\hline
\multicolumn{6}{|c|}{Optimal Number of Layers, $L$} \\
\hline
\textbf{N} & \textbf{2D} & \textbf{3D} & \textbf{4D} & \textbf{5D} & \textbf{6D} \\ \hline \hline
1,600      & 4           & 3           & 2           & 2           & 2           \\ \hline
3,200      & 4           & 3           & 3           & 2           & 2           \\ \hline
6,400      & 5           & 3           & 3           & 3           & 2           \\ \hline
12,800     & 5           & 4           & 3           & 3           & 2           \\ \hline
25,600     & 6           & 4           & 3           & 3           & 3           \\ \hline
51,200     & 6           & 4           & 3           & 3           & 3           \\ \hline
102,400    & 7           & 4           & 3           & 3           & 3           \\ \hline
204,800    & 7           & 4           & 4           & 4           & 3           \\ \hline
409,600    & 8           & 5           & 4           & 4           &             \\ \hline
819,200    & 8           & 5           & 4           &             &             \\ \hline
1,638,400  & 9           & 5           & 4           &             &             \\ \hline
3,276,800  & 9           & 5           &             &             &             \\ \hline
6,553,600  & 9           & 6           &             &             &             \\ \hline
13,107,200 & 10          &             &             &             &             \\ \hline
26,214,400 & 10          &             &             &             &             \\ \hline
\end{tabular}
        }
        \scalebox{0.64}{%
            \begin{tabular}{|c|c|c|c|c|c|c|}
\hline
\multicolumn{7}{|c|}{Index Construction Distance Computations} \\
\hline
\textbf{N} & \textbf{N(N-1)/2}   & \textbf{2D}    & \textbf{3D}    & \textbf{4D}    & \textbf{5D}   & \textbf{6D}   \\ \hline
1,600      & 1,279,200           & 264,911        & 516,628        & 871,615        & 1,265,981     & 1,424,701     \\ \hline
3,200      & 5,118,400           & 685,955        & 1,529,411      & 2,545,681      & 4,380,391     & 5,389,184     \\ \hline
6,400      & 20,476,800          & 1,808,727      & 4,237,192      & 7,647,949      & 11,573,618    & 20,194,814    \\ \hline
12,800     & 81,913,600          & 4,418,773      & 12,527,095     & 22,838,137     & 36,807,770    & 69,953,472    \\ \hline
25,600     & 327,667,200         & 10,999,945     & 32,634,271     & 65,019,984     & 112,659,381   & 176,000,348   \\ \hline
51,200     & 1,310,694,400       & 25,725,358     & 83,754,842     & 182,367,161    & 333,940,276   & 545,109,215   \\ \hline
102,400    & 5,242,828,800       & 61,217,847     & 209,606,677    & 499,604,440    & 967,623,862   & 1,653,233,653 \\ \hline
204,800    & 20,971,417,600      & 138,971,047    & 517,808,692    & 1,334,902,118  & 2,959,923,254 & 5,150,670,019 \\ \hline
409,600    & 83,885,875,200      & 317,976,622    & 1,308,344,233  & 3,942,824,032  & 8,199,558,499 &               \\ \hline
819,200    & 335,543,910,400     & 707,049,296    & 3,074,992,726  & 9,863,795,760  &               &               \\ \hline
1,638,400  & 1,342,176,460,800   & 1,580,489,249  & 7,196,074,192  & 24,264,742,122 &               &               \\ \hline
3,276,800  & 5,368,707,481,600   & 3,451,125,580  & 16,680,031,216 &                &               &               \\ \hline
6,553,600  & 21,474,833,203,200  & 7,495,962,620  & 39,314,844,606 &                &               &               \\ \hline
13,107,200 & 85,899,339,366,400  & 16,340,695,641 &                &                &               &               \\ \hline
26,214,400 & 343,597,370,572,800 & 35,074,351,743 &                &                &               &               \\ \hline
\end{tabular}
        }
    \end{minipage}
    
    \vspace{0.25cm}
    
    \begin{minipage}{\textwidth}
        \scalebox{0.75}{%
            \begin{tabular}{|c|c|c|c|c|c|}
\hline
\multicolumn{6}{|c|}{Index Construction Time (hr)} \\
\hline
\textbf{N} & \textbf{2D} & \textbf{3D} & \textbf{4D} & \textbf{5D} & \textbf{6D} \\ \hline \hline
1,600      & 2.306E-04   & 4.368E-04   & 6.376E-04   & 1.665E-03   & 7.062E-03   \\ \hline
3,200      & 5.619E-04   & 1.171E-03   & 2.966E-03   & 6.863E-03   & 1.566E-02   \\ \hline
6,400      & 3.079E-03   & 2.794E-03   & 1.024E-02   & 3.564E-02   & 9.357E-02   \\ \hline
12,800     & 4.362E-03   & 1.594E-02   & 3.132E-02   & 9.312E-02   & 0.353       \\ \hline
25,600     & 1.378E-02   & 5.056E-02   & 0.101       & 0.399       & 1.768       \\ \hline
51,200     & 3.004E-02   & 0.110       & 0.296       & 0.928       & 2.158       \\ \hline
102,400    & 0.135       & 0.232       & 0.868       & 2.607       & 8.542       \\ \hline
204,800    & 1.025       & 0.585       & 2.293       & 14.793      & 21.961      \\ \hline
409,600    & 1.106       & 2.546       & 14.219      & 39.001      &             \\ \hline
819,200    & 2.664       & 3.936       & 23.547      &             &             \\ \hline
1,638,400  & 2.181       & 10.055      & 58.863      &             &             \\ \hline
3,276,800  & 21.782      & 21.594      &             &             &             \\ \hline
6,553,600  & 37.572      & 58.467      &             &             &             \\ \hline
13,107,200 & 48.889      &             &             &             &             \\ \hline
26,214,400 & 98.506      &             &             &             &             \\ \hline
\end{tabular}
        }
        \scalebox{0.75}{%
            \begin{tabular}{|c|c|c|c|c|c|}
\hline
\multicolumn{6}{|c|}{Memory Usage (GB)} \\
\hline
\textbf{N} & \textbf{2D} & \textbf{3D} & \textbf{4D} & \textbf{5D} & \textbf{6D} \\ \hline \hline
1,600      & 6.969E-03   & 8.556E-03   & 1.033E-02   & 1.785E-02   & 2.692E-02   \\ \hline
3,200      & 1.107E-02   & 1.443E-02   & 2.731E-02   & 3.832E-02   & 6.510E-02   \\ \hline
6,400      & 2.230E-02   & 2.674E-02   & 4.927E-02   & 9.863E-02   & 0.159       \\ \hline
12,800     & 4.011E-02   & 6.742E-02   & 0.115       & 0.220       & 0.345       \\ \hline
25,600     & 9.842E-02   & 0.121       & 0.238       & 0.481       & 0.928       \\ \hline
51,200     & 0.179       & 0.235       & 0.469       & 1.044       & 2.165       \\ \hline
102,400    & 0.407       & 0.453       & 0.966       & 2.299       & 5.000       \\ \hline
204,800    & 0.814       & 0.877       & 1.964       & 6.864       & 12.637      \\ \hline
409,600    & 1.794       & 2.292       & 5.754       & 14.385      &             \\ \hline
819,200    & 3.476       & 4.420       & 11.410      &             &             \\ \hline
1,638,400  & 7.612       & 8.559       & 22.185      &             &             \\ \hline
3,276,800  & 14.775      & 16.697      &             &             &             \\ \hline
6,553,600  & 28.952      & 42.420      &             &             &             \\ \hline
13,107,200 & 67.396      &             &             &             &             \\ \hline
26,214,400 & 132.520     &             &             &             &             \\ \hline
\end{tabular}
        }
    \end{minipage}

\end{table}
\fi

%
\begin{table}[h!]
    \centering
    \caption{\footnotesize Results for real world datasets. (top) Corel, $N=68,040$ in 57D, (middle) MNIST, $N=60,000$ instances with 64D embeddings obtained through a neural network, and (bottom) LA, $N=1,073,727$ instances in 2D. Accuracy is established by comparison to the brute-force construction for the first two datasets, but for the last dataset, both the brute-force method and the algorithm by \textit{Hacid et al.} are impractical to run on a dataset of such size. The accuracy of Rayar \textit{et al.} in this case is found by comparing to our method. The last 100 data points are reserved as a test set for search.}
    
\scalebox{0.8}{%
\begin{tabular}{|c|c|ccccc|}
\hline
\textbf{dataset}                                                          & \textbf{Algorithm} & \multicolumn{1}{c|}{\textbf{Total Links}} &\multicolumn{1}{c|}{\textbf{\begin{tabular}[c]{@{}c@{}}Extra (+) \& \\ Missing (-) Links\end{tabular}}} & \multicolumn{1}{c|}{\textbf{Average Degree}} & \multicolumn{1}{c|}{\textbf{\begin{tabular}[c]{@{}c@{}}Search \\ Distances\end{tabular}}} & \textbf{\begin{tabular}[c]{@{}c@{}}Index Construction \\ Distances\end{tabular}} \\ \hline
\multirow{3}{*}{\begin{tabular}[c]{@{}c@{}}Corel \\ $N=68$k\end{tabular}} & Hacid \textit{et. al}       & \multicolumn{1}{c|}{212,211}              & \multicolumn{1}{c|}{+21,802/-4}                                                                        & \multicolumn{1}{c|}{6.2378}                  & \multicolumn{1}{c|}{177,972.36}                                                           & 9,823,840,198,726                                                                \\ \cline{2-7} 
                                                                          & Rayar \textit{et. al}       & \multicolumn{1}{c|}{190,908}              & \multicolumn{1}{c|}{+535/-40}                                                                          & \multicolumn{1}{c|}{5.6116}                  & \multicolumn{1}{c|}{169,575.08}                                                           & 6,432,673,175                                                                    \\ \cline{2-7} 
                                                                          & Ours               & \multicolumn{1}{c|}{190,413}              & \multicolumn{1}{c|}{+0/-0}                                                                             & \multicolumn{1}{c|}{5.5971}                  & \multicolumn{1}{c|}{43,729.20}                                                            & 1,611,369,217                                                                    \\ \hline
\multirow{3}{*}{\begin{tabular}[c]{@{}c@{}}MNIST \\ $N=60$k\end{tabular}} & Hacid \textit{et. al}       & \multicolumn{1}{c|}{118,248}              & \multicolumn{1}{c|}{+3,778/-3}                                                                         & \multicolumn{1}{c|}{3.9416}                  & \multicolumn{1}{c|}{87,713.10}                                                            & 1,430,022,984,523                                                                \\ \cline{2-7} 
                                                                          & Rayar \textit{et. al}       & \multicolumn{1}{c|}{114,893}              & \multicolumn{1}{c|}{+865/-445}                                                                         & \multicolumn{1}{c|}{3.8298}                  & \multicolumn{1}{c|}{88,172.04}                                                            & 2,639,416,420                                                                    \\ \cline{2-7} 
                                                                          & Ours               & \multicolumn{1}{c|}{114,473}              & \multicolumn{1}{c|}{+0/-0}                                                                             & \multicolumn{1}{c|}{3.8158}                  & \multicolumn{1}{c|}{10,058.90}                                                            & 407,689,553                                                                      \\ \hline
\multirow{3}{*}{\begin{tabular}[c]{@{}c@{}}LA \\ $N=1$M\end{tabular}}     & Hacid \textit{et. al}       & \multicolumn{5}{c|}{Impractical}                                                                                                                                                                                                                                                                                                                                                 \\ \cline{2-7} 
                                                                          & Rayar \textit{et. al}       & \multicolumn{1}{c|}{1,277,369}            & \multicolumn{1}{c|}{+3,254/-33,706}                                                                    & \multicolumn{1}{c|}{2.3793}                  & \multicolumn{1}{c|}{2,147,498.42}                                                         & 1,153,035,099,784                                                                \\ \cline{2-7} 
                                                                          & Ours               & \multicolumn{1}{c|}{1,307,821}            & \multicolumn{1}{c|}{-}                                                                                 & \multicolumn{1}{c|}{2.4360}                  & \multicolumn{1}{c|}{1,020.71}                                                             & 1,042,175,220                                                                    \\ \hline
\end{tabular}
}

    
    \label{tab:results_real}
\end{table}

\section{Experiments} \label{sec:experiments}
Experiments on uniformly distributed and clustered synthetic data in $\mathcal{R}^d$ show the effectiveness of the proposed approach. Note that for all datasets where brute-force is possible, the RNG has been validated for exactness. Figure~\ref{fig:grng_savings_bruteForce}(a) shows that our method is effective in uniformly distributed data and a hierarchy helps, although the optimal number of layers depends on $N$. Figure~Figure~\ref{fig:grng_savings_bruteForce}(b) shows that search is extremely efficient and is essentially logarithmic in $N$. Figure~\ref{fig:results_synthetic} shows that construction costs are exponential in $N$ and dimension $d$ for uniform data (but search remains logarithmic), in contrast to clustered data where both construction and search costs are well-behaved, Figure~\ref{fig:results_synthetic}(c,d). Figure~\ref{fig:results_synthetic}(e) shows that the connectivity of RNG is effectively linear in intrinsic dimension of the data. 

Experiments on several real-world datasets, namely, COREL, MNIST, and LA. For MNIST, a neural network trained using triplet loss was used to reduce the 784D Euclidean representation into 64D. The results are shown in Table \ref{tab:results_real}. These results show that our method is significantly more efficient while also producing the exact RNG. 

%
%
\bibliographystyle{splncs04}
\bibliography{bibliography}

\end{document}